\documentclass[utf8]{frontiersSCNS} 

\usepackage{url,hyperref,lineno,microtype,subcaption}
\usepackage[onehalfspacing]{setspace}

\DeclareMathAlphabet{\mathsc}{OT1}{cmr}{m}{sc}
\def\testbx{bx}%
\DeclareRobustCommand{\ion}[2]{%
\relax\ifmmode
\ifx\testbx\f@series
{\mathbf{#1\,\mathsc{#2}}}\else
{\mathrm{#1\,\mathsc{#2}}}\fi
\else\textup{#1\,{\mdseries\textsc{#2}}}%
\fi}

\newcommand{\arcsec}{\hbox{$^{\prime\prime}$}}

\def\keyFont{\fontsize{8}{11}\helveticabold }
\def\firstAuthorLast{G. Del Zanna {et~al.}} 
\def\Authors{Giulio Del Zanna\,$^{1,*}$, Vincenzo Andretta\,$^{2}$, Peter J. Cargill\,$^{3,4}$,
  Alain J. Corso\,$^{5}$, Adrian N. Daw\,$^{6}$, Leon Golub\,$^{7}$, James A. Klimchuk\,$^{6}$,
  and Helen E. Mason\,$^{1}$}
\def\Address{$^{1}$ DAMTP,  University of Cambridge, Cambridge, UK \\
  $^{2}$ INAF, Osservatorio Astronomico di Capodimonte,  Naples, Italy \\
  $^{3}$ Space and Atmospheric Physics, The Blackett Laboratory, Imperial College, London, UK \\
   $^{4}$ School of Mathematics and Statistics, University of St Andrews, St
Andrews, Fife, KY16 9SS, UK \\
  $^{5}$ National Research Council - Institute for Photonics and Nanotechnologies, Padova, Italy \\
  $^{6}$ NASA Goddard Space Flight Center,  Greenbelt, MD, USA \\
  $^{7}$ Harvard-Smithsonian Center for Astrophysics,  Cambridge, MA, USA 
}
\def\corrAuthor{DAMTP, University of Cambridge, Wilberforce Road, Cambridge CB3 0WA, UK}
\def\corrEmail{gd232@cam.ac.uk}

\begin{document}
\onecolumn
\firstpage{1}

\title[Soft X-rays]{High resolution soft X-ray spectroscopy and the quest for the hot (5--10 MK) plasma in
  solar active regions}

\author[\firstAuthorLast ]{\Authors} 
\address{} 
\correspondance{} 

\extraAuth{}

\maketitle

\begin{abstract}
We discuss the diagnostics available to study  the 5--10 MK plasma in the solar corona, 
which is key to understanding the heating in  the cores of 
solar active regions.
We present  several simulated spectra, and show that
excellent diagnostics are available in the soft X-rays, around 100~\AA,
as six ionisation stages of Fe  can simultaneously be observed, and electron densities
derived, within a narrow spectral region. 
As this spectral range is almost  unexplored,
we present an analysis of available and simulated spectra,
to compare the hot emission with the cooler component.
We adopt recently designed multilayers to present estimates
of count rates in the hot lines, with a baseline spectrometer design.
Excellent count rates are found, opening up the exciting opportunity
to obtain high-resolution spectroscopy of hot plasma.
\tiny
\keyFont{ \section{Keywords:} Techniques: spectroscopy \and  Sun: corona \and Sun: X-rays}
\end{abstract}

\section{Introduction}


The main aim of the present paper is to present  the scientific case
for a soft X-ray (SXR, 90--150~\AA) 
spectrometer with high  resolving power
(capability to measure 5 km s$^{-1}$ Doppler shifts),  high sensitivity and  moderate (1\arcsec) resolution. 
We believe that such an instrument would provide breakthroughs
in understanding various magnetic energy
conversion processes in the solar corona, in particular within:
A) non-flaring active region (AR) cores; B)  flares of all sizes.

As discussed below, the SXR wavelengths contain many 
spectral lines formed over temperatures from 0.1 to 12 MK, and
are particularly suited to measure the physical state of
`hot' 5--10 MK plasma, in particular mass and turbulent flows,
electron densities, departures from ionization equilibrium,
and chemical abundances. 
Such SXR spectroscopic observations of this hot plasma are needed
because:

A) for non-flaring ARs, impulsive heating probably associated with small-scale reconnection
was predicted in the early 1990s to produce  hot emission (see, e.g. \citealt{cargill:94}).
There is some observational
evidence for this (see below), but a systematic study is badly needed.
The intensity and temporal behaviour
of the hot emission can be used to constrain the cadence and energy released in the heating
events, their duration and the background plasma conditions. Such observations will go a long
way to finally resolving the `coronal heating problem'.

B) For microflares and flares, the compelling scientific question is how a large number of
particles can be accelerated. The acceleration is likely to arise as a consequence of magnetic
reconnection in a hot plasma. Various models exist involving strong or weak turbulence,
shocks, collapsing traps, and all are associated with different types of mass motions.
Observations of line shifts and broadenings at high temperatures are required to resolve which
of these are important. Also, the energy transfer and the physical processes
causing chromospheric evaporation during the impulsive phase are not well understood.
High-sensitivity, time-dependent measurements with high spectral
resolution in a range of hot lines are needed.

Despite significant observational and
theoretical advances in the past few decades, the solution 
to these and similar long standing problems related to  plasma heating
in the solar corona 
 remains elusive \citep[see, e.g.][]{klimchuk:06,klimchuk:2015,reale:2014_lr}. 
This is partly because we have been  missing  key spectroscopic observations
of the 5--10 MK  emission:
the vast majority of the solar coronal observations 
have been in the EUV and UV of low $T$ (0.1-3 MK)  plasma  that is 
in the process of  cooling, so essential information about
the energy release process has been lost.
The 5-10 MK temperature regime is largely unexplored in solar physics.
This point was made in two white papers
(written by GDZ and JAK in 2016)  in response to an international call  
to provide suggestions for a Next Generation JAXA/NASA/ESA Solar Physics Mission.

There have been many imaging observations in the  X-rays, such as
the interesting and puzzling 
\ion{Mg}{xii} images, produced by the  CORONAS-I  \citep{sobelman_etal:1996},
 CORONAS-F  \citep[see, e.g.][]{zhitnik_etal:03}, and CORONAS-PHOTON missions \citep{kuzin_etal:2009}.
  There have  also been many spectral observations of Bremsstrahlung emission,
  but with either no spatial
  resolution or with limited  imaging capability and sensitivity.
 Also, the spectral resolution has been typically low, so only the He-like Fe complex
becomes visible during large flares.  
   For example, many  full-Sun X-ray spectra of large flares were produced by the
Solar X-ray Spectrometer (SOXS) Mission, see e.g. \cite{jain_etal:2006}.  
RHESSI \citep{lin_etal:2002} produced many observations but only of larger flares
and with indirect  imaging capability.
Smaller flares were observed by SphinX \citep{sylwester_etal:08} on board the CORONAS-PHOTON mission.
High-resolution full-disk spectra of large flares have been produced by
RESIK \citep{sylwester_etal:2005} on board the the CORONAS-F mission.
\cite{caspi_etal:2015ApJ...802L...2C} presented
 full-Sun X-ray spectra obtained during two sounding rocket flights.
 The spectra had a higher resolution than e.g. RHESSI and SphinX,
 but much less than RESIK, so only a few of the H- and He-like complexes
 are visible in the spectra.
 Similar full-Sun spectra have been produced since 2019 by the
 XSM on board Chandrayaan-2  \citep{xsm_paper}.
 The  Focusing Optics X-ray Solar Imager
(FOXSI) sounding rocket flights \citep[cf][]{krucker_etal:2014},
provided some active region observations with improved sensitivity and direct
imaging, although with low spatial and spectral resolution.
The Nuclear Spectroscopic Telescope ARray (NuSTAR) \citep{harrison_etal:2013}
has provided  excellent observations of weak  Bremsstrahlung emission
from some active regions, but the instrument is not designed
to make regular solar observations as flares could damage it. 

As the above list of observations
demonstrate, to carry out detailed studies of magnetic energy conversion,
high-resolution observations of directly heated 5--10 MK spectral lines are needed.

Mass motions (measured via line profile bulk flows and broadenings) of hot plasma
are a very important diagnostic of impulsive heating.
Also, absolute density measurements are required. This provides information on
plasma fine structure that either influences, or is a consequence of, the reconnection process,
as well as accurate column depths through which accelerated particles must move.

Spectroscopy provides extra spatial information hidden in the line profiles.
For example, as shown by \cite{delzanna_etal:2011_flare},
in the kernels of chromospheric evaporation during the impulsive phase
of a flare, coronal line profiles were a superposition of a stationary
foreground active region component and a blue-shifted component
originating from  a thin layer. This enabled the measurement of not only velocities,
but also electron densities and the thickness of the evaporating layer. 

In general, measurements of hot plasma  have been
very difficult, at its emission is very weak, for a variety of reasons:
1) densities are low if the energy release is in the
corona -- values around  10$^{8}$--10$^{9}$  cm$^{-3}$ could be expected;
2) hot plasma cools rapidly; 
3) non-equilibrium ionisation could reduce the ion populations
\citep[see, e.g.][]{golub_etal:1989SoPh..122..245G,bradshaw_cargill:2006, reale_orlando:2008,dudik_etal:2017_review}.
Measuring non-equilibrium ionisation  requires  observations
 of several successive stages of ionisation of the same element and
 simultaneous measurements of the electron densities in high-$T$ lines,
 to estimate ionisation/recombination timescales.
{ Such measurements have never been obtained, although some information from
a few ions or different elements has been available.
As an example, the ionization or recombination timescales for
\ion{Fe}{xix} at a coronal density of 10$^{8}$  cm$^{-3}$ and 10 MK are
over 300 seconds, a very long time for most of the short-lived events
seen in active regions. Therefore, departures from ionization equilibrium
should be common. At a higher density of 10$^{11}$  cm$^{-3}$, the
timescales are instead about 0.3 seconds. 
}

Finally, measurements of absolute chemical abundances (relative to hydrogen)
are important as their  variability 
could be directly related to the heating processes, as discussed by
\cite{laming:2015}.
As we discuss in this paper, the best diagnostics of 5--10 MK  lines are  to be found in the
soft X-rays, although some are also available in the X-rays (1--50~\AA),
EUV (150--900~\AA), and UV (900--2000~\AA). 

Clearly, to make progress, such
detailed spectral diagnostics need to be  combined
with detailed simulations  (with forward modelling) of the
plasma heating/cooling processes.

The paper is organised as follows. 
Section~2 gives a short review on the requirements and science background,
pointing out some of the limitations of previous, current, and upcoming observations. 
Section~3 presents a 10 MK spectrum and radiances of the main lines,
from the X-rays to the UV. It then discusses some of the pros and cons of the
various spectral ranges, with emphasis on the SXR. Also, it summarises
available density diagnostics of hot lines in the XUV.
 Section~4 presents a straw-man design and estimates of  achievable count rates in the soft X-rays
 adopting newly developed multilayers, for several simulations of hot plasma.
Section~5   draws the  conclusions.
Details of various observations and simulations used to
assess the completeness of the atomic data, line blending and
identifications are given in an extended Supplement.

\section{A short review on the requirements and science background}

As active region cores have a strong emission around 3 MK,
ions such as  \ion{Fe}{xvii},  \ion{Ca}{xvii}, \ion{Fe}{xviii}  are mostly formed
around these temperatures, rather than the temperature of peak
ion abundance in equilibrium. Therefore, they are not necessarily useful by
themselves for  probing  the presence
of hotter plasma within AR cores  \citep[see,e.g.][]{delzanna:2013_multithermal,parenti_etal:2017}.
Lines from higher-$T$ ions need to also be observed.
Also,  as shown e.g. by
\cite{parenti_etal:2006} with a multi-stranded loop simulation,
such hot lines from e.g. \ion{Fe}{xix} or higher ionization stages need to
be observed,  to study the heating.
Lower-$T$ ions such as \ion{Fe}{xv}  are in fact formed during the cooling phase,
and the signatures of the input heating function are completely lost.
Furthermore, the emission comes mostly from evaporated plasma,
not the plasma that was heated directly in the corona.

\subsection{Non-flaring ARs}

The quiescent 3 MK emission in non-flaring ARs 
could be due to a range of processes, involving for example
magnetic reconnection, turbulence,  nanoflare storms
\citep[see, e.g.][and references therein]{cargill:2014},
or  Alfven waves \citep[see, e.g.][]{van_ballegooijen_etal:2011}.
Most theoretical models predict the presence of some 
hot plasma above 3 MK (the `smoking gun'),
 see e.g. \cite{cargill:94,cargill_klimchuk:2004}.
The presence of such
hot plasma has been a matter of much debate in the literature, as its emission is  
very weak and difficult to observe. For a recent list of relevant
references, see e.g. Sect.6.2.5 in \cite{hinode_review:2019}.

The EUV  has excellent diagnostics for lower-temperature
plasma, up to about 4~MK, but only a few `hot'  lines.
For example, the Hinode EIS instrument \citep{culhane_etal:2007} provided excellent
EUV observations of 1--4 MK plasma with e.g. 
\ion{Fe}{xvii},  \ion{Ca}{xvii} lines, but  is
 `blind' in the 5--10 MK range. Only 
the very hot (about 15 MK) flare lines  from \ion{Fe}{xxiii} and \ion{Fe}{xxiv} are
observed \citep[see, e.g.][]{winebarger_etal:2012}.

  There are some studies of the hot emission based on full-disk spectra
  \citep[see, e.g.][]{2010A&A...514A..82S,miceli_etal:2012} 
or imaging \citep[in e.g. Mg XII][]{reva_etal:2018}, but  few  spatially-resolved
 spectroscopic observations of the 5--10 MK plasma in AR cores exist. They were at the limits
of the instruments and provided an unclear picture.
For example, EUNIS-13 observed significant signal in the  \ion{Fe}{xix} 592.2~\AA\ line 
\citep{brosius_etal:2014}, in the core of AR11726.
However, unpublished analysis
of EUNIS-13  observations of  AR11724 and AR11723 (by A.Daw) indicated a
much weaker or no  signal in the same \ion{Fe}{xix} line.

The Solar Maximum Mission  (SMM)  X-ray polychromator \citep{acton_etal:1980}
Flat  Crystal Spectrometers (FCS), had a collimator of about 15\arcsec\ $\times$14\arcsec\
and provided some observations of quiescent ARs, which were analysed by
\cite{delzanna_mason:2014}. It was only possible to
put an upper limit  to the emission measure at 7--10~MK of 
about three orders of magnitude lower than the peak value at 3~MK.

\cite{parenti_etal:2017}    found 
a few places where  faint  \ion{Fe}{xix} emission was observed by SoHO SUMER, but 
it was necessary to integrate for two hours and average spatially to achieve enough signal.
The \ion{Fe}{xix} intensity implied in some places
an emission measure around  2.5--3 orders of magnitude below the peak.
 As SUMER only observed one hot line from \ion{Fe}{xix}, it was not possible to
  establish the temperature (or the distribution of temperatures)
  producing the weak signal in the line, which in principle is formed
 between 6 and 12  MK in ionization equilibrium.

Observations of Bremsstrahlung emission in quiescent ARs with 
FOXSI and NuSTAR have confirmed the FCS and SUMER results, indicating
 little emission at high temperatures, although also in these cases
the actual temperature distribution of the hot plasma could not be established.
For example, \cite{ishikawa_etal:2014} used FOXSI to place
an upper limit in the 4--15 MK range, while the peak emission around 3 MK was constrained by
Hinode XRT and EIS observations.
\cite{hannah_etal:2016} used  NuSTAR observations to also
find upper limits to the temperature distribution between 3 and 12 MK.
The upper limit of the emission measure at 10 MK is about three orders of magnitude
lower than the peak at 3 MK.

\cite{reva_etal:2018} used CORONAS-F/SPIRIT \ion{Mg}{xii} images to estimate
        an upper limit of the emission measure around 10 MK about four orders
        of magnitude lower than the peak value around 3 MK,
        which was constrained by SoHO EIT imaging. However, 
        the information from    \ion{Mg}{xii} is somewhat limited as
        this ion in equilibrium is formed over a very broad temperature range, from 4 to well over 15 MK.
        
On the other hand, some evidence of hot emission 
was found by \cite{marsh_etal:2018ApJ...864....5M}, also using FOXSI and
NuSTAR observations. The lower temperatures were constrained using SDO AIA and
Hinode XRT images.  Nanoflare modelling was able to reproduce in
some cases the FOXSI and NuSTAR observations.

{ 
  To make progress, we need new observations to be combined with the
predictions of nanoflare modelling, see e.g.
\cite{2015ApJ...799..128L,barnes_etal:2016ApJ...829...31B,barnes_etal:2016ApJ...833..217B,athiray_etal:2019}.
Regarding such nanoflare models, it is important to point out that they often 
tend to over-estimate the strength of the hot emission, although 
they can also predict no emission, depending on what assumptions are made
\citep{barnes_etal:2016ApJ...829...31B}.}
Given these  uncertainties (observational and theoretical) on the
strength of the hot emission, we provide below two simulations,
one based on the FCS and SUMER observations, one on the
\cite{2015ApJ...799..128L} simulations, just to show
what signal we might expect to observe. 

{ In summary, to constrain the cadence and energy release in the heating events,
  we require measurements of 
  strong unblended hot lines with a high sensitivity
  (to measure the weak emission) and high spectral resolution (a few km s$^{-1}$). 
  Measurements of the electron density in the hot lines in the
  10$^{8}$-10$^{11}$ cm$^{-3}$ regime would also be needed. 
High spatial  resolution (1\arcsec\ or higher) is  not required 
to characterize the hot emission, as we expect it to be well below
such resolutions, and  as there would
always be several individual loop structures (strands) along the line of sight.

}

\subsection{From  flares to microflares}

{
There is ample literature on observations and models of flares 
of GOES C class and above \citep[see, e.g. the reviews by][]{shibata:2011LRSP,fletcher_etal:2011,benz:2017}.
}
Most observations, however, have been of the flare loops
formed as a by-product of chromospheric evaporation.
It is generally thought that magnetic
reconnection occurs in the corona, but the mechanisms by which
energy is transferred and deposited into the chromosphere are not clear.
Thermal conduction by electrons and non-thermal electrons have been
considered for a long time, but other processes could be at play,
as e.g.  large-scale Alfv{\'e}n waves  \citep{fletcher_hudson:2008}
or high-energy protons.
A significant amount of particles need to be accelerated in the corona,
but how is not clear.  The most interesting but poorly observed
aspects are those related to the reconnection region.
Significant progress has been made on chromospheric evaporation.

{
Very few spectral observations showing strong
Doppler flows in hot lines possibly associated with the magnetic reconnection
region exist, see e.g.
\cite{imada_etal:2013,tian_etal:2014ApJ...797L..14T,warren_etal:2018,polito_etal:2018ApJ...865..161P}.
}
The likely reason is that the emission is weak, because reconnection
is  occurring  in low-density plasma and on spatial scales well below
current resolutions. Also, as we have mentioned, if the plasma is out of
ionization equilibrium, very different line intensities can be expected.
This was shown e.g. by \cite{imada_etal:2011ApJ...742...70I}.
The effects can easily be of one order of magnitude, and depend critically on the local
electron density and the timescale of the heating.

Early X-ray observations (e.g. from Skylab, SMM/UVSP, SMM/BCS)
indicated strong upflows and non-thermal
broadenings during the impulsive phase, but did not provide stigmatic
images. 
The upflows were usually  a weak component, compared to a strong
stationary component, contrary to the predictions from
hydrodynamic modelling. 
Only few spatially-resolved observations from SoHO SUMER
in hot lines (mostly \ion{Fe}{xix} and \ion{Fe}{xxi}) exist,
showing interesting behaviour in the line profiles
\cite[see e.g.][]{kliem_etal:2002} during the peak phase. 

Spatially-resolved observations in \ion{Fe}{xix} from
SoHO CDS  during the impulsive phase of two M-class
flares showed that line profiles
were symmetric and blue-shifted by about 150 km s$^{-1}$, decreasing with time
\citep[][]{brosius:2003,delzanna_etal:06_m1_flare}.
CDS also observed some lines formed around 1-3 MK, which showed
weaker upflows. Non-thermal broadenings  in \ion{Fe}{xix} of about 50 km s$^{-1}$ 
were also decreasing with time, following the upflows. 
The pattern of upflows appeared to follow  the predictions from
hydrodynamic modelling \citep{delzanna_etal:06_m1_flare}.
A few other CDS observations followed. 
Such features are hard to observe as they are short-lived (of the order of 
minutes) and the  \ion{Fe}{xix} intensities are weak, typically a few times up to 
one order of magnitude weaker than the intensities in the post-flare
loops. 

Several  Hinode EIS observations  of chromospheric evaporation have been
published, see e.g.
\cite{2009ApJ...699..968M,delzanna_etal:2011_flare,brosius:2013ApJ...762..133B,young_etal:2013_flare}.
We also have many IRIS \citep{depontieu_etal:2014b} observations of chromospheric evaporation,
but they only included  \ion{Fe}{xxi} for  large (C-class) flares, and
 low-temperature lines. 
EIS  observed 1-3 MK lines and
hot lines from only \ion{Fe}{xxiii} and \ion{Fe}{xxiv}, formed above 10 MK.
Asymmetric  profiles were often observed, which was puzzling.
On the other hand,  IRIS  observations
of \ion{Fe}{xxi} \citep[see e.g.][]{young_etal:2015,polito_etal:2015,brosius_daw:2015}
 have normally  shown  symmetric profiles,  with
 temporal evolutions following the CDS results of \cite{delzanna_etal:06_m1_flare},
 i.e. decreasing upflows and non-thermal widths with time.
Simultaneous observations from EIS ($\simeq$ 3--4\arcsec)  and IRIS ($\simeq$ 0.33\arcsec)
clarified that some of these asymmetries could be due to 
a superposition of different components along the line of sight
\citep{polito_etal:2016a}. 
The kernels of chromospheric evaporation appear in fact to
be small in size, about 1--2\arcsec\ as seen with IRIS and AIA
\citep[see e.g.][]{young_etal:2015}.
A superposition of different flows during long exposure times
could also explain asymmetric line profiles, see e.g. \cite{mandage_bradshaw:2020}.
For larger flares,  high-cadence IRIS observations provide an
indication that a cadence of tens of seconds would be sufficient to
observe the fastest flows at the start of the evaporation
\citep[see, e.g.][]{graham_cauzzi:2015}. 

M-class and X-class flares often show upflows of a few hundreds of km s$^{-1}$ in
hot lines.
Smaller flares, however, have shown weaker upflows.
For example, upflows in \ion{Fe}{xxiii} of only  50 km s$^{-1}$
were observed during the impulsive phase of a B-class flare
\citep{delzanna_etal:2011_flare}. Interestingly, stronger
upflows of about 170 km s$^{-1}$ were seen in \ion{Fe}{xvi} (formed around 3 MK),
before any signal could be seen in \ion{Fe}{xxiii}.
It could well be that stronger upflows were present in the
hot lines but the sensitivity was not sufficient to observe them.

Generally, plasma diagnostics of flares, from the smallest to the bigger events
have  been limited by the lack of observations of lines
formed in the 5--10 MK range 
and  of measurements of electron densities at such temperatures.
Measurements of time-dependent ionization have also been lacking, although some
evidence that non-equilibrium ionization is present durgin flares
has been found \citep[see, e.g.][]{kawate_etal:2016}.

Measurements of hot line profiles in the kernels of chromospheric evaporation
are needed, as well as their temporal evolution during the formation of the
flare loops. We have hydrodynamic models such as
HYDRAD \citep[see, e.g.][and following updates]{bradshaw_mason:03} where we can predict
flows with time-dependent ionization, following energy
deposition in the chromosphere by thermal and non-thermal particles,
but are missing the key observations to constrain the models.

 A statistical study of flares within AR cores using RHESSI  Bremsstrahlung emission
 in the 6--12 keV energy range
 showed peak temperatures of 10-15 MK
 and total estimated energies of 10$^{28}$--10$^{30}$ erg
 \citep[][]{hannah_etal:2008ApJ...677..704H}.
 They were called microflares but were actually  mostly  B- and C-class.
 These measurements typically assume an isothermal plasma,
   because the observations are not generally adequate to distinguish between
   this and a distribution of temperatures, which is more likely.
    That would easily be assessed with
measurements of spectral lines formed at different temperatures.

  Within AR cores, weaker  `microflares', e.g. flares of A-class or below are
a lot more frequent than larger ones. They also have lower temperatures.
This has been clearly shown with recent X-ray
irradiance spectrometers such as SphinX  on board the CORONAS-PHOTON mission 
\citep[see, e.g.][]{mrozek_etal:2018} 
and in 2019/2020 by 
XSM on board Chandrayaan-2. Further, XSM has shown that
microflares occur frequently also outside ARs, and their energies were found to be
in the range 4 $\times$ 10$^{26}$--10$^{28}$ erg \citep{xsm_paper}.
\cite{kirichenko_bogahev:2017} performed a statistical study of
  microflares of GOES class A0.01 to B
  using the SphinX full-Sun X-ray spectra and the
  Mg XII images from CORONAS-PHOTON, showing that they have a different relationship
between X-ray flux and temperatures, compared to larger flares. 

An understanding of the physics of microflares remains elusive, as key
spatially-resolved spectroscopic  observations have been lacking, and given
that they have peak temperatures in the
4--8~MK range
{
 \citep[see, e.g.][]{feldman_etal:1996,mitra-kraev_delzanna:2019,hannah_etal:2019,cooper_etal:2020ApJ...893L..40C,xsm_paper},
which have largely been unexplored by previous and current imaging
spectrometers.
 Consequently, only a few models of microflare loops and associated events
 have been developed \citep[see e.g.][]{testa_reale:2020,joshi_etal:2021}.
}

Some information  has been provided
with imaging spectroscopy of Bremsstrahlung emission with e.g.
NuSTAR  and FOXSI-2.
A recent    NuSTAR observation of a microflare
was published by \cite{cooper_etal:2020ApJ...893L..40C}.
The microflare  was estimated to be approximately
equivalent to a GOES 0.005 A-class flare,  i.e. much weaker  than the
0.1 A class microflares recently observed by  FOXSI-2 \citep{athiray_etal:2020}.
This very weak  NuSTAR event had an energy content of about 10$^{26}$ erg, i.e. close to those
thought to occur in nanoflares, often quoted to be in the range
10$^{23}$--10$^{25}$ erg (although its definition is a bit artificial,
as what really matters is energy per unit cross sectional area).

We note that microflares often are  composed of a few
loop structures which appear resolved  at 1\arcsec\ resolution in
Fe XVIII emission within the AIA 94~\AA\ band
\citep{delzanna:2013_multithermal,delzanna:12_sxr1, mitra-kraev_delzanna:2019}.
Therefore, although higher spatial resolution
could show unresolved structures (if present), 1\arcsec\ resolution would be sufficient
to follow the evolution of the main structures.

  To summarize, we need observations in hot 5--10 MK lines with:
  1) high spectral resolution (to resolve the hot lines from the
  background signal and measure Doppler shifts of the order of 10 km s$^{-1}$ );
  2) high sensitivity (to capture the faint emission during the impulsive phase
  and allow temporal resolutions of tens of seconds);
  3) moderate/high spatial resolution (1\arcsec\  or better);
  4) several ionization stages of the same element;
  5) measurements of electron densities in the  10$^{8}$-10$^{13}$ cm$^{-3}$ range.
  Plus of course a  spatial coverage large enough to observe events.
  As microflares are normally composed of single loops
  with typical lengths of 50~\arcsec, they are easier to observe than
 bigger flares, which can be ten times (or more) larger.

\subsection{Additional considerations}

 A related important science problem is the cycle of
 evaporation and condensation of mass in the corona,
 and in particular in quiescent AR loops. 
 Chromospheric evaporation signatures are expected to be in high-temperature
 lines, short-lived  and  very weak  \citep[see, e.g.][]{patsourakos_klimchuk:09}.
 { Such signatures (enhanced emission in the blue wing) have not been unambiguously observed yet.
 Therefore, also
 in this case high-sensitivity  spectral observations of hot lines are needed.

 An important point to make regards the 'background' cooler emission.
 In the case of non-flaring emission, the intensity and temporal behaviour
 of the hot and  'background' emission will need to be combined with
 forward models.
 For the flaring emission, there is ample evidence
that the cooler lines, e.g. those formed below 3 MK, are mostly not affected 
during the heating and initial cooling phase of an event. The post-flare loops
are seen to be progressively filled in by the hot plasma, and it is only
during the following cooling of the plasma that  lower-temperature
lines are observed, as e.g. shown in the case of a microflare by
\cite{mitra-kraev_delzanna:2019}.
Therefore, with spatially-resolved
spectroscopy, we do not expect the background emission to interfere
with the emission of the 5--10 MK lines.
The situation is more complex and unclear in the kernels of chromospheric evaporation, during the
impulsive phase of a flare. 
Strongly enhanced emission in transition-region lines
\citep[see e.g.][]{testa_etal:2014} or coronal lines \citep[see, e.g.][]{delzanna_etal:2011_flare}
has been observed alongside hot emission. As we have mentioned,
Doppler flows and non-thermal broadenings are also present, so careful
analyses will be required to disentangle any foreground emission for the
cooler lines, and remove any cooler component from the few hot lines
which may become blended.
}

\section{Where are the hot lines and their density diagnostics? }

In order to illustrate where the hot lines fall in the XUV spectrum,
and discuss the pros and cons of the different wavelength ranges,
we present in this Section estimated radiances of a 10 MK plasma.
We used  CHIANTI version 10 \citep{dere_etal:97,chianti_v10}
  and assumed ionisation equilibrium.
We assumed an isothermal emission at $T=$ 10 MK,
 a low  density of 1$\times 10^9$ (cm$^{-3}$), typical
of an active region core, a column emission measure $EM = 10^{25.5}$ (cm$^{-5}$),
 and the  active region core `coronal' abundances
 of \cite{delzanna:2013_multithermal,delzanna_mason:2014}.
 Such emission measure is nearly  three orders of magnitude below
the usual peak emission (around 3 MK) $EM \simeq  10^{28}$ (cm$^{-5}$)
of an active region core, and is representing possible weak
emission caused by nanoflares,
consistent with the SUMER results of \cite{parenti_etal:2017}.

\begin{figure*}[!htbp]
  \centerline{
     \includegraphics[width=16cm, angle=0]{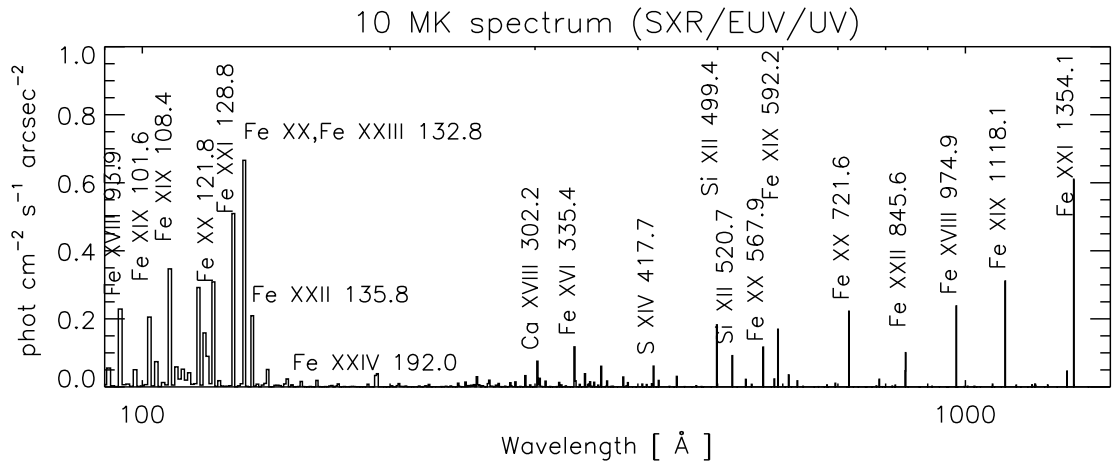} }
\caption{A simulated CHIANTI version 9 isothermal spectrum at 10~MK. 
Note that the spectrum has a bin size of 1~\AA.}
 \label{fig:phot_spectrum}
\end{figure*}

The resulting XUV spectrum 
is shown in Figure~\ref{fig:phot_spectrum}, and a list of the strongest
lines with their radiances is given in Table~\ref{tab:list}. 
 One might think at first sight that there are plenty of strong emission lines.
 However, it turns out that most of them 
 have even stronger contributions from much lower temperatures. 
 An 8~MK spectrum looks similar, with  stronger lower-$T$ lines
 such as \ion{Fe}{xvi},\ion{Fe}{xvii}, \ion{Fe}{xviii}, \ion{Si}{xii},
 and weaker `flare' lines such as \ion{Fe}{xxiv}.
 We highlight in the Table which lines are relatively strong and `hot' 
 and which ones are weak in terms of number of photons emitted. 
It is clear from the Table that the strongest hot lines are in the SXR and in the UV.
These lines   have peak formation temperatures  around 10 MK, but they still have significant
emission at lower and higher temperatures.
We stress that the radiances of the hot lines are only indicative of what might be
  observed. We recall that a temperature or a distribution
  of temperatures could not be established. Higher or lower
  temperature emission could produce the same 
  SUMER \ion{Fe}{xix} intensity, but different radiances
  for the lines from the other ionization stages.
Below we briefly discuss pros and cons of each spectral range.

{
One issue that has often been overlooked is the absorption of
the radiation due to photoionization of neutral H, He and ionised He,
with thresholds at 912, 504, and 228~\AA, plus inner-shell
photoionization of metals.  This absorption is very common in active region
cores as cool material such as filaments is ubiquitous.
As filament activation and heating to 10 MK  is a common feature of
larger flares \citep[see, e.g.][]{dudik_etal:2014}, it would be useful to
have diagnostics not affected by absorption. 
Such absorption  can be substantial
in the EUV/UV  below 912~\AA, is negligible in the X-rays and is much attenuated in the SXR.
The absorption can affect measurement in at least two ways:
attenuating the total intensity, hence reducing the chance of observing already faint
hot emission and by changing the ratios of lines at different
wavelengths, hence affecting e.g. density-sensitive line ratios.
Obviously, the attenuation could be modeled or measured in some circumstances,
depending on the plasma dstribution along the line of sight.
}

\subsection{X-rays}

The X-rays (5--20~\AA) are  rich in spectral lines emitted by hot plasma,
from 3 to 15 MK, although most hot lines are blended,
hence extremely high resolution spectroscopy is needed.
That might not be enough to resolve the main lines whenever large 
non-thermal widths are present, as we have seen from early flare observations.

{
The X-rays are
excellent for measuring the chemical abundances of hot plasma.
They also allow estimates of non-thermal electrons via line
ratio techniques \citep{dudik_etal:2019}, although better diagnostics
are available at shorter wavelengths, involving satellite lines
\citep[see e.g. the review by][]{delzanna_mason:2018}.
The X-rays also provide 
electron density diagnostics but for the hot (10 MK) lines
they are limited to high values (above 10$^{12}$  cm$^{-3}$).
Lines formed below 3 MK are not present in this wavelength range.
}
Previous high-resolution spectroscopy in the X-rays has provided many
important observations of flares and chromospheric evaporation, but has been limited by the lack
of  stigmatic imaging and relatively low sensitivity.
The last  solar spectra in this wavelength range were obtained 
by the  SMM  X-ray polychromator \citep{acton_etal:1980}
Flat  Crystal Spectrometers (FCS), which had a collimator of about 15\arcsec\ $\times$14\arcsec.

The  sounding rocket  MaGIXS (led by A. Winebarger, MSFC, USA; 
see \cite{kobayashi_etal:2011}),  will provide for the first time 
stigmatic imaging spectroscopy in the same spectral region, though
at the expense of a small geometrical area.
The design employs a Wolter-type grazing incidence telescope with mirrors developed 
at MSFC, the same as those used by the successful  FOXSI flights.
One of the limitations of these
focusing mirrors is the moderate  spatial resolution, in the range 5--10\arcsec.

\subsection{Soft X-rays}

Many hot lines from six ionisation stages of iron, from \ion{Fe}{xviii} to \ion{Fe}{xxiii},
are  available in the soft X-rays (SXR) within a narrow (30~\AA) spectral range.
Most readers would be familiar with the  \ion{Fe}{xviii} observed with the
AIA 94~\AA\ band, and  the \ion{Fe}{xxi} in the AIA 131~\AA\ band.
The Table  highlights
those that we consider our `primary' SXR lines,  the strongest
resonance lines from six ionisation stages of iron, plus some density diagnostics.

The SXR spectral region has been used extensively in studies of laboratory
plasma. It also  showed its diagnostic power to perform time-resolved
spectroscopy of  stellar flares  \citep[see, e.g.][]{delzanna_tesi,monsignori_etal:96}.
However, the SXR have been largely unexplored in solar physics.

\begin{figure*}[!htbp]
  \centerline{
     \includegraphics[width=16cm, angle=0]{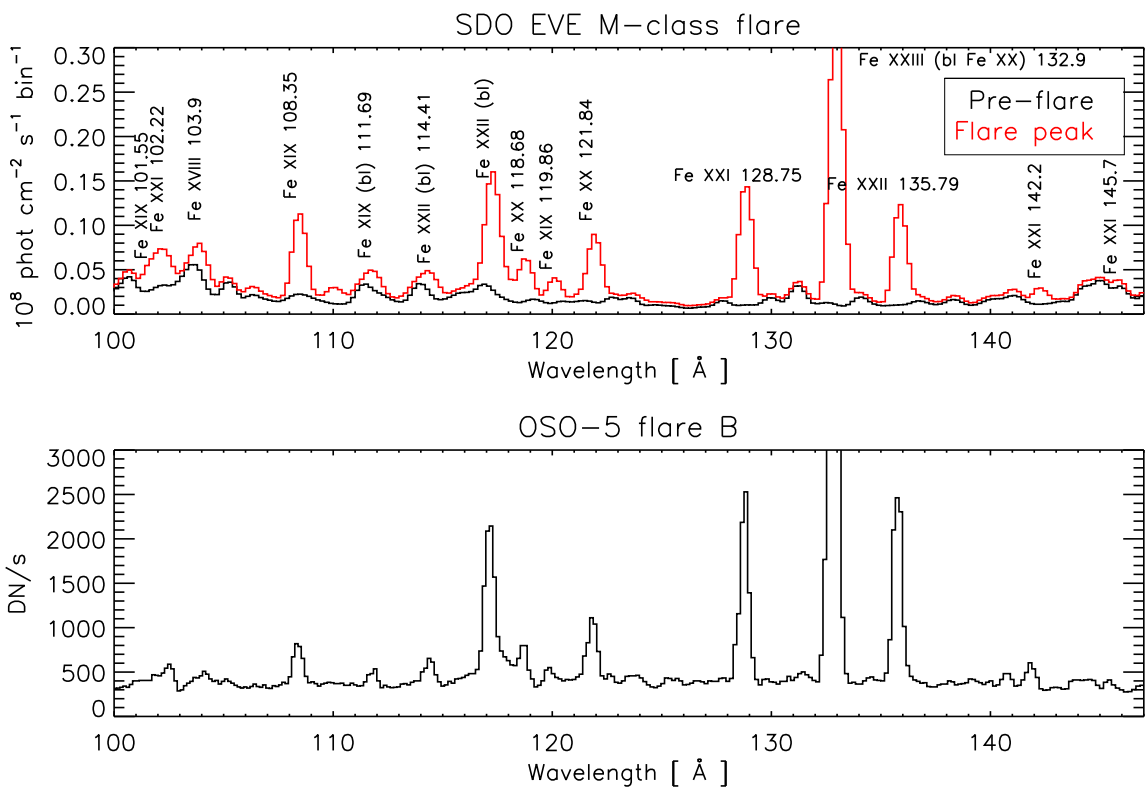} }
  \caption{SDO EVE SXR spectra of an M-class flare (top), and the OSO-5 flare B
    spectrum by \citep{kastner_etal:74}. Note that the OSO-5 spectrum is uncalibrated.}
 \label{fig:sxr}
\end{figure*}

The hot  soft X-ray lines were observed on the Sun for the first time
in 1969 by the grazing incidence instrument on board
OSO-5 \citep{kastner_etal:74}. A portion of their flare B is shown in  Figure~\ref{fig:sxr}.
After OSO-5, the same lines were observed during 2010--2014 by the SDO EVE
MEGS-A spectrometer \citep{woods_etal:12}. Both instruments observed the full Sun.
As significant low-$T$ background emission from the quiet Sun and active regions
is present, these high-$T$ lines are clearly observed only in  EVE spectra of larger flares.
As an example, Figure~\ref{fig:sxr}  shows a portion of the
SXR spectra during the peak phase of an M5-class flare (red) on 2010 Nov 6, with the
pre-flare background spectrum in black.

The SXR also provide excellent electron density diagnostics
around  10$^{11}$  cm$^{-3}$  of high-$T$ plasma,
as  discussed by e.g. \cite{mason_etal:84}.
{The EVE medium (1~\AA) resolution made electron density measurements
  achievable although difficult, 
  as discussed e.g. by \cite{milligan_etal:12_dens,delzanna_woods:2013,keenan_etal:2017}.
  }
OSO-5 and EVE measurements indicate
densities of solar flares
in the range 10$^{11.5}$--10$^{12}$  cm$^{-3}$, as reviewed in \cite{delzanna_mason:2018}.
However, spatially resolved measurements  of 
such a fundamental parameter obtained from high-$T$ spectral lines  have been lacking.

As with the X-rays, the SXR allow line to continuum measurements, i.e.
diagnostics of absolute chemical abundances (relative to hydrogen) of hot plasma  during flares, as shown
e.g. by \cite{warren:2014} using EVE spectra of large flares.
Such measurements are  in principle also available in the UV
\citep{feldman_etal:2003}.

The SXR contain lines from dozens of ions formed from 0.1 to 12 MK,
plus all Iron ionisation stages  from \ion{Fe}{viii} to \ion{Fe}{xxiii}.
There are many
diagnostics for the lower temperatures that are not discussed here,
since this is not the focus of the paper.
Covering temperatures from 0.1 to 12 MK in such a narrow region
is a significant advantage over the other spectral ranges.
We made a significant effort to review all the SXR lines taking into account a revision of
available observations, to indicate which hot lines are likely to be blended
with cooler lines. The main results are summarised in the Supplementary material.

The identifications and atomic data for the cooler ions
in the SXR still require a significant effort. We note that the
strongest solar iron lines in the SXR were only identified recently
by  \cite{delzanna:12_sxr1} when benchmarking against laboratory and
solar data a series of 
large-scale scattering calculations for the SXR. The new 
atomic data and identifications  were  introduced in
CHIANTI version 8 by \cite{delzanna_chianti_v8},
but a large fraction of weaker lines still await firm identifications.
Renewed efforts on measurements of laboratory plasma such as those
of \cite{traebert_etal:2014_131} and on further atomic calculations
should enable us to improve the completeness of  the atomic data.

 Doppler shifts of about 5 km s$^{-1}$ can easily be measured in the
SXR, and large broadenings are relatively easy to measure in the
against the background of cooler lines (which typically retain their
small widths).

The first solar observations of the SXR range with an imaging spectrograph
will be the upcoming flight of the Extreme Ultraviolet Normal Incidence
Spectrograph (EUNIS).
EUNIS will cover the wavelength range
92-112~\AA\  employing a dual multilayer PdB$_4$C stack with
two  reflectivity peaks (of about 0.1) centred on 
the \ion{Fe}{xviii} 94~\AA\ and \ion{Fe}{xix} 108~\AA\ lines, and
a few \AA\ wide.
Note
that multilayer coatings designed specifically to target a few hot lines
can provide higher reflectivities, such as proposed for the Multi-slit
Solar Explorer (MUSE) and in this paper.
MUSE, described in \cite{depontieu_etal:2020}, is designed  to observe the SXR
\ion{Fe}{xix} 108.36~\AA\ and \ion{Fe}{xxi} 108.12\AA\ lines
(in addition to the strong coronal lines from \ion{Fe}{ix} 171 and \ion{Fe}{xv}
284~\AA)
with a large FOV (170\arcsec $\times$ 170\arcsec) and 
a high resolution (0.4\arcsec).
The innovative and multiplexing design of this instrument will allow high
cadence (12s) spectral observations in these four lines through the use of 37 slits
(each spaced about 4.5\arcsec\ apart) covering a region of 170\arcsec\ $\times$170\arcsec.

It is important to note that high spatial resolution can be achieved in the SXR
with normal incidence. This has been shown  by the  
two SXR channels of the
SDO AIA  \citep{lemen_etal:2012} telescopes,  at  94 and 131~\AA. 
Their sensitivity is high, due to good  peak reflectances of  the SXR multilayers
(0.4 and 0.7 respectively, see \citealt{soufli_etal:2005}), although obtained at the expense of
 narrow spectral bands.
 An additional advantage of the SXR is that the Zr front filters,
 employed for  AIA  for the first time in space, 
 have shown minimal in-flight degradation \citep[see, e.g.][]{boerner_etal:2014},
 unlike nearly every filter adopted for EUV/UV instruments flown in space
 \citep[see, e.g.][]{benmoussa_etal:2013,delzanna_mason:2018}.
The main limitation of the SXR for a spectrometer has been the lack of multilayers
with sufficiently high reflectances and wide spectral bands.
A significant improvement has  recently been obtained
by \cite{corso_etal:2020} with  multilayers of 
higher reflectances (0.25--0.4)  at the specific wavelengths of the primary lines selected here.

\subsection{EUV}

The EUV  has excellent diagnostics for lower-temperature
plasma, up to about 4~MK, but there are only few `hot'  lines.
For example, the Hinode EIS instrument  has provided excellent
EUV observations of 1--4 MK plasma, but  is
 `blind' until 
the very hot (about 15 MK) flare lines  from \ion{Fe}{xxiii} and \ion{Fe}{xxiv} are
observed \citep[see, e.g.][]{winebarger_etal:2012}.
These lines, as well as the low-temperature \ion{Fe}{xvii} and \ion{Ca}{xvii} lines
require careful deblending from cooler lines, when their intensity is weak
\citep[see, e.g.][]{young_etal:07a,delzanna:08_bflare,warren_etal:08,delzanna_ishikawa:09,delzanna_etal:2011_flare}. 

Within the EUV, there is  also 
the strong \ion{Ca}{xviii} 302.2~\AA\ line,  observed by e.g. 
Skylab NRL-SO82A  \citep[see, e.g.][]{dere:78} and
the SPIRIT slitless spectroheliograph on board CORONAS-F \citep{shestov_etal:2014}. 
Another important hot line in the EUV is 
 the \ion{Fe}{xix}  592.2~\AA\ line,
which was first observed by the Skylab Harvard SO55 instrument,
then by SoHO CDS, and more recently
by the EUNIS-13 rocket flight \citep{brosius_etal:2014}.

Currently, one advantage of the EUV over the X-rays is  that high spatial resolution can be
achieved with  normal incidence, as shown e.g. with the
 Hi-C sounding rocket, which obtained  a spatial resolution of about
 0.25\arcsec\ \citep{kobayashi_etal:2014}.
 Even higher resolutions are  achievable in the UV, as e.g. shown  by IRIS.

\subsection{UV}

The main UV hot lines are \ion{Fe}{xix}  1118.1~\AA, \ion{Fe}{xx} 721.5~\AA,
and \ion{Fe}{xxi} 1354.1~\AA.
Several flare observations have been obtained 
with high-resolution imaging spectroscopy from e.g. SoHO SUMER
(mostly of \ion{Fe}{xix} and \ion{Fe}{xxi})
and IRIS   (\ion{Fe}{xxi}), see e.g.
\cite{kliem_etal:2002,polito_etal:2015}. \ion{Fe}{xxi} was
also observed earlier by SMM/UVSP. 
The \ion{Fe}{xviii} 974~\AA\ is another  strong line, which was
observed by SUMER  \citep[see, e.g.][]{teriaca_etal:2012},
and is available to  Solar Orbiter SPICE \citep{anderson_etal:2020}.
However, as we have mentioned,  \ion{Fe}{xviii}
has a significant contribution from 3--4 MK emission and so its use
for measuring  high temperatures by itself is  limited.

A significant improvement in terms of sensitivity and resolution over SUMER is the 
EUV Spectroscopic Telescope (EUVST, see \citealt{shimizu_etal:2019}), an M-class mission
recently selected by the Japanese Space Agency (JAXA).
The EUVST has a high throughput and a high spatial resolution of 0.4\arcsec,
with a design  based on the LEMUR instrument
\citep{teriaca_etal:2012_lemur}: the optical components have a 
standard Mo/Si ML for the EUV: 170-–215~\AA, and a B$_4$C top layer
providing good reflectances in three UV bands: 690–-850,
925–-1085, and 1115–-1275~\AA.
The strongest lines in the wavelength regions
463--542 and 557--637~\AA\ would be observed in second order. 

{ EUVST has been designed to address a broad range of science
  questions.
  The key requirement for EUVST is to obtain high-cadence, high-resolution
  observations in spectral lines formed from photospheric to flare temperatures.
  It will also be  able to provide observations of iron lines formed
  in the 5--10 MK range, hence  will be able to provide important
  contributions to the science topics mentioned here.
However,  with the exception of \ion{Fe}{xix}, the lines from the
other iron ions are intrinsically weaker than the resonance SXR lines,
are observed in regions with lower sensitivities, and could be
  affected by photoabsorption in active regions. The planned
EUVST spectral range is
  also  limited in electron density diagnostics (see below).
On the other hand, EUVST is excellent for measuring non-thermal widths and Doppler flows of a 
few km s$^{-1}$ (as it also has photospheric lines to measure
rest wavelengths). 
}

\subsection{Density diagnostics for hot (10 MK) plasma}

 Finally, a few comments about the important issue of measuring electron densities
 from line ratios. There are plenty of diagnostics and measurements across
 different temperatures, as summarised in the review by \cite{delzanna_mason:2018}, but very few 
 for hot (10 MK) plasma.
 We provide here a summary and further details on the hot plasma
diagnostics, which were not all included in the review.
 There are  measurements with SMM FCS from \ion{Fe}{xxi},
 \ion{Fe}{xxii} lines in the X-rays around 9~\AA\ \citep{phillips_etal:1996}, which
 indicated densities of  10$^{13}$  cm$^{-3}$ during a flare. These lines are
 very weak and are difficult to measure though.
Still within the X-rays, there are potentially a few density diagnostic
ratios from  \ion{Fe}{xix}, \ion{Fe}{xxi}, and \ion{Fe}{xxii},
but these involve weak and often blended lines in a spectral
  region that is over-crowded even for the best crystal spectrometers such as
SMM FCS  or the P78-1 SOLEX  \citep{mckenzie_etal:80}. 
 Other density diagnostics involving  satellite lines are
 available, but at  wavelengths shorter than 2~\AA.
The He-like hot ions in the X-rays do provide density diagnostics, but only for 
densities above 10$^{13}$  cm$^{-3}$. At lower temperatures around 4 MK,
Ne He-like lines have indicated densities of 10$^{12}$  cm$^{-3}$ at the start of a flare
\citep{wolfson_etal:1983ApJ...269..319W}.

\begin{figure}[!htbp]
  \centerline{
     \includegraphics[width=8cm, angle=0]{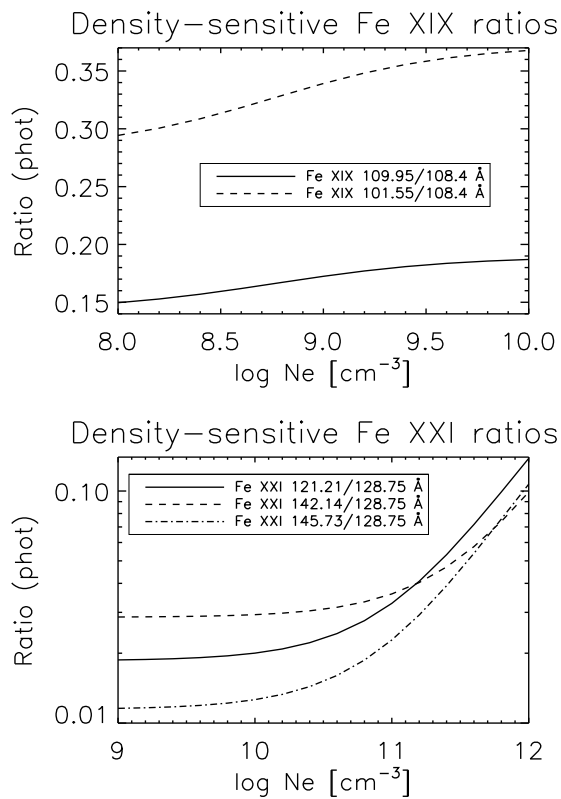} }
\caption{A few of the available SXR density-sensitive ratios. }
 \label{fig:ratios}
\end{figure}

Aside from the X-rays, the only other measurements of densities of hot plasma are those
in the SXR, from OSO-5 \citep{mason_etal:1979,mason_etal:84}
and EVE. These density diagnostics  are well-known and well-studied
in laboratory plasma, see e.g. \cite{stratton_etal:1984}.
The atomic data and identifications
for these $n$=2$\to$ $n$=2 transitions are well established.
The best diagnostic
 ratios are those with the \ion{Fe}{xxi} 102.2, 121.2, 142.2, 145.7~\AA\
 lines vs. the resonance line at 128.75~\AA, as shown in Figure~\ref{fig:ratios}
 (see also Table~1). They provide excellent measurements
 at relatively low densities, above 10$^{10}$  cm$^{-3}$.
 Other ratios involve  \ion{Fe}{xix}, \ion{Fe}{xx}, and \ion{Fe}{xxii} lines.
 There are also interesting ratios involving \ion{Fe}{xix} lines,
 as shown in Figure~\ref{fig:ratios}: they are sensitive to very low densities,
 between 10$^{8}$ and 10$^{10}$ cm$^{-3}$, which would be particularly
 important to investigate further. The variation in the \ion{Fe}{xix} ratios is only 25\%  but measurable,
 especially if the multi-layers are  fine-tuned to increase the signal in the
 weaker density-sensitive lines.
 It would therefore be possible to measure any densities from 10$^{8}$ to
10$^{13}$ cm$^{-3}$ and above, 
observing  both the \ion{Fe}{xix} and \ion{Fe}{xxi}  line ratios.
That would be quite an achievement. 


{
 There are in principle other density diagnostics at longer wavelengths.
 Within the EUV, a good density diagnostic above 10$^{12}$  cm$^{-3}$ available
 at the EUVST wavelengths is the \ion{Fe}{xx}
 567.8/721.5~\AA\ ratio, although it 
could be affected by neutral hydrogen
 absorption in active region observations.
 The 567.8~\AA\ is intrinsically a relatively strong line,
 but will be observed in second order, where the sensitivity is low.
 Also, as the thermal width at 10 MK of the
 567.8 line is 0.17~\AA, the line will be
 blended with the strong \ion{Si}{vii} 1135.4~\AA\ transition.
}

 There are also two other options outside the EUVST spectral range.
 One is the \ion{Fe}{xxi} 786/1354~\AA\
  ratio, useful for densities higher than 10$^{11}$  cm$^{-3}$. 
%
%
The other one, \ion{Fe}{xix}
  1328/1118~\AA,  is an excellent  diagnostic for
  densities of  10$^{11}$  cm$^{-3}$ or lower
  \citep[see, e.g.][]{feldman_etal:2000}, although
  the 1328~\AA\ line is intrinsically very weak: at
  10$^{10}$ cm$^{-3}$, its intensity is only 2\% that of the
  1118~\AA\ line, and decreases with density.
  


\section{Observed and predicted count rates in the SXR}

\subsection{Straw-man design}

To illustrate the current capability for an imaging SXR spectrograph, an example
(straw-man) instrument is presented.
For high collecting area and resolution, the
telescope mirror is a 20 cm diameter off-axis paraboloid with a focal length of
2 m that feeds a spectrograph with a magnification of 1.4,
along the lines of the successful Hinode EIS.   A back-illuminated
CCD (or CMOS array) with 13.5 $\mu$m pixels at a distance of 1.4 m from the grating
provides an image scale of 1\arcsec/pixel.
Reducing the spatial resolution from e.g. the 0.3\arcsec\ of
the original  LEMUR design not only  increases the throughput but also significantly 
simplifies the thermal requirements for a more compact instrument.
Clearly, a higher spatial resolution would be desirable, but would have
to be evaluated in a trade-off study, to ensure that, depending on the
size of the primary, sufficient signal can be obtained to
achieve the science goals of a specific mission.

In a recent  study, \cite{corso_etal:2020} produced
a few new  multilayers (ML), tuned to 
have peak reflectances  in our primary SXR lines.
We adopt two MLs: a  three-fold Mo/Si standard ML for the
126--150~\AA\ range, and a new aperiodic B$_4$C/Y/B$_4$C/Pd for the
100--126~\AA\ range. 
As  the SXR hot primary lines  cannot be observed with a
single multilayer,  we envisage that both the mirror and the grating would be 
segmented in two halves, each with a separate ML, as in the EIS instrument.

The spectral dispersion can be chosen to
use one or two 2048$\times$2048 pixel detectors to cover the wavelength region of
interest.
Holographic gratings with excellent micro-roughness on the scales relevant to
obtaining high SXR reflectivity (0.3 nm on spatial scales of 0.01-1 um) have
been demonstrated with line densities of 4000/mm and a blaze to maximize
throughput (the EUNIS flight grating has a density of 3800 l/mm, the EIS
one 4200 l/mm).
If used in first or second order, this would provide a dispersion
of 12 or 24 m\AA/pixel and a wavelength range of 24 or 49~\AA\ per detector,
respectively.
At the Nyquist limit, that provides a resolving power of R=2500
in first order (comparable to previous EUV spectrographs), or R=5000 in second
order. We select the second option here, and use the grating efficiencies in second order
 calculated by \cite{corso_etal:2020}, which are close to 40\%.
The second order with two 2048 pixel detectors (as the Hinode EIS) provides 
a pixel size of 0.01~\AA, corresponding to about 20 km s$^{-1}$.
This means that Dopplershifts of 2-5 km s$^{-1}$ are measurable.
We note that reconnection and associated flows are expected to be very fast
(on short time scales) so even a 
resolution of a few tens of km s$^{-1}$ for the hot lines could be sufficient.

With a 0.01\AA\  slit, a 0.03\AA\ (or better) spectral  resolution  is achievable.
The thermal FWHM  of the Fe ions in the ranges
6-10 MK and 108-135~\AA\  is 0.025-0.04~\AA, so with such resolution
the thermal width would be resolved.
{ However, considering that
  significant non-thermal widths in the hot lines are likely present, a
  lower spectral  resolution could be sufficient.
In the plots that follow, we have adopted a 0.01~\AA\ pixel resolution,
included  a  thermal width of the lines (using their peak formation temperature),
and added an instrumental FWHM of 0.025~\AA.}

An alternative option would be a
4000/mm grating in first order, which would have higher efficiency (better than 50\%)
and would reduce the spectrograph size by 80\%, at the expense of a  0.02~\AA\ pixel
resolution, which would still be acceptable. 


%
For the  detector, we have assumed an 
efficiency of 0.8, achievable with standard CCDs, such as the 4kx4k thinned
back-illuminated CCDs used by AIA. We note that these CCDs have been
proven very stable and  higher efficiencies (0.85) have been achieved (Hi-C).

We have included a front filter,  the same one used
for the  AIA 94 and 131~\AA\ channels:  a Zr filter, which has 
a  transmission across the soft X-rays of about 0.4.
We note that  such filter includes  a reduction of 15\% due to the supporting mesh,
although   a mesh with a 95\% transmission has been flown aboard
Hi-C \citep[see, e.g.][]{kobayashi_etal:2014}, so better transmission is achievable.
We also note that the inclusion of a safety redundant filter
in front of the detector would reduce by about 50\%
the signal we predict here. An alternative, adopted for the
EUNIS-13 flight, would be not to use a front filter and use a KBr coated
micro-channel plate (MCP) detector (not affected by visible light)
instead of a CCD detector. In this case, the detected signal would be much higher than the
values presented here.

\begin{figure*}[!htbp]
\centerline{\includegraphics[width=16cm, angle=0]{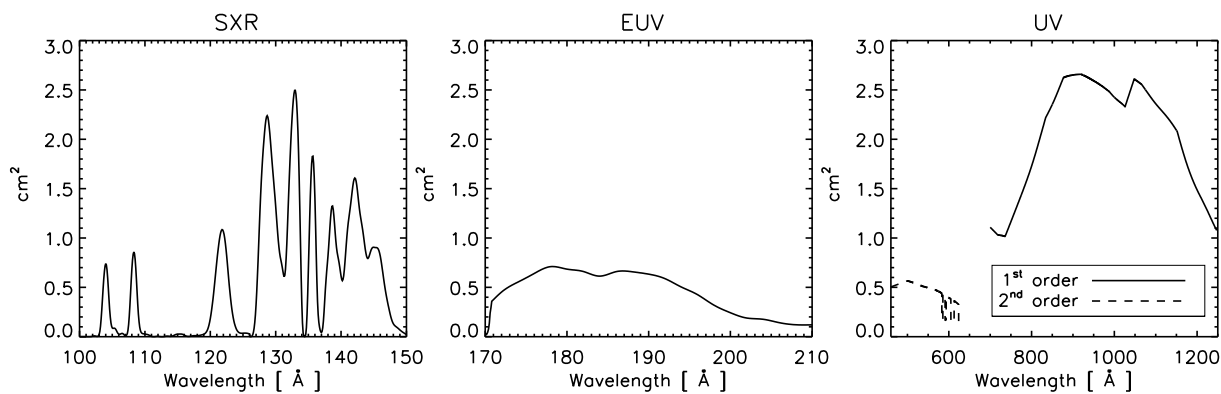}}
\caption{Effective areas for the soft X-ray channels (left),
  with those of a scaled-down version of the LEMUR instrument in the  EUV/UV
  (see text). 
}
\label{fig:effa} 
 \end{figure*}

Fig.~\ref{fig:effa}  shows the resulting effective area, which is the product
of the filter transmission, the reflectivity of the primary,
that of the grating, the grating efficiency, the geometrical area and
the quantum efficiency of the detector.
 For comparison purposes, the same figure also shows the
 LEMUR \citep{teriaca_etal:2012_lemur} effective area in the EUV and UV,
 scaled to the same geometrical area of each SXR channel.
 The figure clearly indicates that similar effective areas are
 achievable in the SXR and UV, for an equivalent aperture. As we have seen that
 the photons emitted by the hot lines in the SXR and UV are comparable,
 this indicates that similar numbers of detected photons  are achievable either in
 the SXR or the UV.
We note however  that the  EUVST baseline design assumes a large primary mirror
with a 28 cm diameter, so its effective area is 
actually about a factor of  four  higher than what is shown in the EUV and UV
panels in  Fig.~\ref{fig:effa}.
We also note that MUSE has a comparable  effective area of 2 cm$^2$,
while MaGIXS has a peak
effective area of 0.03 cm$^2$ \citep{athiray_etal:2019}.

To estimate the signal $S$ detected (data numbers per second, DN/s)
in a pixel we use:
\begin{equation}
S =  I_{\rm r} \, A_{\rm eff} \, {12398.5 \over 3.65 \, \lambda \, G } \, \Omega 
\end{equation}
\noindent
where the  terms convert the number 
of electrons produced in the CCD by a photon of wavelength $\lambda$ (\AA)
into data numbers DN. $I_{\rm r}$ is the incident radiance,
$A_{\rm eff}$ is the effective area,
$G$ is the gain of the camera, and  $\Omega$ is the solid angle subtended by a
pixel. 
We have assumed a gain of 6.3, the same as the EIS CCD.

\subsection{Count rates for the 10 MK emission}

We now return to the 10 MK simulation. The first radiance column
in Table~1 shows that  the \ion{Fe}{xix} 1118.0~\AA\  radiance is  0.23 erg cm$^{-2}$ s$^{-1}$ sr$^{-1}$,
which is very close to the minimum values (0.22--0.5) recorded
(with a total exposure of two hours  and only in some locations) 
by SUMER \citep{parenti_etal:2017}.

The \ion{Fe}{xix} 592.2~\AA\ radiance is 
 0.24  erg cm$^{-2}$ s$^{-1}$ sr$^{-1}$, a value over  20 times  lower than
 that  measured  with EUNIS-13 by \cite{brosius_etal:2014}, but not
 inconsistent with other EUNIS-13 observations, as we have mentioned.

\begin{figure*}[!htbp]
\centerline{\includegraphics[width=16cm, angle=0]{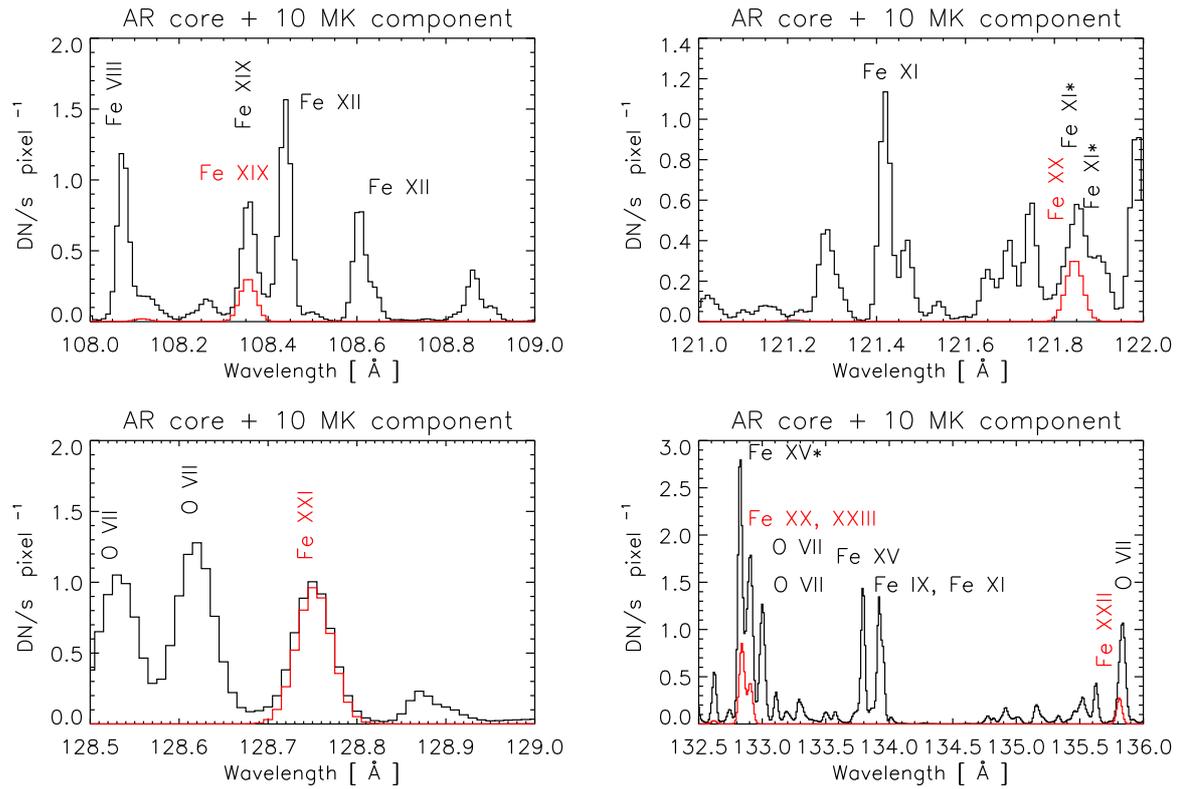}}
\caption{Simulated count rates (1\arcsec\ pixel) for the  emission from 
  a very weak isothermal plasma at 10~MK (red spectrum), added to 
  the emission from a quiescent active region core (black spectrum).
}
\label{fig:sp_10mk} 
\end{figure*}

Fig.~\ref{fig:sp_10mk}  shows the expected SXR count rates for the
very  weak 10 MK emission of  Figure~\ref{fig:phot_spectrum}.  
The 10 MK spectrum has been added to  that of an active region core, 
discussed in the Supplement. Note that the units in the spectra are
per pixel (0.01~\AA) resolution, while  Table~1 provides the total
count rates in the lines.

As we have discussed, a good proxy for the 5--10 MK emission are
\ion{Fe}{xix} and higher ionisation stages.
The  count rates in the
\ion{Fe}{xix}-- \ion{Fe}{xxiii} strongest lines are in the range 1.4--6.5.
Such signals  are easily  measurable with a cadence of a few seconds
and spatial averaging over $\simeq$ 4\arcsec.
This is a major improvement over the two hours exposures and
spatial averaging obtained with SUMER in  \ion{Fe}{xix}
\citep{parenti_etal:2017}.

 As we have mentioned, none of the previous observations of hot emission
  in AR quiescent cores were able to provide an indication of the distribution
  of temperatures above 5 MK. Upper limits in the 4--15 MK range
  have been provided by continuum emission or the \ion{Mg}{xii} images.
  The few \ion{Fe}{xix} observations from SUMER and CDS provide a similarly  unclear picture. 
  So it is possible that the count rates we estimate at 10 MK are either
  an over- or an under-estimate by a large margin. If say they were over-estimated
  by two orders of magnitude,  a measureable signal of $\simeq$ 100 DN/s
could be obtained by an acceptable 
  spatial averaging over $\simeq$ 20\arcsec\ and a two-minute exposure.

The \ion{Fe}{xix} 108.36~\AA, \ion{Fe}{xx}  121.85~\AA,
\ion{Fe}{xxi} 128.75~\AA,  and \ion{Fe}{xxii} 135.81~\AA\
resonance lines are all excellent candidates.
The \ion{Fe}{xix} 108.36~\AA\ has a small contribution from the
quiescent AR core, which is questionable.
The \ion{Fe}{xx}  121.85~\AA\ is super-imposed on an
unidentified weak  \ion{Fe}{xi} transition which 
is currently expected at the same wavelength,
although the quiet Sun spectra suggest that this is not the case
(see the Supplementary material).
As there are many \ion{Fe}{xi} transitions
within the two SXR channels, it would be easy to deblend the  \ion{Fe}{xi}
contribution, if the line was at that wavelength. 

The \ion{Fe}{xxi} 128.75~\AA\ is unblended, while the \ion{Fe}{xxii} 135.81~\AA\
is blended on its red wing with a \ion{O}{vii} 135.83~\AA\ line. The \ion{O}{vii}
can  accurately be estimated measuring other \ion{O}{vii}
transitions, such as the strong \ion{O}{vii} 120.33~\AA\ self-blend.
Finally, the \ion{Fe}{xxiii} could be deblended from the
\ion{O}{vii} and other lines.

\subsection{Count rates from a nanoflare simulation}

\begin{figure*}[!htbp]
\centerline{\includegraphics[width=16cm, angle=0]{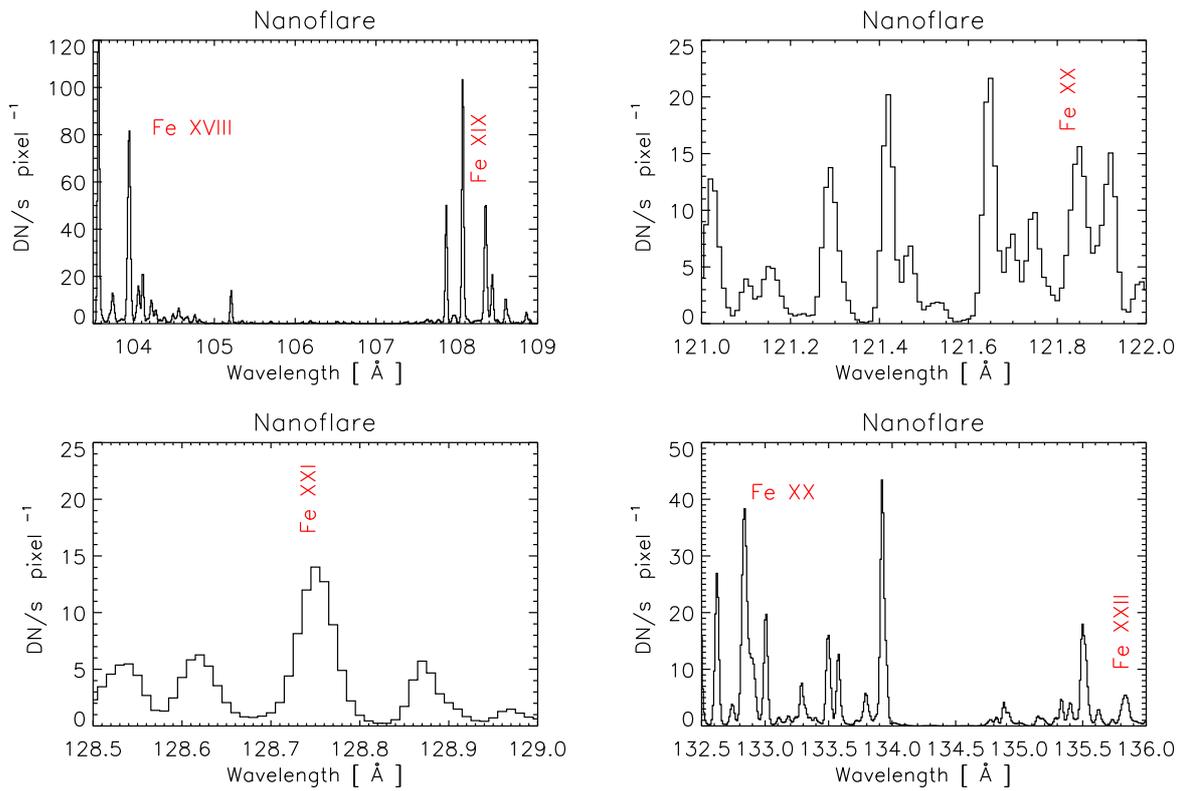}}
\caption{Simulated count rates (1\arcsec\ pixel) for the  emission from 
  a nanoflare simulation \citep{2015ApJ...799..128L}. The main
  hot lines are labelled.
}
\label{fig:sp_nano} 
\end{figure*}

To provide an estimate based on a numerical
nanoflare simulation, we took the 
DEM distribution from the  L = 120 Mm case as described by
\cite{2015ApJ...799..128L}.
This simulation is realistic in that it includes a
variety of magnitudes and frequencies.
The DEM distribution has large values at low temperatures, hence the
simulation also naturally includes all the transition-region lines,
as well as the coronal and the hot lines.
As we would expect low densities around 10$^{8}$  cm$^{-3}$
for the hotter plasma and higher ones (10$^{9}$  cm$^{-3}$) for the coronal lines,
we have adopted a constant pressure of
10$^{15.5}$ cm$^{-3}$ K for the simulated spectra.
Fig.~\ref{fig:sp_nano} and Table~1 show the expected SXR count rates.
There is clearly a very strong signal in all the hot primary lines
(i.e. from \ion{Fe}{xviii}--\ion{Fe}{xxiii}).

\subsection{Count rates for a sub-A class microflare, additional broadening and densities}

\begin{figure*}[!htbp]
\centerline{\includegraphics[width=16cm, angle=0]{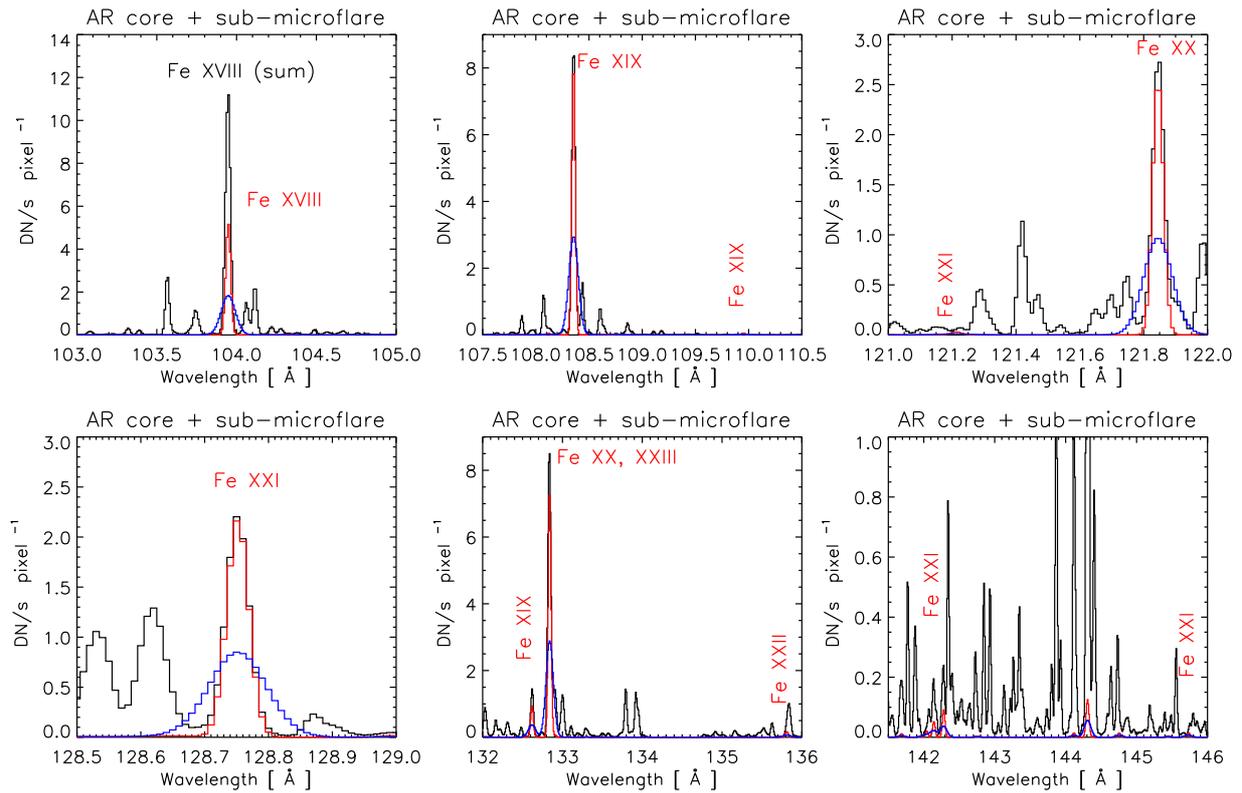}}
\caption{Simulated SXR count rates (1\arcsec\ pixel) 
  for a sub-A class microflare of 6.7 MK (red spectrum).
  The black spectrum is the sum of the
microflare and active region core.
The blue spectrum shows the hot emission from the sub-A class microflare
but with an additional  broadening of 200 km s$^{-1}$ FWHM.
}
\label{fig:sp_submflare} 
 \end{figure*}

To estimate the signal for a very weak microflare, we consider the
0.005 A-class  NuSTAR observations discussed by \cite{cooper_etal:2020ApJ...893L..40C}.
The NuSTAR X-ray Bremsstrahlung emission was well fitted with an
isothermal emission of 6.7 MK.  The signal in the AIA 94~\AA\ band, due to \ion{Fe}{xviii},
was at the limit of detection. 
We converted  the volume EM
to a column emission measure  $EM = 1.18  \times $10$^{27}$ (cm$^{-5}$).
Using our 'coronal' abundances we expect about 3 DN/s/pixel
due to \ion{Fe}{xviii} in the AIA 94~\AA\ band, close to what was observed.
Fig.~\ref{fig:sp_submflare} shows sections of our predicted
SXR count rates for a density of 10$^{11}$  cm$^{-3}$,
added to those of the active region core 
discussed in the Supplement. The total count rates are given in
Table~1.

There is plenty of signal in all the resonance lines.
They are not affected by the 'background' emission, with the
exception of \ion{Fe}{xxii}, which is very weak given such a low
temperature, and \ion{Fe}{xviii}, as significant quiescent
emission is expected to be present.
Time-dependent ionisation on a 1\arcsec spatial scale could be studied
with exposure times of 1--10s.

The Figure also shows the effect of an additional
broadening of 200 km s$^{-1}$ FWHM, which would be easily measurable for the main
\ion{Fe}{xviii}, \ion{Fe}{xix}, \ion{Fe}{xx}, and \ion{Fe}{xxi} lines,
as no significant background emission from cooler lines is present.
Much stronger broadenings would become difficult to observe, for
this very weak hot emission.

Regarding density measurements, the comparisons with irradiance
spectra shown in the Supplement indicate that the
\ion{Fe}{xix} 119.98~\AA\ and  \ion{Fe}{xxi} 121.21, 142.2~\AA\ density-sensitive
lines fall in regions relatively free of background emission.
The \ion{Fe}{xxi}  145.73~\AA, on the other hand, is close to a \ion{Ni}{x}
line and would require background subtraction.
The expected count rates indicate that even for this
extremely weak microflare, densities could be measured with
exposure times of about 100 seconds  at 1\arcsec\ resolution.

\subsection{Count rates for an A class microflare}

As representative of a weak A class microflare, we have considered
an isothermal plasma emission at 8 MK, an
electron density of $N_{\rm e}$=10$^{11}$  cm$^{-3}$,
the  \cite{delzanna:2013_multithermal} coronal abundances and
an emission measure $EM = 10^{29}$ cm$^{-5}$.
The microflare radiances are listed in Table~\ref{tab:list}.
These parameters  were chosen so as to reproduce
 the \ion{Fe}{xvii} and \ion{Ca}{xvii} radiances during
 the peak emission of a  microflare  recorded by EIS \citep{mitra-kraev_delzanna:2019}.
 We note that the actual peak temperature of the post-flare loops was about 5 MK,
 although short-lived  higher temperatures were probably present during
 the impulsive phase in one of the footpoints
\citep[see also][]{testa_reale:2020}.
Using the effective areas of the instruments, we found out that 
 such a weak 8~MK  microflare  would  be invisible to EIS in
 \ion{Fe}{xxiii} and at the  limit of detection  for 
IRIS in the \ion{Fe}{xxi} 1354.1~\AA\ line
(we estimate 3 IRIS  DN/s in the line). On the other hand, it  
would have been well observed by the SDO AIA SXR bands, with 
about 200 DN/s in the 
\ion{Fe}{xviii} 94~\AA\ channel and 126 DN/s in \ion{Fe}{xxi}
within the 131~\AA\ band.

\begin{figure*}[!htbp]
\centerline{\includegraphics[width=16cm, angle=0]{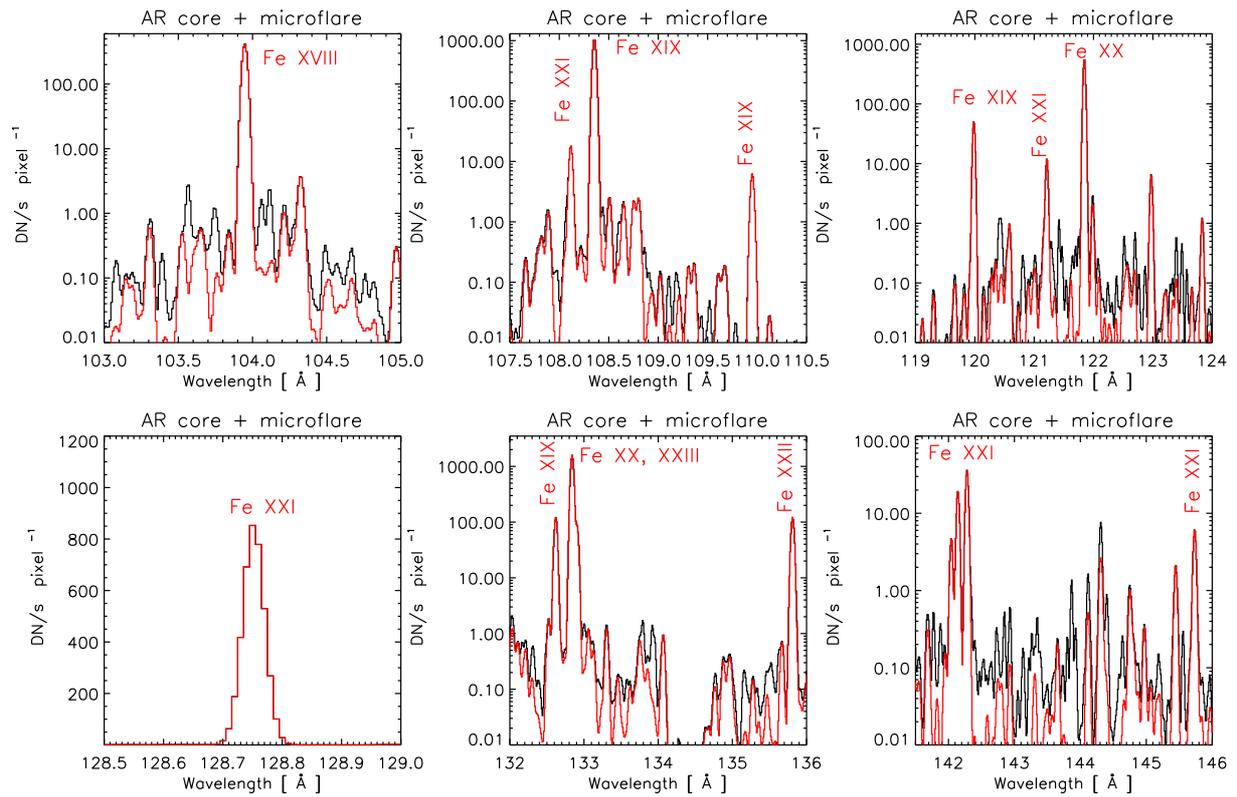}}
\caption{Simulated SXR count rates (1\arcsec\ pixel) 
  for an 8 MK A-class  microflare (red spectrum), added to those of 
  an active region core. The black spectrum is the sum of the two.  
}
\label{fig:sp_mflare} 
 \end{figure*}

Fig.~\ref{fig:sp_mflare} shows the simulated SXR count rates (1\arcsec\ pixel) 
for the  microflare case study, added to those of the active region core 
discussed in the Supplement. 
The SXR channel provides very large count rates in all the primary lines,
as  listed in Table~1. 
For example, the \ion{Fe}{xxi} 128.75~\AA\ resonance line
would produce 3.0$\times$10$^{3}$ DN/s.
The \ion{Fe}{xix} and \ion{Fe}{xxi}  density-sensitive lines are also strong, 
providing excellent density measurements
with very short timescales of a second or so at  1\arcsec\ resolution. 
Larger (e.g. B-class) flares would require sub-second exposure times to avoid saturation.

The significant differences with the spectra of the
sub-A class microflare are due not just to the increased emission measure,
but also to the increased temperature, from 6.7 to 8 MK.

\section{Conclusions}

Understanding a range of heating/cooling events in active regions and probing
for the presence of time-dependent ionisation requires high-resolution line
spectroscopy of 5--10 MK plasma, with observations of multiple ionisation
stages of an element and simultaneous observations of the electron densities.
This has never been achieved in solar physics. There are plenty of detailed
observations of cooler or hotter plasma, but very few around these
temperatures.

{%
  In this paper, we have demonstrated that the soft X-ray around 100~\AA\ is
  the best range to carry out such investigations, since it provides, in a
  relatively narrow wavelength range of $\sim$50 \AA, six ionisation stages of
  iron to probe 5--10 MK plasma, together with excellent density diagnostics
  above 10$^{10}$ cm$^{-3}$, plus some above 10$^{8}$ cm$^{-3}$.
  %
As this spectral region is practically unexplored, } we
have provided here and in the Supplementary material an overview of the main
spectral features for different solar conditions, from the quiet Sun to active
regions, nanoflares and microflares. The primary SXR hot lines are very
strong, close in wavelength and very sensitive to temperatures in the 5--12 MK
range.

We have presented estimated count rates with a straw-man imaging spectrometer,
similar in size to the successful Hinode EIS instrument.  The technology for a
soft X-ray instrument is mature. With the exception of the new multilayers,
which would have to be fabricated and tested, all the components are standard,
have flown on previous missions, and have proven to be long-lived.
However, as discussed in \cite{corso_etal:2020}, multilayers of the type
adopted here have already been fabricated and have shown good stability to thermal
effects and over time, see e.g. \cite{windt_gullikson:2015}.

We have shown that effective areas comparable to those in the UV (for
equivalent telescope aperture) can be achieved with high spatial resolutions
of 1\arcsec\ or better.  Time-dependent ionisation, heating and cooling cycles
can be studied at such resolutions with a cadence of seconds for a wide range
of sub A-class microflares. Flows can be studied with a few km s$^{-1}$
resolutions. The unresolved hot emission expected to result from nanoflares
can be studied with a cadence of a few seconds with spatial averaging.

{%
  The present concept for a SXR spectrometer is designed to demonstrate the
potential for  discovery in  that largely unexplored wavelength range, with the
primary science goal of understanding the physics of
 the hot (5-10 MK) plasma in active
  regions.  As such, this straw-man instrument is complementary to other
  future or proposed missions designed to address, for instance, the top-level
  science objective of the formation mechanisms of the hot and dynamic outer
  solar atmosphere, as described in the report of the Next Generation Solar
  Physics Mission Science Objectives Team, NGSPM-SOT.
}

{%
  Once the first SXR observations from EUNIS and further laboratory
  observations  become available, it will be possible to confirm the
  present predictions, thus allowing to perform trade-off studies for specific
  science goals.  In particular, the
}
multi-layers
adopted here were designed to maximize the signal in the primary resonance
lines in the six ionisation stages of iron, but could be adjusted to
increase the signal in the much weaker density-sensitive lines.
{%
  Also, a scaled-down version of the proposed straw-man design would
  result in a very compact and cost-effective instrument,
producing novel observations in this unexplored region.
}

In summary, the 5-10 MK temperature regime is a largely unexplored discovery
space, precisely where magnetic energy conversion is occurring.
High-resolution spectroscopy in this regime can be expected to provide
breakthroughs.  Although in this paper we have focused on the almost
unexplored 5--10 MK plasma emission, the SXR instrument is also sensitive to a
broad range of temperatures, from 0.1 to 5 MK, with many diagnostics not
discussed here. Also, the proposed SXR instrument is sensitive to larger
(B-class and over) flares and higher temperatures, up to around 15--20 MK with
\ion{Fe}{xxiii} and even higher with the continuum emission, with timescales
much shorter than a second.

\section*{Conflict of Interest Statement}

The authors declare that the research was conducted in the absence of any commercial or financial relationships that could be construed as a potential conflict of interest.

\section*{Author Contributions}

GDZ wrote the article and produced the figures
but received contributions to the text from all the co-authors.
AJC also provided the reflectivities of the multilayers and the grating.
AND (PI of EUNIS-13) also contributed to the figures for the
straw-man design. JAK also provided the results of the nanoflare simulations.

\section*{Funding}

GDZ and HEM acknowledge support from STFC (UK) via the consolidated grants 
to the atomic astrophysics group (AAG) at DAMTP,
University of Cambridge (ST/P000665/1. and ST/T000481/1).
The work of JAK was supported by the GSFC Internal Scientist Funding Model (competitive work package) program.
AND acknowledges support through NASA Heliophysics awards 13-HTIDS13 2-0074
and 16-HTIDS16\_2-0064.

\section*{Acknowledgments}

We would like to thank Bart De Pontieu (USA),  Vanessa Polito (USA), and Luca Teriaca (Germany)
for providing useful comments on the manuscript, as well as the reviewers.

\section*{Supplemental Data}

The supplementary material presents an analysis of
quiet Sun SXR spectra, and predicted spectra for the
quiet Sun and quiescent active region cores, to show
the expected `background'  emission.




\newcommand\aj{{AJ}}%
\newcommand\araa{{ARA\&A}}%
\newcommand\apj{{ApJ}}%
\newcommand\apjl{{ApJ}}%
\newcommand\apjs{{ApJS}}%
\newcommand\ao{{Appl.~Opt.}}%
\newcommand\apss{{Ap\&SS}}%
\newcommand\aap{{A\&A}}%
\newcommand\aapr{{A\&A~Rev.}}%
\newcommand\aaps{{A\&AS}}%
\newcommand\azh{{AZh}}%
\newcommand\baas{{BAAS}}%
\newcommand\jrasc{{JRASC}}%
\newcommand\memras{{MmRAS}}%
\newcommand\mnras{{MNRAS}}%
\newcommand\pra{{Phys.~Rev.~A}}%
\newcommand\prb{{Phys.~Rev.~B}}%
\newcommand\prc{{Phys.~Rev.~C}}%
\newcommand\prd{{Phys.~Rev.~D}}%
\newcommand\pre{{Phys.~Rev.~E}}%
\newcommand\prl{{Phys.~Rev.~Lett.}}%
\newcommand\pasp{{PASP}}%
\newcommand\pasj{{PASJ}}%
\newcommand\qjras{{QJRAS}}%
\newcommand\skytel{{S\&T}}%
\newcommand\solphys{{Sol.~Phys.}}%
\newcommand\sovast{{Soviet~Ast.}}%
\newcommand\ssr{{Space~Sci.~Rev.}}%
\newcommand\zap{{ZAp}}%
\newcommand\nat{{Nature}}%
\newcommand\iaucirc{{IAU~Circ.}}%
\newcommand\aplett{{Astrophys.~Lett.}}%
\newcommand\apspr{{Astrophys.~Space~Phys.~Res.}}%
\newcommand\bain{{Bull.~Astron.~Inst.~Netherlands}}%
\newcommand\fcp{{Fund.~Cosmic~Phys.}}%
\newcommand\gca{{Geochim.~Cosmochim.~Acta}}%
\newcommand\grl{{Geophys.~Res.~Lett.}}%
\newcommand\jcp{{J.~Chem.~Phys.}}%
\newcommand\jgr{{J.~Geophys.~Res.}}%
\newcommand\jqsrt{{J.~Quant.~Spec.~Radiat.~Transf.}}%
\newcommand\memsai{{Mem.~Soc.~Astron.~Italiana}}%
\newcommand\nphysa{{Nucl.~Phys.~A}}%
\newcommand\physrep{{Phys.~Rep.}}%
\newcommand\physscr{{Phys.~Scr}}%
\newcommand\planss{{Planet.~Space~Sci.}}%
\newcommand\procspie{{Proc.~SPIE}}%


\bibliographystyle{frontiersinSCNS_ENG_HUMS} 


\bibliography{paper}

\begin{thebibliography}{13}
\providecommand{\natexlab}[1]{#1}
\expandafter\ifx\csname urlstyle\endcsname\relax
  \providecommand{\doi}[1]{doi:\discretionary{}{}{}#1}\else
  \providecommand{\doi}{doi:\discretionary{}{}{}\begingroup
  \urlstyle{rm}\Url}\fi
\providecommand{\selectlanguage}[1]{\relax}
\providecommand{\bibAnnoteFile}[1]{%
  \IfFileExists{#1}{\begin{quotation}\noindent\textsc{Key:} #1\\
  \textsc{Annotation:}\ \input{#1}\end{quotation}}{}}
\providecommand{\bibAnnote}[2]{%
  \begin{quotation}\noindent\textsc{Key:} #1\\
  \textsc{Annotation:}\ #2\end{quotation}}

\bibitem[{{Andretta} and {Del Zanna}(2014)}]{andretta_delzanna:2014}
{Andretta}, V. and {Del Zanna}, G. (2014).
\newblock {The EUV spectrum of the Sun: SOHO CDS NIS radiances during solar
  cycle 23}.
\newblock \emph{\aap} 563, A26.
\newblock \doi{10.1051/0004-6361/201322841}
\bibAnnoteFile{andretta_delzanna:2014}

\bibitem[{{Andretta} et~al.(2003){Andretta}, {Del Zanna}, and
  {Jordan}}]{andretta_etal:03}
{Andretta}, V., {Del Zanna}, G., and {Jordan}, S.~D. (2003).
\newblock {The EUV helium spectrum in the quiet Sun: A by-product of coronal
  emission?}
\newblock \emph{\aap} 400, 737--752
\bibAnnoteFile{andretta_etal:03}

\bibitem[{{Behring} et~al.(1972){Behring}, {Cohen}, and
  {Feldman}}]{behring_etal:72}
{Behring}, W.~E., {Cohen}, L., and {Feldman}, U. (1972).
\newblock {The Solar Spectrum: Wavelengths and Identifications from 60 TO 385
  Angstroms}.
\newblock \emph{\apj} 175, 493--+
\bibAnnoteFile{behring_etal:72}

\bibitem[{{Beiersdorfer} et~al.(2014){Beiersdorfer}, {Lepson}, {Desai},
  {D{\'\i}az}, and {Ishikawa}}]{beiersdorfer_etal:2014_procyon}
{Beiersdorfer}, P., {Lepson}, J.~K., {Desai}, P., {D{\'\i}az}, F., and
  {Ishikawa}, Y. (2014).
\newblock {New Identifications of Fe IX, Fe X, Fe XI, Fe XII, and Fe XIII Lines
  in the Spectrum of Procyon Observed with the Chandra X-Ray Observatory}.
\newblock \emph{\apjs} 210, 16.
\newblock \doi{10.1088/0067-0049/210/2/16}
\bibAnnoteFile{beiersdorfer_etal:2014_procyon}

\bibitem[{Del~Zanna(2012)}]{delzanna:12_sxr1}
Del~Zanna, G. (2012).
\newblock Benchmarking atomic data for astrophysics: a first look at the soft
  x-ray lines.
\newblock \emph{\aap} 546, A97.
\newblock \doi{10.1051/0004-6361/201219923}
\bibAnnoteFile{delzanna:12_sxr1}

\bibitem[{Del~Zanna(2013)}]{delzanna:2013_multithermal}
Del~Zanna, G. (2013).
\newblock The multi-thermal emission in solar active regions.
\newblock \emph{\aap} 558, A73.
\newblock \doi{10.1051/0004-6361/201321653}
\bibAnnoteFile{delzanna:2013_multithermal}

\bibitem[{Del~Zanna et~al.(2011)Del~Zanna, Mitra-Kraev, Bradshaw, Mason, and
  Asai}]{delzanna_etal:2011_flare}
Del~Zanna, G., Mitra-Kraev, U., Bradshaw, S.~J., Mason, H.~E., and Asai, A.
  (2011).
\newblock The 22 may 2007 b-class flare: new insights from hinode observations.
\newblock \emph{\aap} 526, A1.
\newblock \doi{10.1051/0004-6361/201014906}
\bibAnnoteFile{delzanna_etal:2011_flare}

\bibitem[{{Lepson} et~al.(2002){Lepson}, {Beiersdorfer}, {Brown}, {Liedahl},
  {Utter}, {Brickhouse} et~al.}]{lepson_etal:2002}
{Lepson}, J.~K., {Beiersdorfer}, P., {Brown}, G.~V., {Liedahl}, D.~A., {Utter},
  S.~B., {Brickhouse}, N.~S., et~al. (2002).
\newblock {Emission Lines of Fe VII-Fe X in the Extreme Ultraviolet Region,
  60-140 {\AA}}.
\newblock \emph{\apj} 578, 648--656.
\newblock \doi{10.1086/342274}
\bibAnnoteFile{lepson_etal:2002}

\bibitem[{{Malinovsky} and {Heroux}(1973)}]{malinovsky_heroux:73}
{Malinovsky}, L. and {Heroux}, M. (1973).
\newblock {An Analysis of the Solar Extreme-Ultraviolet Between 50 and 300 A}.
\newblock \emph{\apj} 181, 1009--1030
\bibAnnoteFile{malinovsky_heroux:73}

\bibitem[{{Manson}(1972)}]{manson:72}
{Manson}, J.~E. (1972).
\newblock {Measurements of the Solar Spectrum between 30 and 128 {\AA}}.
\newblock \emph{\solphys} 27, 107--129.
\newblock \doi{10.1007/BF00151774}
\bibAnnoteFile{manson:72}

\bibitem[{{Mitra-Kraev} and {Del Zanna}(2019)}]{mitra-kraev_delzanna:2019}
{Mitra-Kraev}, U. and {Del Zanna}, G. (2019).
\newblock {Solar microflares: a case study on temperatures and the Fe XVIII
  emission}.
\newblock \emph{\aap} 628, A134.
\newblock \doi{10.1051/0004-6361/201834856}
\bibAnnoteFile{mitra-kraev_delzanna:2019}

\bibitem[{{Tr{\"a}bert} et~al.(2014){Tr{\"a}bert}, {Beiersdorfer},
  {Brickhouse}, and {Golub}}]{traebert_etal:2014_131}
{Tr{\"a}bert}, E., {Beiersdorfer}, P., {Brickhouse}, N.~S., and {Golub}, L.
  (2014).
\newblock {High-resolution Laboratory Spectra on the {$\lambda$}131 Channel of
  the AIA Instrument On Board the Solar Dynamics Observatory}.
\newblock \emph{\apjs} 211, 14.
\newblock \doi{10.1088/0067-0049/211/1/14}
\bibAnnoteFile{traebert_etal:2014_131}

\bibitem[{{Woods} et~al.(2009){Woods}, {Chamberlin}, {Harder}, {Hock}, {Snow},
  {Eparvier} et~al.}]{woods_etal:09}
{Woods}, T.~N., {Chamberlin}, P.~C., {Harder}, J.~W., {Hock}, R.~A., {Snow},
  M., {Eparvier}, F.~G., et~al. (2009).
\newblock {Solar Irradiance Reference Spectra (SIRS) for the 2008 Whole
  Heliosphere Interval (WHI)}.
\newblock \emph{\grl} 36, 1101--+.
\newblock \doi{10.1029/2008GL036373}
\bibAnnoteFile{woods_etal:09}

\end{thebibliography}


\begin{thebibliography}{122}
\providecommand{\natexlab}[1]{#1}
\expandafter\ifx\csname urlstyle\endcsname\relax
  \providecommand{\doi}[1]{doi:\discretionary{}{}{}#1}\else
  \providecommand{\doi}{doi:\discretionary{}{}{}\begingroup
  \urlstyle{rm}\Url}\fi
\providecommand{\selectlanguage}[1]{\relax}
\providecommand{\bibAnnoteFile}[1]{%
  \IfFileExists{#1}{\begin{quotation}\noindent\textsc{Key:} #1\\
  \textsc{Annotation:}\ \input{#1}\end{quotation}}{}}
\providecommand{\bibAnnote}[2]{%
  \begin{quotation}\noindent\textsc{Key:} #1\\
  \textsc{Annotation:}\ #2\end{quotation}}

\bibitem[{{Acton} et~al.(1980){Acton}, {Finch}, {Gilbreth}, {Culhane},
  {Bentley}, {Bowles} et~al.}]{acton_etal:1980}
{Acton}, L.~W., {Finch}, M.~L., {Gilbreth}, C.~W., {Culhane}, J.~L., {Bentley},
  R.~D., {Bowles}, J.~A., et~al. (1980).
\newblock {The soft X-ray polychromator for the Solar Maximum Mission}.
\newblock \emph{\solphys} 65, 53--71.
\newblock \doi{10.1007/BF00151384}
\bibAnnoteFile{acton_etal:1980}

\bibitem[{{Anderson} et~al.(2019){Anderson}, {Appourchaux}, {Auch{\`e}re},
  {Aznar Cuadrado}, {Barbay}, {Baudin} et~al.}]{anderson_etal:2020}
{Anderson}, M., {Appourchaux}, T., {Auch{\`e}re}, F., {Aznar Cuadrado}, R.,
  {Barbay}, J., {Baudin}, F., et~al. (2019).
\newblock {The Solar Orbiter SPICE instrument -- An extreme UV imaging
  spectrometer}.
\newblock \emph{A\&A} \doi{https://doi.org/10.1051/0004-6361/201935574}
\bibAnnoteFile{anderson_etal:2020}

\bibitem[{{Athiray} et~al.(2020){Athiray}, {Vievering}, {Glesener}, {Ishikawa},
  {Narukage}, {Buitrago-Casas} et~al.}]{athiray_etal:2020}
{Athiray}, P.~S., {Vievering}, J., {Glesener}, L., {Ishikawa}, S.-n.,
  {Narukage}, N., {Buitrago-Casas}, J.~C., et~al. (2020).
\newblock {FOXSI-2 Solar Microflares. I. Multi-instrument Differential Emission
  Measure Analysis and Thermal Energies}.
\newblock \emph{\apj} 891, 78.
\newblock \doi{10.3847/1538-4357/ab7200}
\bibAnnoteFile{athiray_etal:2020}

\bibitem[{{Athiray} et~al.(2019){Athiray}, {Winebarger}, {Barnes}, {Bradshaw},
  {Savage}, {Warren} et~al.}]{athiray_etal:2019}
{Athiray}, P.~S., {Winebarger}, A.~R., {Barnes}, W.~T., {Bradshaw}, S.~J.,
  {Savage}, S., {Warren}, H.~P., et~al. (2019).
\newblock {Solar Active Region Heating Diagnostics from High-temperature
  Emission Using the MaGIXS}.
\newblock \emph{\apj} 884, 24.
\newblock \doi{10.3847/1538-4357/ab3eb4}
\bibAnnoteFile{athiray_etal:2019}

\bibitem[{{Barnes} et~al.(2016{\natexlab{a}}){Barnes}, {Cargill}, and
  {Bradshaw}}]{barnes_etal:2016ApJ...829...31B}
{Barnes}, W.~T., {Cargill}, P.~J., and {Bradshaw}, S.~J. (2016{\natexlab{a}}).
\newblock {Inference of Heating Properties from ``Hot'' Non-flaring Plasmas in
  Active Region Cores. I. Single Nanoflares}.
\newblock \emph{\apj} 829, 31.
\newblock \doi{10.3847/0004-637X/829/1/31}
\bibAnnoteFile{barnes_etal:2016ApJ...829...31B}

\bibitem[{{Barnes} et~al.(2016{\natexlab{b}}){Barnes}, {Cargill}, and
  {Bradshaw}}]{barnes_etal:2016ApJ...833..217B}
{Barnes}, W.~T., {Cargill}, P.~J., and {Bradshaw}, S.~J. (2016{\natexlab{b}}).
\newblock {Inference of Heating Properties from ``Hot'' Non-flaring Plasmas in
  Active Region Cores. II. Nanoflare Trains}.
\newblock \emph{\apj} 833, 217.
\newblock \doi{10.3847/1538-4357/833/2/217}
\bibAnnoteFile{barnes_etal:2016ApJ...833..217B}

\bibitem[{{BenMoussa} et~al.(2013){BenMoussa}, {Gissot}, {Sch{\"u}hle}, {Del
  Zanna}, {Auch{\`e}re}, {Mekaoui} et~al.}]{benmoussa_etal:2013}
{BenMoussa}, A., {Gissot}, S., {Sch{\"u}hle}, U., {Del Zanna}, G.,
  {Auch{\`e}re}, F., {Mekaoui}, S., et~al. (2013).
\newblock {On-Orbit Degradation of Solar Instruments}.
\newblock \emph{\solphys} 288, 389--434.
\newblock \doi{10.1007/s11207-013-0290-z}
\bibAnnoteFile{benmoussa_etal:2013}

\bibitem[{{Benz}(2017)}]{benz:2017}
{Benz}, A.~O. (2017).
\newblock {Flare Observations}.
\newblock \emph{Living Reviews in Solar Physics} 14, 2.
\newblock \doi{10.1007/s41116-016-0004-3}
\bibAnnoteFile{benz:2017}

\bibitem[{{Boerner} et~al.(2014){Boerner}, {Testa}, {Warren}, {Weber}, and
  {Schrijver}}]{boerner_etal:2014}
{Boerner}, P.~F., {Testa}, P., {Warren}, H., {Weber}, M.~A., and {Schrijver},
  C.~J. (2014).
\newblock {Photometric and Thermal Cross-calibration of Solar EUV Instruments}.
\newblock \emph{\solphys} 289, 2377--2397.
\newblock \doi{10.1007/s11207-013-0452-z}
\bibAnnoteFile{boerner_etal:2014}

\bibitem[{{Bradshaw} and {Cargill}(2006)}]{bradshaw_cargill:2006}
{Bradshaw}, S.~J. and {Cargill}, P.~J. (2006).
\newblock {Explosive heating of low-density coronal plasma}.
\newblock \emph{\aap} 458, 987--995.
\newblock \doi{10.1051/0004-6361:20065691}
\bibAnnoteFile{bradshaw_cargill:2006}

\bibitem[{{Bradshaw} and {Mason}(2003)}]{bradshaw_mason:03}
{Bradshaw}, S.~J. and {Mason}, H.~E. (2003).
\newblock {A self-consistent treatment of radiation in coronal loop modelling}.
\newblock \emph{\aap} 401, 699--709.
\newblock \doi{10.1051/0004-6361:20030089}
\bibAnnoteFile{bradshaw_mason:03}

\bibitem[{{Brosius}(2003)}]{brosius:2003}
{Brosius}, J.~W. (2003).
\newblock {Chromospheric Evaporation and Warm Rain during a Solar Flare
  Observed in High Time Resolution with the Coronal Diagnostic Spectrometer
  aboard the Solar and Heliospheric Observatory}.
\newblock \emph{\apj} 586, 1417--1429.
\newblock \doi{10.1086/367958}
\bibAnnoteFile{brosius:2003}

\bibitem[{{Brosius}(2013)}]{brosius:2013ApJ...762..133B}
{Brosius}, J.~W. (2013).
\newblock {Chromospheric Evaporation in Solar Flare Loop Strands Observed with
  the Extreme-ultraviolet Imaging Spectrometer on Board Hinode}.
\newblock \emph{\apj} 762, 133.
\newblock \doi{10.1088/0004-637X/762/2/133}
\bibAnnoteFile{brosius:2013ApJ...762..133B}

\bibitem[{{Brosius} and {Daw}(2015)}]{brosius_daw:2015}
{Brosius}, J.~W. and {Daw}, A.~N. (2015).
\newblock {Quasi-periodic Fluctuations and Chromospheric Evaporation in a Solar
  Flare Ribbon Observed by IRIS}.
\newblock \emph{\apj} 810, 45.
\newblock \doi{10.1088/0004-637X/810/1/45}
\bibAnnoteFile{brosius_daw:2015}

\bibitem[{{Brosius} et~al.(2014){Brosius}, {Daw}, and
  {Rabin}}]{brosius_etal:2014}
{Brosius}, J.~W., {Daw}, A.~N., and {Rabin}, D.~M. (2014).
\newblock {Pervasive Faint Fe XIX Emission from a Solar Active Region Observed
  with EUNIS-13: Evidence for Nanoflare Heating}.
\newblock \emph{\apj} 790, 112.
\newblock \doi{10.1088/0004-637X/790/2/112}
\bibAnnoteFile{brosius_etal:2014}

\bibitem[{{Cargill}(1994)}]{cargill:94}
{Cargill}, P.~J. (1994).
\newblock {Some implications of the nanoflare concept}.
\newblock \emph{\apj} 422, 381--393.
\newblock \doi{10.1086/173733}
\bibAnnoteFile{cargill:94}

\bibitem[{{Cargill}(2014)}]{cargill:2014}
{Cargill}, P.~J. (2014).
\newblock {Active Region Emission Measure Distributions and Implications for
  Nanoflare Heating}.
\newblock \emph{\apj} 784, 49.
\newblock \doi{10.1088/0004-637X/784/1/49}
\bibAnnoteFile{cargill:2014}

\bibitem[{{Cargill} and {Klimchuk}(2004)}]{cargill_klimchuk:2004}
{Cargill}, P.~J. and {Klimchuk}, J.~A. (2004).
\newblock {Nanoflare Heating of the Corona Revisited}.
\newblock \emph{\apj} 605, 911--920.
\newblock \doi{10.1086/382526}
\bibAnnoteFile{cargill_klimchuk:2004}

\bibitem[{{Caspi} et~al.(2015){Caspi}, {Woods}, and
  {Warren}}]{caspi_etal:2015ApJ...802L...2C}
{Caspi}, A., {Woods}, T.~N., and {Warren}, H.~P. (2015).
\newblock {New Observations of the Solar 0.5-5 keV Soft X-Ray Spectrum}.
\newblock \emph{\apjl} 802, L2.
\newblock \doi{10.1088/2041-8205/802/1/L2}
\bibAnnoteFile{caspi_etal:2015ApJ...802L...2C}

\bibitem[{{Cooper} et~al.(2020){Cooper}, {Hannah}, {Grefenstette}, {Glesener},
  {Krucker}, {Hudson} et~al.}]{cooper_etal:2020ApJ...893L..40C}
{Cooper}, K., {Hannah}, I.~G., {Grefenstette}, B.~W., {Glesener}, L.,
  {Krucker}, S., {Hudson}, H.~S., et~al. (2020).
\newblock {NuSTAR Observation of a Minuscule Microflare in a Solar Active
  Region}.
\newblock \emph{\apjl} 893, L40.
\newblock \doi{10.3847/2041-8213/ab873e}
\bibAnnoteFile{cooper_etal:2020ApJ...893L..40C}

\bibitem[{{Corso} et~al.(2020){Corso}, {Del Zanna}, and
  {Polito}}]{corso_etal:2020}
{Corso}, A., {Del Zanna}, G., and {Polito}, V. (2020).
\newblock {}.
\newblock \emph{Exp. Astron.}
\bibAnnoteFile{corso_etal:2020}

\bibitem[{{Culhane} et~al.(2007){Culhane}, {Harra}, {James}, {Al-Janabi},
  {Bradley}, {Chaudry} et~al.}]{culhane_etal:2007}
{Culhane}, J.~L., {Harra}, L.~K., {James}, A.~M., {Al-Janabi}, K., {Bradley},
  L.~J., {Chaudry}, R.~A., et~al. (2007).
\newblock {The EUV Imaging Spectrometer for Hinode}.
\newblock \emph{\solphys} , 60--+\doi{10.1007/s01007-007-0293-1}
\bibAnnoteFile{culhane_etal:2007}

\bibitem[{{De Pontieu} et~al.(2020){De Pontieu}, {Mart{\'\i}nez-Sykora},
  {Testa}, {Winebarger}, {Daw}, {Hansteen} et~al.}]{depontieu_etal:2020}
{De Pontieu}, B., {Mart{\'\i}nez-Sykora}, J., {Testa}, P., {Winebarger}, A.~R.,
  {Daw}, A., {Hansteen}, V., et~al. (2020).
\newblock {The Multi-slit Approach to Coronal Spectroscopy with the Multi-slit
  Solar Explorer (MUSE)}.
\newblock \emph{\apj} 888, 3.
\newblock \doi{10.3847/1538-4357/ab5b03}
\bibAnnoteFile{depontieu_etal:2020}

\bibitem[{{De Pontieu} et~al.(2014){De Pontieu}, {Title}, {Lemen}, {Kushner},
  {Akin}, {Allard} et~al.}]{depontieu_etal:2014b}
{De Pontieu}, B., {Title}, A.~M., {Lemen}, J.~R., {Kushner}, G.~D., {Akin},
  D.~J., {Allard}, B., et~al. (2014).
\newblock {The Interface Region Imaging Spectrograph (IRIS)}.
\newblock \emph{\solphys} 289, 2733--2779.
\newblock \doi{10.1007/s11207-014-0485-y}
\bibAnnoteFile{depontieu_etal:2014b}

\bibitem[{{Del Zanna}(1995)}]{delzanna_tesi}
{Del Zanna}, G. (1995).
\newblock Ph.D. thesis, Univ.\ of Florence, Italy
\bibAnnoteFile{delzanna_tesi}

\bibitem[{{Del Zanna}(2008)}]{delzanna:08_bflare}
{Del Zanna}, G. (2008).
\newblock {Flare lines in Hinode EIS spectra}.
\newblock \emph{\aap} 481, L69--L72.
\newblock \doi{10.1051/0004-6361:20079033}
\bibAnnoteFile{delzanna:08_bflare}

\bibitem[{Del~Zanna(2012)}]{delzanna:12_sxr1}
Del~Zanna, G. (2012).
\newblock Benchmarking atomic data for astrophysics: a first look at the soft
  x-ray lines.
\newblock \emph{\aap} 546, A97.
\newblock \doi{10.1051/0004-6361/201219923}
\bibAnnoteFile{delzanna:12_sxr1}

\bibitem[{Del~Zanna(2013)}]{delzanna:2013_multithermal}
Del~Zanna, G. (2013).
\newblock The multi-thermal emission in solar active regions.
\newblock \emph{\aap} 558, A73.
\newblock \doi{10.1051/0004-6361/201321653}
\bibAnnoteFile{delzanna:2013_multithermal}

\bibitem[{{Del Zanna} et~al.(2006){Del Zanna}, {Berlicki}, {Schmieder}, and
  {Mason}}]{delzanna_etal:06_m1_flare}
{Del Zanna}, G., {Berlicki}, A., {Schmieder}, B., and {Mason}, H.~E. (2006).
\newblock {A Multi-Wavelength Study of the Compact M1 Flare on October 22,
  2002}.
\newblock \emph{\solphys} 234, 95--113.
\newblock \doi{10.1007/s11207-006-0016-6}
\bibAnnoteFile{delzanna_etal:06_m1_flare}

\bibitem[{{Del Zanna} et~al.(2020){Del Zanna}, {Dere}, {Young}, and
  {Landi}}]{chianti_v10}
{Del Zanna}, G., {Dere}, K.~P., {Young}, P.~R., and {Landi}, E. (2020).
\newblock {CHIANTI - An atomic database for emission lines. Version 10}.
\newblock \emph{ApJ}
\bibAnnoteFile{chianti_v10}

\bibitem[{{Del Zanna} et~al.(2015){Del Zanna}, {Dere}, {Young}, {Landi}, and
  {Mason}}]{delzanna_chianti_v8}
{Del Zanna}, G., {Dere}, K.~P., {Young}, P.~R., {Landi}, E., and {Mason}, H.~E.
  (2015).
\newblock {CHIANTI - An atomic database for emission lines. Version 8}.
\newblock \emph{\aap} 582, A56.
\newblock \doi{10.1051/0004-6361/201526827}
\bibAnnoteFile{delzanna_chianti_v8}

\bibitem[{{Del Zanna} and {Ishikawa}(2009)}]{delzanna_ishikawa:09}
{Del Zanna}, G. and {Ishikawa}, Y. (2009).
\newblock {Benchmarking atomic data for astrophysics: Fe XVII EUV lines}.
\newblock \emph{\aap} 508, 1517--1526.
\newblock \doi{10.1051/0004-6361/200911729}
\bibAnnoteFile{delzanna_ishikawa:09}

\bibitem[{{Del Zanna} and {Mason}(2014)}]{delzanna_mason:2014}
{Del Zanna}, G. and {Mason}, H.~E. (2014).
\newblock {Elemental abundances and temperatures of quiescent solar active
  region cores from X-ray observations}.
\newblock \emph{\aap} 565, A14.
\newblock \doi{10.1051/0004-6361/201423471}
\bibAnnoteFile{delzanna_mason:2014}

\bibitem[{{Del Zanna} and {Mason}(2018)}]{delzanna_mason:2018}
{Del Zanna}, G. and {Mason}, H.~E. (2018).
\newblock Xuv spectroscopy.
\newblock \emph{Living Reviews in Solar Physics} 15
\bibAnnoteFile{delzanna_mason:2018}

\bibitem[{Del~Zanna et~al.(2011)Del~Zanna, Mitra-Kraev, Bradshaw, Mason, and
  Asai}]{delzanna_etal:2011_flare}
Del~Zanna, G., Mitra-Kraev, U., Bradshaw, S.~J., Mason, H.~E., and Asai, A.
  (2011).
\newblock The 22 may 2007 b-class flare: new insights from hinode observations.
\newblock \emph{\aap} 526, A1.
\newblock \doi{10.1051/0004-6361/201014906}
\bibAnnoteFile{delzanna_etal:2011_flare}

\bibitem[{Del~Zanna and Woods(2013)}]{delzanna_woods:2013}
Del~Zanna, G. and Woods, T.~N. (2013).
\newblock Spectral diagnostics with the sdo eve flare lines.
\newblock \emph{\aap} 555, A59.
\newblock \doi{10.1051/0004-6361/201220988}
\bibAnnoteFile{delzanna_woods:2013}

\bibitem[{{Dere}(1978)}]{dere:78}
{Dere}, K.~P. (1978).
\newblock {Spectral lines observed in solar flares between 171 and 630
  angstroms}.
\newblock \emph{\apj} 221, 1062--1067
\bibAnnoteFile{dere:78}

\bibitem[{{Dere} et~al.(1997){Dere}, {Landi}, {Mason}, {Monsignori Fossi}, and
  {Young}}]{dere_etal:97}
{Dere}, K.~P., {Landi}, E., {Mason}, H.~E., {Monsignori Fossi}, B.~C., and
  {Young}, P.~R. (1997).
\newblock {CHIANTI - an atomic database for emission lines}.
\newblock \emph{\aaps} 125, 149--173.
\newblock \doi{10.1051/aas:1997368}
\bibAnnoteFile{dere_etal:97}

\bibitem[{{Dud{\'\i}k} et~al.(2019){Dud{\'\i}k}, {Dzif{\v{c}}{\'a}kov{\'a}},
  {Del Zanna}, {Mason}, {Golub}, {Winebarger} et~al.}]{dudik_etal:2019}
{Dud{\'\i}k}, J., {Dzif{\v{c}}{\'a}kov{\'a}}, E., {Del Zanna}, G., {Mason},
  H.~E., {Golub}, L.~L., {Winebarger}, A.~R., et~al. (2019).
\newblock {Signatures of the non-Maxwellian {\ensuremath{\kappa}}-distributions
  in optically thin line spectra. II. Synthetic Fe XVII-XVIII X-ray coronal
  spectra and predictions for the Marshall Grazing-Incidence X-ray Spectrometer
  (MaGIXS)}.
\newblock \emph{\aap} 626, A88.
\newblock \doi{10.1051/0004-6361/201935285}
\bibAnnoteFile{dudik_etal:2019}

\bibitem[{{Dud{\'\i}k} et~al.(2017){Dud{\'\i}k}, {Dzif{\v{c}}{\'a}kov{\'a}},
  {Meyer-Vernet}, {Del Zanna}, {Young}, {Giunta}
  et~al.}]{dudik_etal:2017_review}
{Dud{\'\i}k}, J., {Dzif{\v{c}}{\'a}kov{\'a}}, E., {Meyer-Vernet}, N., {Del
  Zanna}, G., {Young}, P.~R., {Giunta}, A., et~al. (2017).
\newblock {Nonequilibrium Processes in the Solar Corona, Transition Region,
  Flares, and Solar Wind (Invited Review)}.
\newblock \emph{\solphys} 292, 100.
\newblock \doi{10.1007/s11207-017-1125-0}
\bibAnnoteFile{dudik_etal:2017_review}

\bibitem[{{Dud{\'{\i}}k} et~al.(2014){Dud{\'{\i}}k}, {Janvier}, {Aulanier},
  {Del Zanna}, {Karlick{\'y}}, {Mason} et~al.}]{dudik_etal:2014}
{Dud{\'{\i}}k}, J., {Janvier}, M., {Aulanier}, G., {Del Zanna}, G.,
  {Karlick{\'y}}, M., {Mason}, H.~E., et~al. (2014).
\newblock {Slipping Magnetic Reconnection during an X-class Solar Flare
  Observed by SDO/AIA}.
\newblock \emph{\apj} 784, 144.
\newblock \doi{10.1088/0004-637X/784/2/144}
\bibAnnoteFile{dudik_etal:2014}

\bibitem[{{Feldman} et~al.(2000){Feldman}, {Curdt}, {Landi}, and
  {Wilhelm}}]{feldman_etal:2000}
{Feldman}, U., {Curdt}, W., {Landi}, E., and {Wilhelm}, K. (2000).
\newblock {Identification of Spectral Lines in the 500-1600 {\AA} Wavelength
  Range of Highly Ionized Ne, Na, Mg, Ar, K, Ca, Ti, Cr, Mn, Fe, Co, and Ni
  Emitted by Flares (Te > 3x10$^{6}$ K) and Their Potential Use in Plasma
  Diagnostics}.
\newblock \emph{\apj} 544, 508--521.
\newblock \doi{10.1086/317203}
\bibAnnoteFile{feldman_etal:2000}

\bibitem[{{Feldman} et~al.(1996){Feldman}, {Doschek}, {Behring}, and
  {Phillips}}]{feldman_etal:1996}
{Feldman}, U., {Doschek}, G.~A., {Behring}, W.~E., and {Phillips}, K.~J.~H.
  (1996).
\newblock {Electron Temperature, Emission Measure, and X-Ray Flux in A2 to X2
  X-Ray Class Solar Flares}.
\newblock \emph{\apj} 460, 1034--+.
\newblock \doi{10.1086/177030}
\bibAnnoteFile{feldman_etal:1996}

\bibitem[{{Feldman} et~al.(2003){Feldman}, {Landi}, {Doschek}, {Dammasch}, and
  {Curdt}}]{feldman_etal:2003}
{Feldman}, U., {Landi}, E., {Doschek}, G.~A., {Dammasch}, I., and {Curdt}, W.
  (2003).
\newblock {Free-Free Emission in the Far-Ultraviolet Spectral Range: A Resource
  for Diagnosing Solar and Stellar Flare Plasmas}.
\newblock \emph{\apj} 593, 1226--1241.
\newblock \doi{10.1086/376680}
\bibAnnoteFile{feldman_etal:2003}

\bibitem[{{Fletcher} et~al.(2011){Fletcher}, {Dennis}, {Hudson}, {Krucker},
  {Phillips}, {Veronig} et~al.}]{fletcher_etal:2011}
{Fletcher}, L., {Dennis}, B.~R., {Hudson}, H.~S., {Krucker}, S., {Phillips},
  K., {Veronig}, A., et~al. (2011).
\newblock {An Observational Overview of Solar Flares}.
\newblock \emph{\ssr} 159, 19--106.
\newblock \doi{10.1007/s11214-010-9701-8}
\bibAnnoteFile{fletcher_etal:2011}

\bibitem[{{Fletcher} and {Hudson}(2008)}]{fletcher_hudson:2008}
{Fletcher}, L. and {Hudson}, H.~S. (2008).
\newblock {Impulsive Phase Flare Energy Transport by Large-Scale Alfv{\'e}n
  Waves and the Electron Acceleration Problem}.
\newblock \emph{\apj} 675, 1645--1655.
\newblock \doi{10.1086/527044}
\bibAnnoteFile{fletcher_hudson:2008}

\bibitem[{{Golub} et~al.(1989){Golub}, {Hartquist}, and
  {Quillen}}]{golub_etal:1989SoPh..122..245G}
{Golub}, L., {Hartquist}, T.~W., and {Quillen}, A.~C. (1989).
\newblock {Comments on the Observability of Coronal Variations}.
\newblock \emph{\solphys} 122, 245--261.
\newblock \doi{10.1007/BF00912995}
\bibAnnoteFile{golub_etal:1989SoPh..122..245G}

\bibitem[{{Graham} and {Cauzzi}(2015)}]{graham_cauzzi:2015}
{Graham}, D.~R. and {Cauzzi}, G. (2015).
\newblock {Temporal Evolution of Multiple Evaporating Ribbon Sources in a Solar
  Flare}.
\newblock \emph{\apjl} 807, L22.
\newblock \doi{10.1088/2041-8205/807/2/L22}
\bibAnnoteFile{graham_cauzzi:2015}

\bibitem[{{Hannah} et~al.(2008){Hannah}, {Christe}, {Krucker}, {Hurford},
  {Hudson}, and {Lin}}]{hannah_etal:2008ApJ...677..704H}
{Hannah}, I.~G., {Christe}, S., {Krucker}, S., {Hurford}, G.~J., {Hudson},
  H.~S., and {Lin}, R.~P. (2008).
\newblock {RHESSI Microflare Statistics. II. X-Ray Imaging, Spectroscopy, and
  Energy Distributions}.
\newblock \emph{\apj} 677, 704--718.
\newblock \doi{10.1086/529012}
\bibAnnoteFile{hannah_etal:2008ApJ...677..704H}

\bibitem[{{Hannah} et~al.(2016){Hannah}, {Grefenstette}, {Smith}, {Glesener},
  {Krucker}, {Hudson} et~al.}]{hannah_etal:2016}
{Hannah}, I.~G., {Grefenstette}, B.~W., {Smith}, D.~M., {Glesener}, L.,
  {Krucker}, S., {Hudson}, H.~S., et~al. (2016).
\newblock {The First X-Ray Imaging Spectroscopy of Quiescent Solar Active
  Regions with NuSTAR}.
\newblock \emph{\apjl} 820, L14.
\newblock \doi{10.3847/2041-8205/820/1/L14}
\bibAnnoteFile{hannah_etal:2016}

\bibitem[{{Hannah} et~al.(2019){Hannah}, {Kleint}, {Krucker}, {Grefenstette},
  {Glesener}, {Hudson} et~al.}]{hannah_etal:2019}
{Hannah}, I.~G., {Kleint}, L., {Krucker}, S., {Grefenstette}, B.~W.,
  {Glesener}, L., {Hudson}, H.~S., et~al. (2019).
\newblock {Joint X-Ray, EUV, and UV Observations of a Small Microflare}.
\newblock \emph{\apj} 881, 109.
\newblock \doi{10.3847/1538-4357/ab2dfa}
\bibAnnoteFile{hannah_etal:2019}

\bibitem[{{Harrison} et~al.(2013){Harrison}, {Craig}, {Christensen}, {Hailey},
  {Zhang}, {Boggs} et~al.}]{harrison_etal:2013}
{Harrison}, F.~A., {Craig}, W.~W., {Christensen}, F.~E., {Hailey}, C.~J.,
  {Zhang}, W.~W., {Boggs}, S.~E., et~al. (2013).
\newblock {The Nuclear Spectroscopic Telescope Array (NuSTAR) High-energy X-Ray
  Mission}.
\newblock \emph{\apj} 770, 103.
\newblock \doi{10.1088/0004-637X/770/2/103}
\bibAnnoteFile{harrison_etal:2013}

\bibitem[{{Hinode Review Team} et~al.(2019){Hinode Review Team}, {Al-Janabi},
  {Antolin}, {Baker}, {Bellot Rubio}, {Bradley} et~al.}]{hinode_review:2019}
{Hinode Review Team}, {Al-Janabi}, K., {Antolin}, P., {Baker}, D., {Bellot
  Rubio}, L.~R., {Bradley}, L., et~al. (2019).
\newblock {Achievements of Hinode in the first eleven years}.
\newblock \emph{\pasj} 71, R1.
\newblock \doi{10.1093/pasj/psz084}
\bibAnnoteFile{hinode_review:2019}

\bibitem[{{Imada} et~al.(2013){Imada}, {Aoki}, {Hara}, {Watanabe}, {Harra}, and
  {Shimizu}}]{imada_etal:2013}
{Imada}, S., {Aoki}, K., {Hara}, H., {Watanabe}, T., {Harra}, L.~K., and
  {Shimizu}, T. (2013).
\newblock {Evidence for Hot Fast Flow above a Solar Flare Arcade}.
\newblock \emph{\apjl} 776, L11.
\newblock \doi{10.1088/2041-8205/776/1/L11}
\bibAnnoteFile{imada_etal:2013}

\bibitem[{{Imada} et~al.(2011){Imada}, {Murakami}, {Watanabe}, {Hara}, and
  {Shimizu}}]{imada_etal:2011ApJ...742...70I}
{Imada}, S., {Murakami}, I., {Watanabe}, T., {Hara}, H., and {Shimizu}, T.
  (2011).
\newblock {Magnetic Reconnection in Non-equilibrium Ionization Plasma}.
\newblock \emph{\apj} 742, 70.
\newblock \doi{10.1088/0004-637X/742/2/70}
\bibAnnoteFile{imada_etal:2011ApJ...742...70I}

\bibitem[{{Ishikawa} et~al.(2014){Ishikawa}, {Glesener}, {Christe},
  {Ishibashi}, {Brooks}, {Williams} et~al.}]{ishikawa_etal:2014}
{Ishikawa}, S.-n., {Glesener}, L., {Christe}, S., {Ishibashi}, K., {Brooks},
  D.~H., {Williams}, D.~R., et~al. (2014).
\newblock {Constraining hot plasma in a non-flaring solar active region with
  FOXSI hard X-ray observations}.
\newblock \emph{\pasj} 66, S15.
\newblock \doi{10.1093/pasj/psu090}
\bibAnnoteFile{ishikawa_etal:2014}

\bibitem[{{Jain} et~al.(2006){Jain}, {Joshi}, {Kayasth}, {Dave}, and
  {Deshpande}}]{jain_etal:2006}
{Jain}, R., {Joshi}, V., {Kayasth}, S.~L., {Dave}, H., and {Deshpande}, M.~R.
  (2006).
\newblock {Solar X-ray Spectrometer (SOXS) Mission - Low Energy Payload - First
  Results}.
\newblock \emph{Journal of Astrophysics and Astronomy} 27, 175--192.
\newblock \doi{10.1007/BF02702520}
\bibAnnoteFile{jain_etal:2006}

\bibitem[{{Joshi} et~al.(2021){Joshi}, {Schmieder}, {Tei}, {Aulanier},
  {L{\"o}rin{\v{c}}{\'\i}k}, {Chandra} et~al.}]{joshi_etal:2021}
{Joshi}, R., {Schmieder}, B., {Tei}, A., {Aulanier}, G.,
  {L{\"o}rin{\v{c}}{\'\i}k}, J., {Chandra}, R., et~al. (2021).
\newblock {Multi-thermal atmosphere of a mini-solar flare during magnetic
  reconnection observed with IRIS}.
\newblock \emph{\aap} 645, A80.
\newblock \doi{10.1051/0004-6361/202039229}
\bibAnnoteFile{joshi_etal:2021}

\bibitem[{{Kastner} et~al.(1974){Kastner}, {Neupert}, and
  {Swartz}}]{kastner_etal:74}
{Kastner}, S.~O., {Neupert}, W.~M., and {Swartz}, M. (1974).
\newblock {Solar-flare emission lines in the range from 66 to 171 A;
  transitions in highly ionized iron.}
\newblock \emph{\apj} 191, 261--270
\bibAnnoteFile{kastner_etal:74}

\bibitem[{{Kawate} et~al.(2016){Kawate}, {Keenan}, and
  {Jess}}]{kawate_etal:2016}
{Kawate}, T., {Keenan}, F.~P., and {Jess}, D.~B. (2016).
\newblock {Departure of High-temperature Iron Lines from the Equilibrium State
  in Flaring Solar Plasmas}.
\newblock \emph{\apj} 826, 3.
\newblock \doi{10.3847/0004-637X/826/1/3}
\bibAnnoteFile{kawate_etal:2016}

\bibitem[{{Keenan} et~al.(2017){Keenan}, {Milligan}, {Mathioudakis}, and
  {Christian}}]{keenan_etal:2017}
{Keenan}, F.~P., {Milligan}, R.~O., {Mathioudakis}, M., and {Christian}, D.~J.
  (2017).
\newblock {An assessment of Fe xx-Fe xxii emission lines in SDO/EVE data as
  diagnostics for high-density solar flare plasmas using EUVE stellar
  observations}.
\newblock \emph{\mnras} 468, 1117--1122.
\newblock \doi{10.1093/mnras/stx525}
\bibAnnoteFile{keenan_etal:2017}

\bibitem[{{Kirichenko} and {Bogachev}(2017)}]{kirichenko_bogahev:2017}
{Kirichenko}, A.~S. and {Bogachev}, S.~A. (2017).
\newblock {Plasma Heating in Solar Microflares: Statistics and Analysis}.
\newblock \emph{\apj} 840, 45.
\newblock \doi{10.3847/1538-4357/aa6c2b}
\bibAnnoteFile{kirichenko_bogahev:2017}

\bibitem[{{Kliem} et~al.(2002){Kliem}, {Dammasch}, {Curdt}, and
  {Wilhelm}}]{kliem_etal:2002}
{Kliem}, B., {Dammasch}, I.~E., {Curdt}, W., and {Wilhelm}, K. (2002).
\newblock {Correlated Dynamics of Hot and Cool Plasmas in the Main Phase of a
  Solar Flare}.
\newblock \emph{\apjl} 568, L61--L65.
\newblock \doi{10.1086/340136}
\bibAnnoteFile{kliem_etal:2002}

\bibitem[{{Klimchuk}(2006)}]{klimchuk:06}
{Klimchuk}, J.~A. (2006).
\newblock {On Solving the Coronal Heating Problem}.
\newblock \emph{\solphys} 234, 41--77.
\newblock \doi{10.1007/s11207-006-0055-z}
\bibAnnoteFile{klimchuk:06}

\bibitem[{{Klimchuk}(2015)}]{klimchuk:2015}
{Klimchuk}, J.~A. (2015).
\newblock {Key aspects of coronal heating}.
\newblock \emph{Philosophical Transactions of the Royal Society of London
  Series A} 373, 20140256--20140256.
\newblock \doi{10.1098/rsta.2014.0256}
\bibAnnoteFile{klimchuk:2015}

\bibitem[{{Kobayashi} et~al.(2011){Kobayashi}, {Cirtain}, {Golub},
  {Winebarger}, {Hertz}, {Cheimets} et~al.}]{kobayashi_etal:2011}
{Kobayashi}, K., {Cirtain}, J., {Golub}, L., {Winebarger}, A., {Hertz}, E.,
  {Cheimets}, P., et~al. (2011).
\newblock {The Marshall Grazing Incidence X-ray Spectrograph (MaGIXS)}.
\newblock In \emph{Society of Photo-Optical Instrumentation Engineers (SPIE)
  Conference Series}. vol. 8147 of \emph{\procspie}, 81471M.
\newblock \doi{10.1117/12.894071}
\bibAnnoteFile{kobayashi_etal:2011}

\bibitem[{{Kobayashi} et~al.(2014){Kobayashi}, {Cirtain}, {Winebarger},
  {Korreck}, {Golub}, {Walsh} et~al.}]{kobayashi_etal:2014}
{Kobayashi}, K., {Cirtain}, J., {Winebarger}, A.~R., {Korreck}, K., {Golub},
  L., {Walsh}, R.~W., et~al. (2014).
\newblock {The High-Resolution Coronal Imager (Hi-C)}.
\newblock \emph{\solphys} 289, 4393--4412.
\newblock \doi{10.1007/s11207-014-0544-4}
\bibAnnoteFile{kobayashi_etal:2014}

\bibitem[{{Krucker} et~al.(2014){Krucker}, {Christe}, {Glesener}, {Ishikawa},
  {Ramsey}, {Takahashi} et~al.}]{krucker_etal:2014}
{Krucker}, S., {Christe}, S., {Glesener}, L., {Ishikawa}, S.-n., {Ramsey}, B.,
  {Takahashi}, T., et~al. (2014).
\newblock {First Images from the Focusing Optics X-Ray Solar Imager}.
\newblock \emph{\apjl} 793, L32.
\newblock \doi{10.1088/2041-8205/793/2/L32}
\bibAnnoteFile{krucker_etal:2014}

\bibitem[{{Kuzin} et~al.(2009){Kuzin}, {Bogachev}, {Zhitnik}, {Pertsov},
  {Ignatiev}, {Mitrofanov} et~al.}]{kuzin_etal:2009}
{Kuzin}, S.~V., {Bogachev}, S.~A., {Zhitnik}, I.~A., {Pertsov}, A.~A.,
  {Ignatiev}, A.~P., {Mitrofanov}, A.~M., et~al. (2009).
\newblock {TESIS experiment on EUV imaging spectroscopy of the Sun}.
\newblock \emph{Advances in Space Research} 43, 1001--1006.
\newblock \doi{10.1016/j.asr.2008.10.021}
\bibAnnoteFile{kuzin_etal:2009}

\bibitem[{{Laming}(2015)}]{laming:2015}
{Laming}, J.~M. (2015).
\newblock {The FIP and Inverse FIP Effects in Solar and Stellar Coronae}.
\newblock \emph{Living Reviews in Solar Physics} 12, 2.
\newblock \doi{10.1007/lrsp-2015-2}
\bibAnnoteFile{laming:2015}

\bibitem[{{Lemen} et~al.(2012){Lemen}, {Title}, {Akin}, {Boerner}, {Chou},
  {Drake} et~al.}]{lemen_etal:2012}
{Lemen}, J.~R., {Title}, A.~M., {Akin}, D.~J., {Boerner}, P.~F., {Chou}, C.,
  {Drake}, J.~F., et~al. (2012).
\newblock {The Atmospheric Imaging Assembly (AIA) on the Solar Dynamics
  Observatory (SDO)}.
\newblock \emph{\solphys} 275, 17--40.
\newblock \doi{10.1007/s11207-011-9776-8}
\bibAnnoteFile{lemen_etal:2012}

\bibitem[{{Lin} et~al.(2002){Lin}, {Dennis}, {Hurford}, {Smith}, {Zehnder},
  {Harvey} et~al.}]{lin_etal:2002}
{Lin}, R.~P., {Dennis}, B.~R., {Hurford}, G.~J., {Smith}, D.~M., {Zehnder}, A.,
  {Harvey}, P.~R., et~al. (2002).
\newblock {The Reuven Ramaty High-Energy Solar Spectroscopic Imager (RHESSI)}.
\newblock \emph{\solphys} 210, 3--32.
\newblock \doi{10.1023/A:1022428818870}
\bibAnnoteFile{lin_etal:2002}

\bibitem[{{L{\'o}pez Fuentes} and {Klimchuk}(2015)}]{2015ApJ...799..128L}
{L{\'o}pez Fuentes}, M. and {Klimchuk}, J.~A. (2015).
\newblock {Two-dimensional Cellular Automaton Model for the Evolution of Active
  Region Coronal Plasmas}.
\newblock \emph{\apj} 799, 128.
\newblock \doi{10.1088/0004-637X/799/2/128}
\bibAnnoteFile{2015ApJ...799..128L}

\bibitem[{{Mandage} and {Bradshaw}(2020)}]{mandage_bradshaw:2020}
{Mandage}, R.~S. and {Bradshaw}, S.~J. (2020).
\newblock {Asymmetries and Broadenings of Spectral Lines in Strongly Charged
  Iron Produced during Solar Flares}.
\newblock \emph{\apj} 891, 122.
\newblock \doi{10.3847/1538-4357/ab7340}
\bibAnnoteFile{mandage_bradshaw:2020}

\bibitem[{{Marsh} et~al.(2018){Marsh}, {Smith}, {Glesener}, {Klimchuk},
  {Bradshaw}, {Vievering} et~al.}]{marsh_etal:2018ApJ...864....5M}
{Marsh}, A.~J., {Smith}, D.~M., {Glesener}, L., {Klimchuk}, J.~A., {Bradshaw},
  S.~J., {Vievering}, J., et~al. (2018).
\newblock {Hard X-Ray Constraints on Small-scale Coronal Heating Events}.
\newblock \emph{\apj} 864, 5.
\newblock \doi{10.3847/1538-4357/aad380}
\bibAnnoteFile{marsh_etal:2018ApJ...864....5M}

\bibitem[{{Mason} et~al.(1984){Mason}, {Bhatia}, {Neupert}, {Swartz}, and
  {Kastner}}]{mason_etal:84}
{Mason}, H.~E., {Bhatia}, A.~K., {Neupert}, W.~M., {Swartz}, M., and {Kastner},
  S.~O. (1984).
\newblock {Diagnostic application of highly ionized iron lines in the XUV
  spectrum of a solar flare}.
\newblock \emph{\solphys} 92, 199--216
\bibAnnoteFile{mason_etal:84}

\bibitem[{Mason et~al.(1979)Mason, Doschek, Feldman, and
  Bhatia}]{mason_etal:1979}
Mason, H.~E., Doschek, G.~A., Feldman, U., and Bhatia, A.~K. (1979).
\newblock Fe xxi as an electron density diagnostic in solar flares.
\newblock \emph{\aap} 73, 74--81
\bibAnnoteFile{mason_etal:1979}

\bibitem[{{McKenzie} et~al.(1980){McKenzie}, {Landecker}, {Broussard}, {Rugge},
  {Young}, {Feldman} et~al.}]{mckenzie_etal:80}
{McKenzie}, D.~L., {Landecker}, P.~B., {Broussard}, R.~M., {Rugge}, H.~R.,
  {Young}, R.~M., {Feldman}, U., et~al. (1980).
\newblock {Solar flare X-ray spectra between 7.8 and 23.0 angstroms}.
\newblock \emph{\apj} 241, 409--416
\bibAnnoteFile{mckenzie_etal:80}

\bibitem[{{Miceli} et~al.(2012){Miceli}, {Reale}, {Gburek}, {Terzo}, {Barbera},
  {Collura} et~al.}]{miceli_etal:2012}
{Miceli}, M., {Reale}, F., {Gburek}, S., {Terzo}, S., {Barbera}, M., {Collura},
  A., et~al. (2012).
\newblock {X-ray emitting hot plasma in solar active regions observed by the
  SphinX spectrometer}.
\newblock \emph{\aap} 544, A139.
\newblock \doi{10.1051/0004-6361/201219670}
\bibAnnoteFile{miceli_etal:2012}

\bibitem[{{Milligan} and {Dennis}(2009)}]{2009ApJ...699..968M}
{Milligan}, R.~O. and {Dennis}, B.~R. (2009).
\newblock {Velocity Characteristics of Evaporated Plasma Using Hinode/EUV
  Imaging Spectrometer}.
\newblock \emph{\apj} 699, 968--975.
\newblock \doi{10.1088/0004-637X/699/2/968}
\bibAnnoteFile{2009ApJ...699..968M}

\bibitem[{{Milligan} et~al.(2012){Milligan}, {Kennedy}, {Mathioudakis}, and
  {Keenan}}]{milligan_etal:12_dens}
{Milligan}, R.~O., {Kennedy}, M.~B., {Mathioudakis}, M., and {Keenan}, F.~P.
  (2012).
\newblock {Time-dependent Density Diagnostics of Solar Flare Plasmas Using
  SDO/EVE}.
\newblock \emph{\apjl} 755, L16.
\newblock \doi{10.1088/2041-8205/755/1/L16}
\bibAnnoteFile{milligan_etal:12_dens}

\bibitem[{{Mitra-Kraev} and {Del Zanna}(2019)}]{mitra-kraev_delzanna:2019}
{Mitra-Kraev}, U. and {Del Zanna}, G. (2019).
\newblock {Solar microflares: a case study on temperatures and the Fe XVIII
  emission}.
\newblock \emph{\aap} 628, A134.
\newblock \doi{10.1051/0004-6361/201834856}
\bibAnnoteFile{mitra-kraev_delzanna:2019}

\bibitem[{{Monsignori Fossi} et~al.(1996){Monsignori Fossi}, {Landini}, {Del
  Zanna}, and {Bowyer}}]{monsignori_etal:96}
{Monsignori Fossi}, B.~C., {Landini}, M., {Del Zanna}, G., and {Bowyer}, S.
  (1996).
\newblock {A Time-resolved Extreme-Ultraviolet Spectroscopic Study of the
  Quiescent and Flaring Corona of the Flare Star AU Microscopii}.
\newblock \emph{\apj} 466, 427
\bibAnnoteFile{monsignori_etal:96}

\bibitem[{{Mrozek} et~al.(2018){Mrozek}, {Gburek}, {Siarkowski}, {Sylwester},
  {Sylwester}, {Kepa} et~al.}]{mrozek_etal:2018}
{Mrozek}, T., {Gburek}, S., {Siarkowski}, M., {Sylwester}, B., {Sylwester}, J.,
  {Kepa}, A., et~al. (2018).
\newblock {Solar Microflares Observed by SphinX and RHESSI}.
\newblock \emph{\solphys} 293, 101.
\newblock \doi{10.1007/s11207-018-1319-0}
\bibAnnoteFile{mrozek_etal:2018}

\bibitem[{{Parenti} et~al.(2006){Parenti}, {Buchlin}, {Cargill}, {Galtier}, and
  {Vial}}]{parenti_etal:2006}
{Parenti}, S., {Buchlin}, E., {Cargill}, P.~J., {Galtier}, S., and {Vial},
  J.-C. (2006).
\newblock {Modeling the Radiative Signatures of Turbulent Heating in Coronal
  Loops}.
\newblock \emph{\apj} 651, 1219--1228.
\newblock \doi{10.1086/507594}
\bibAnnoteFile{parenti_etal:2006}

\bibitem[{{Parenti} et~al.(2017){Parenti}, {Del Zanna}, {Petralia}, {Reale},
  {Teriaca}, {Testa} et~al.}]{parenti_etal:2017}
{Parenti}, S., {Del Zanna}, G., {Petralia}, A., {Reale}, F., {Teriaca}, L.,
  {Testa}, P., et~al. (2017).
\newblock {Spectroscopy of Very Hot Plasma in Non-flaring Parts of a Solar Limb
  Active Region: Spatial and Temporal Properties}.
\newblock \emph{\apj} 846, 25.
\newblock \doi{10.3847/1538-4357/aa835f}
\bibAnnoteFile{parenti_etal:2017}

\bibitem[{{Patsourakos} and {Klimchuk}(2009)}]{patsourakos_klimchuk:09}
{Patsourakos}, S. and {Klimchuk}, J.~A. (2009).
\newblock {Spectroscopic Observations of Hot Lines Constraining Coronal Heating
  in Solar Active Regions}.
\newblock \emph{\apj} 696, 760--765.
\newblock \doi{10.1088/0004-637X/696/1/760}
\bibAnnoteFile{patsourakos_klimchuk:09}

\bibitem[{{Phillips} et~al.(1996){Phillips}, {Bhatia}, {Mason}, and
  {Zarro}}]{phillips_etal:1996}
{Phillips}, K.~J.~H., {Bhatia}, A.~K., {Mason}, H.~E., and {Zarro}, D.~M.
  (1996).
\newblock {High Coronal Electron Densities in a Solar Flare from Fe XXI and Fe
  XXII X-Ray Line Measurements}.
\newblock \emph{\apj} 466, 549.
\newblock \doi{10.1086/177531}
\bibAnnoteFile{phillips_etal:1996}

\bibitem[{{Polito} et~al.(2018){Polito}, {Galan}, {Reeves}, and
  {Musset}}]{polito_etal:2018ApJ...865..161P}
{Polito}, V., {Galan}, G., {Reeves}, K.~K., and {Musset}, S. (2018).
\newblock {Possible Signatures of a Termination Shock in the 2014 March 29
  X-class Flare Observed by IRIS}.
\newblock \emph{\apj} 865, 161.
\newblock \doi{10.3847/1538-4357/aadada}
\bibAnnoteFile{polito_etal:2018ApJ...865..161P}

\bibitem[{{Polito} et~al.(2016){Polito}, {Reep}, {Reeves}, {Sim{\~o}es},
  {Dud{\'{\i}}k}, {Del Zanna} et~al.}]{polito_etal:2016a}
{Polito}, V., {Reep}, J.~W., {Reeves}, K.~K., {Sim{\~o}es}, P.~J.~A.,
  {Dud{\'{\i}}k}, J., {Del Zanna}, G., et~al. (2016).
\newblock {Simultaneous IRIS and Hinode/EIS Observations and Modelling of the
  2014 October 27 X2.0 Class Flare}.
\newblock \emph{\apj} 816, 89.
\newblock \doi{10.3847/0004-637X/816/2/89}
\bibAnnoteFile{polito_etal:2016a}

\bibitem[{{Polito} et~al.(2015){Polito}, {Reeves}, {Del Zanna}, {Golub}, and
  {Mason}}]{polito_etal:2015}
{Polito}, V., {Reeves}, K.~K., {Del Zanna}, G., {Golub}, L., and {Mason}, H.~E.
  (2015).
\newblock {Joint High Temperature Observation of a Small C6.5 Solar Flare With
  Iris/Eis/Aia}.
\newblock \emph{\apj} 803, 84.
\newblock \doi{10.1088/0004-637X/803/2/84}
\bibAnnoteFile{polito_etal:2015}

\bibitem[{{Reale}(2014)}]{reale:2014_lr}
{Reale}, F. (2014).
\newblock {Coronal Loops: Observations and Modeling of Confined Plasma}.
\newblock \emph{Living Reviews in Solar Physics} 11, 4.
\newblock \doi{10.12942/lrsp-2014-4}
\bibAnnoteFile{reale:2014_lr}

\bibitem[{{Reale} and {Orlando}(2008)}]{reale_orlando:2008}
{Reale}, F. and {Orlando}, S. (2008).
\newblock {Nonequilibrium of Ionization and the Detection of Hot Plasma in
  Nanoflare-heated Coronal Loops}.
\newblock \emph{\apj} 684, 715--724.
\newblock \doi{10.1086/590338}
\bibAnnoteFile{reale_orlando:2008}

\bibitem[{{Reva} et~al.(2018){Reva}, {Ulyanov}, {Kirichenko}, {Bogachev}, and
  {Kuzin}}]{reva_etal:2018}
{Reva}, A., {Ulyanov}, A., {Kirichenko}, A., {Bogachev}, S., and {Kuzin}, S.
  (2018).
\newblock {Estimate of the Upper Limit on Hot Plasma Differential Emission
  Measure (DEM) in Non-Flaring Active Regions and Nanoflare Frequency Based on
  the Mg xii Spectroheliograph Data from CORONAS-F/SPIRIT}.
\newblock \emph{\solphys} 293, 140.
\newblock \doi{10.1007/s11207-018-1363-9}
\bibAnnoteFile{reva_etal:2018}

\bibitem[{{Shestov} et~al.(2014){Shestov}, {Reva}, and
  {Kuzin}}]{shestov_etal:2014}
{Shestov}, S., {Reva}, A., and {Kuzin}, S. (2014).
\newblock {Extreme Ultraviolet Spectra of Solar Flares from the Extreme
  Ultraviolet Spectroheliograph SPIRIT Onboard the CORONAS-F Satellite}.
\newblock \emph{\apj} 780, 15.
\newblock \doi{10.1088/0004-637X/780/1/15}
\bibAnnoteFile{shestov_etal:2014}

\bibitem[{{Shibata} and {Magara}(2011)}]{shibata:2011LRSP}
{Shibata}, K. and {Magara}, T. (2011).
\newblock {Solar Flares: Magnetohydrodynamic Processes}.
\newblock \emph{Living Reviews in Solar Physics} 8, 6.
\newblock \doi{10.12942/lrsp-2011-6}
\bibAnnoteFile{shibata:2011LRSP}

\bibitem[{{Shimizu} et~al.(2019){Shimizu}, {Imada}, {Kawate}, {Ichimoto},
  {Suematsu}, {Hara} et~al.}]{shimizu_etal:2019}
{Shimizu}, T., {Imada}, S., {Kawate}, T., {Ichimoto}, K., {Suematsu}, Y.,
  {Hara}, H., et~al. (2019).
\newblock {The Solar-C\_EUVST mission}.
\newblock In \emph{UV, X-Ray, and Gamma-Ray Space Instrumentation for Astronomy
  XXI}. vol. 11118 of \emph{Society of Photo-Optical Instrumentation Engineers
  (SPIE) Conference Series}, 1111807.
\newblock \doi{10.1117/12.2528240}
\bibAnnoteFile{shimizu_etal:2019}

\bibitem[{{Sobel'Man} et~al.(1996){Sobel'Man}, {Zhitnik}, {Ignat'ev},
  {Korneev}, {Klepikov}, {Krutov} et~al.}]{sobelman_etal:1996}
{Sobel'Man}, I.~I., {Zhitnik}, I.~A., {Ignat'ev}, A.~P., {Korneev}, V.~V.,
  {Klepikov}, V.~Y., {Krutov}, V.~V., et~al. (1996).
\newblock {X-ray spectroscopy of the Sun in the 0.84-30.4 nm band in the
  TEREK-K and RES-K experiments on the KORONAS-I satellite}.
\newblock \emph{Astronomy Letters} 22, 539--554
\bibAnnoteFile{sobelman_etal:1996}

\bibitem[{{Soufli} et~al.(2005){Soufli}, {Windt}, {Robinson}, {Baker},
  {Spiller}, {Dollar} et~al.}]{soufli_etal:2005}
{Soufli}, R., {Windt}, D.~L., {Robinson}, J.~C., {Baker}, S.~L., {Spiller}, E.,
  {Dollar}, F.~J., et~al. (2005).
\newblock {Development and testing of EUV multilayer coatings for the
  atmospheric imaging assembly instrument aboard the Solar Dynamics
  Observatory}.
\newblock In \emph{Solar Physics and Space Weather Instrumentation}, eds.
  S.~{Fineschi} and R.~A. {Viereck}. vol. 5901 of \emph{\procspie}, 173--183.
\newblock \doi{10.1117/12.617370}
\bibAnnoteFile{soufli_etal:2005}

\bibitem[{{Stratton} et~al.(1984){Stratton}, {Moos}, and
  {Finkenthal}}]{stratton_etal:1984}
{Stratton}, B.~C., {Moos}, H.~W., and {Finkenthal}, M. (1984).
\newblock {Electron density-dependent intensity ratios of highly ionized iron
  lines - A comparison of theory and experiment}.
\newblock \emph{\apjl} 279, L31--L34.
\newblock \doi{10.1086/184249}
\bibAnnoteFile{stratton_etal:1984}

\bibitem[{{Sylwester} et~al.(2010){Sylwester}, {Sylwester}, and
  {Phillips}}]{2010A&A...514A..82S}
{Sylwester}, B., {Sylwester}, J., and {Phillips}, K.~J.~H. (2010).
\newblock {Soft X-ray coronal spectra at low activity levels observed by
  RESIK}.
\newblock \emph{\aap} 514, A82.
\newblock \doi{10.1051/0004-6361/200912907}
\bibAnnoteFile{2010A&A...514A..82S}

\bibitem[{{Sylwester} et~al.(2005){Sylwester}, {Gaicki}, {Kordylewski},
  {Kowali{\'n}ski}, {Nowak}, {P{\l}ocieniak} et~al.}]{sylwester_etal:2005}
{Sylwester}, J., {Gaicki}, I., {Kordylewski}, Z., {Kowali{\'n}ski}, M.,
  {Nowak}, S., {P{\l}ocieniak}, S., et~al. (2005).
\newblock {Resik: A Bent Crystal X-ray Spectrometer for Studies of Solar
  Coronal Plasma Composition}.
\newblock \emph{\solphys} 226, 45--72.
\newblock \doi{10.1007/s11207-005-6392-5}
\bibAnnoteFile{sylwester_etal:2005}

\bibitem[{{Sylwester} et~al.(2008){Sylwester}, {Kowalinski}, {Szymon},
  {Bakala}, {Kuzin}, {Kotov} et~al.}]{sylwester_etal:08}
{Sylwester}, J., {Kowalinski}, M., {Szymon}, G., {Bakala}, J., {Kuzin}, S.,
  {Kotov}, Y., et~al. (2008).
\newblock {The Soft X-ray Spectrophotometer SphinX for the CORONAS-Photon
  Mission}.
\newblock In \emph{37th COSPAR Scientific Assembly}. vol.~37 of \emph{COSPAR,
  Plenary Meeting}, 3111--+
\bibAnnoteFile{sylwester_etal:08}

\bibitem[{{Teriaca} et~al.(2012{\natexlab{a}}){Teriaca}, {Andretta},
  {Auch{\`e}re}, {Brown}, {Buchlin}, {Cauzzi} et~al.}]{teriaca_etal:2012_lemur}
{Teriaca}, L., {Andretta}, V., {Auch{\`e}re}, F., {Brown}, C.~M., {Buchlin},
  E., {Cauzzi}, G., et~al. (2012{\natexlab{a}}).
\newblock {LEMUR: Large European module for solar Ultraviolet Research.
  European contribution to JAXA's Solar-C mission}.
\newblock \emph{Experimental Astronomy} 34, 273--309.
\newblock \doi{10.1007/s10686-011-9274-x}
\bibAnnoteFile{teriaca_etal:2012_lemur}

\bibitem[{{Teriaca} et~al.(2012{\natexlab{b}}){Teriaca}, {Warren}, and
  {Curdt}}]{teriaca_etal:2012}
{Teriaca}, L., {Warren}, H.~P., and {Curdt}, W. (2012{\natexlab{b}}).
\newblock {Spectroscopic Observations of Fe XVIII in Solar Active Regions}.
\newblock \emph{\apjl} 754, L40.
\newblock \doi{10.1088/2041-8205/754/2/L40}
\bibAnnoteFile{teriaca_etal:2012}

\bibitem[{{Testa} et~al.(2014){Testa}, {De Pontieu}, {Allred}, {Carlsson},
  {Reale}, {Daw} et~al.}]{testa_etal:2014}
{Testa}, P., {De Pontieu}, B., {Allred}, J., {Carlsson}, M., {Reale}, F.,
  {Daw}, A., et~al. (2014).
\newblock {Evidence of nonthermal particles in coronal loops heated impulsively
  by nanoflares}.
\newblock \emph{Science} 346, 1255724.
\newblock \doi{10.1126/science.1255724}
\bibAnnoteFile{testa_etal:2014}

\bibitem[{{Testa} and {Reale}(2020)}]{testa_reale:2020}
{Testa}, P. and {Reale}, F. (2020).
\newblock {On the Coronal Temperature in Solar Microflares}.
\newblock \emph{\apj} 902, 31.
\newblock \doi{10.3847/1538-4357/abb36e}
\bibAnnoteFile{testa_reale:2020}

\bibitem[{{Tian} et~al.(2014){Tian}, {Li}, {Reeves}, {Raymond}, {Guo}, {Liu}
  et~al.}]{tian_etal:2014ApJ...797L..14T}
{Tian}, H., {Li}, G., {Reeves}, K.~K., {Raymond}, J.~C., {Guo}, F., {Liu}, W.,
  et~al. (2014).
\newblock {Imaging and Spectroscopic Observations of Magnetic Reconnection and
  Chromospheric Evaporation in a Solar Flare}.
\newblock \emph{\apjl} 797, L14.
\newblock \doi{10.1088/2041-8205/797/2/L14}
\bibAnnoteFile{tian_etal:2014ApJ...797L..14T}

\bibitem[{{Tr{\"a}bert} et~al.(2014){Tr{\"a}bert}, {Beiersdorfer},
  {Brickhouse}, and {Golub}}]{traebert_etal:2014_131}
{Tr{\"a}bert}, E., {Beiersdorfer}, P., {Brickhouse}, N.~S., and {Golub}, L.
  (2014).
\newblock {High-resolution Laboratory Spectra on the {$\lambda$}131 Channel of
  the AIA Instrument On Board the Solar Dynamics Observatory}.
\newblock \emph{\apjs} 211, 14.
\newblock \doi{10.1088/0067-0049/211/1/14}
\bibAnnoteFile{traebert_etal:2014_131}

\bibitem[{{Vadawale} et~al.(2021){Vadawale}, {Mithun}, and
  {Mondal}}]{xsm_paper}
{Vadawale}, S., {Mithun}, N., and {Mondal}, a., B.~{et} (2021).
\newblock {Non-Active Region Microflares and Abundances in Quietest Solar
  Corona of the Space Age}.
\newblock \emph{ApJ}
\bibAnnoteFile{xsm_paper}

\bibitem[{{van Ballegooijen} et~al.(2011){van Ballegooijen}, {Asgari-Targhi},
  {Cranmer}, and {DeLuca}}]{van_ballegooijen_etal:2011}
{van Ballegooijen}, A.~A., {Asgari-Targhi}, M., {Cranmer}, S.~R., and {DeLuca},
  E.~E. (2011).
\newblock {Heating of the Solar Chromosphere and Corona by Alfv{\'e}n Wave
  Turbulence}.
\newblock \emph{\apj} 736, 3.
\newblock \doi{10.1088/0004-637X/736/1/3}
\bibAnnoteFile{van_ballegooijen_etal:2011}

\bibitem[{{Warren}(2014)}]{warren:2014}
{Warren}, H.~P. (2014).
\newblock {Measurements of Absolute Abundances in Solar Flares}.
\newblock \emph{\apjl} 786, L2.
\newblock \doi{10.1088/2041-8205/786/1/L2}
\bibAnnoteFile{warren:2014}

\bibitem[{{Warren} et~al.(2018){Warren}, {Brooks}, {Ugarte-Urra}, {Reep},
  {Crump}, and {Doschek}}]{warren_etal:2018}
{Warren}, H.~P., {Brooks}, D.~H., {Ugarte-Urra}, I., {Reep}, J.~W., {Crump},
  N.~A., and {Doschek}, G.~A. (2018).
\newblock {Spectroscopic Observations of Current Sheet Formation and
  Evolution}.
\newblock \emph{\apj} 854, 122.
\newblock \doi{10.3847/1538-4357/aaa9b8}
\bibAnnoteFile{warren_etal:2018}

\bibitem[{{Warren} et~al.(2008){Warren}, {Feldman}, and
  {Brown}}]{warren_etal:08}
{Warren}, H.~P., {Feldman}, U., and {Brown}, C.~M. (2008).
\newblock {Solar Observations of High-Temperature Emission with the
  Extreme-Ultraviolet Imaging Spectrometer on Hinode}.
\newblock \emph{\apj} 685, 1277--1285.
\newblock \doi{10.1086/591075}
\bibAnnoteFile{warren_etal:08}

\bibitem[{Windt and Gullikson(2015)}]{windt_gullikson:2015}
Windt, D.~L. and Gullikson, E.~M. (2015).
\newblock Pd/b4c/y multilayer coatings for extreme ultraviolet applications
  near 10 nm wavelength.
\newblock \emph{Appl. Opt.} 54, 5850--5860.
\newblock \doi{10.1364/AO.54.005850}
\bibAnnoteFile{windt_gullikson:2015}

\bibitem[{{Winebarger} et~al.(2012){Winebarger}, {Warren}, {Schmelz},
  {Cirtain}, {Mulu-Moore}, {Golub} et~al.}]{winebarger_etal:2012}
{Winebarger}, A.~R., {Warren}, H.~P., {Schmelz}, J.~T., {Cirtain}, J.,
  {Mulu-Moore}, F., {Golub}, L., et~al. (2012).
\newblock {Defining the ``Blind Spot'' of Hinode EIS and XRT Temperature
  Measurements}.
\newblock \emph{\apjl} 746, L17.
\newblock \doi{10.1088/2041-8205/746/2/L17}
\bibAnnoteFile{winebarger_etal:2012}

\bibitem[{{Wolfson} et~al.(1983){Wolfson}, {Leibacher}, {Doyle}, and
  {Phillips}}]{wolfson_etal:1983ApJ...269..319W}
{Wolfson}, C.~J., {Leibacher}, J.~W., {Doyle}, J.~G., and {Phillips}, K.~J.~H.
  (1983).
\newblock {X-ray line ratios from helium-like ions - Updated theory and SMM
  flare observations}.
\newblock \emph{\apj} 269, 319--328.
\newblock \doi{10.1086/161045}
\bibAnnoteFile{wolfson_etal:1983ApJ...269..319W}

\bibitem[{{Woods} et~al.(2012){Woods}, {Eparvier}, {Hock}, {Jones},
  {Woodraska}, {Judge} et~al.}]{woods_etal:12}
{Woods}, T.~N., {Eparvier}, F.~G., {Hock}, R., {Jones}, A.~R., {Woodraska}, D.,
  {Judge}, D., et~al. (2012).
\newblock {Extreme Ultraviolet Variability Experiment (EVE) on the Solar
  Dynamics Observatory (SDO): Overview of Science Objectives, Instrument
  Design, Data Products, and Model Developments}.
\newblock \emph{\solphys} 275, 115--143.
\newblock \doi{10.1007/s11207-009-9487-6}
\bibAnnoteFile{woods_etal:12}

\bibitem[{{Young} et~al.(2007){Young}, {Del Zanna}, {Mason}, {Dere}, {Landi},
  {Landini} et~al.}]{young_etal:07a}
{Young}, P.~R., {Del Zanna}, G., {Mason}, H.~E., {Dere}, K.~P., {Landi}, E.,
  {Landini}, M., et~al. (2007).
\newblock {EUV emission lines and diagnostics observed with Hinode/EIS}.
\newblock \emph{PASJ} 59, 857
\bibAnnoteFile{young_etal:07a}

\bibitem[{{Young} et~al.(2013){Young}, {Doschek}, {Warren}, and
  {Hara}}]{young_etal:2013_flare}
{Young}, P.~R., {Doschek}, G.~A., {Warren}, H.~P., and {Hara}, H. (2013).
\newblock {Properties of a Solar Flare Kernel Observed by Hinode and SDO}.
\newblock \emph{\apj} 766, 127.
\newblock \doi{10.1088/0004-637X/766/2/127}
\bibAnnoteFile{young_etal:2013_flare}

\bibitem[{{Young} et~al.(2015){Young}, {Tian}, and {Jaeggli}}]{young_etal:2015}
{Young}, P.~R., {Tian}, H., and {Jaeggli}, S. (2015).
\newblock {The 2014 March 29 X-flare: Subarcsecond Resolution Observations of
  Fe XXI {$\lambda$}1354.1}.
\newblock \emph{\apj} 799, 218.
\newblock \doi{10.1088/0004-637X/799/2/218}
\bibAnnoteFile{young_etal:2015}

\bibitem[{{Zhitnik} et~al.(2003){Zhitnik}, {Bugaenko}, {Ignat'ev}, {Krutov},
  {Kuzin}, {Mitrofanov} et~al.}]{zhitnik_etal:03}
{Zhitnik}, I.~A., {Bugaenko}, O.~I., {Ignat'ev}, A.~P., {Krutov}, V.~V.,
  {Kuzin}, S.~V., {Mitrofanov}, A.~V., et~al. (2003).
\newblock {Dynamic 10 MK plasma structures observed in monochromatic full-Sun
  images by the SPIRIT spectroheliograph on the CORONAS-F mission}.
\newblock \emph{\mnras} 338, 67--71.
\newblock \doi{10.1046/j.1365-8711.2003.06014.x}
\bibAnnoteFile{zhitnik_etal:03}

\end{thebibliography}



\begin{table}[!htbp]
\begin{center}
  \caption{List of the strongest lines
    in the XUV for a 5--10 MK plasma.
    Radiances  (in photons cm$^{-2}$ s$^{-1}$ arcsec$^{-2}$)
are for a weak 10 MK plasma (I$_{\rm 10}$), a nanoflare simulation
(I$_{\rm n}$), a  0.005 A-class  microflare (I$_{\rm mf1}$) at 6.7 MK,
and an A-class flare  at 8 MK (I$_{\rm mf2}$).
Primary SXR lines  are noted with *.
We also list total
DN/s per 1\arcsec\ pixel for the SXR straw-man design.
The notes indicate when a line is `hot', i.e. only formed above 5 MK, and
if a line can provide a density-diagnostic ($N_{\rm e}$),
when observed together with the resonance  line of an ion.
(sbl) indicates a self-blend of transitions from the same ion,
(bl) a blend.
\label{tab:list}
}
\small
\begin{tabular}{lrrrrrrrrrl}
\hline\noalign{\smallskip}
Ion &$\lambda$ (\AA) & I$_{\rm 10}$ & I$_{\rm 10}$ & I$_{\rm n}$  & I$_{\rm n}$ &  I$_{\rm mf1}$ & I$_{\rm mf1}$ & I$_{\rm mf2}$ & I$_{\rm mf2}$ & Notes \\
    &               &             & DN/s        &            & DN/s       &              &  DN/s        &            & DN/s &               \\
\noalign{\smallskip}\hline\noalign{\smallskip}

\ion{Si}{xiii} & 6.65  &  1.4$\times$10$^{-2}$ & & 2.7 & &  0.3 & & 36 & &   \\
\ion{Mg}{xii} & 8.42  &  1.4$\times$10$^{-2}$ & &  2.4 & &   0.3 & &  37 & &   \\
\ion{Mg}{xi} & 9.17  &  1.1$\times$10$^{-2}$ & &  14 & &    0.8 & &   57 & &  \\

\ion{Fe}{xxi} & 12.28  &  4.6$\times$10$^{-2}$ & &  0.5 & &   0.05 & &  22 & &  hot weak (bl) \\

\ion{Fe}{xx} & 12.83  &  2.9$\times$10$^{-2}$  & &  0.6 & &   0.25 & &  68 & &   hot weak (sbl) \\

\ion{Fe}{xix} & 13.52  &  3.4$\times$10$^{-2}$ & &  2.5 & &   0.8 & &  136 & &   hot weak (sbl) \\

\ion{Fe}{xviii} & 14.20   &  7.8$\times$10$^{-2}$ & &  31 & &  4  & &   380 & &  (sbl)  \\

\ion{Fe}{xvii} & 15.01  &  6$\times$10$^{-2}$  & &  246 & &   9.1 & &  555 & &   \\


\noalign{\smallskip}\hline\noalign{\smallskip}

\ion{Fe}{xx} & 93.78  &  0.03 & &   0.8 & &   0.2 & &  65 & &  hot weak  \\

\ion{Fe}{xviii} & 93.93  &  0.2 & &  186 & &   15 & &   1210 & &  (bl)  \\

\ion{Fe}{xxi} & 97.86  &  0.04 & &   0.5 & &   0.08 & &  37 & &    \\


\ion{Fe}{xix} & 101.55  &  0.11 & &  14  & &  3 & &  381 & &   hot \\

\ion{Fe}{xxi} & 102.22  &  0.09  & &  1.1 & &   0.2 & &  87 & &   hot weak $N_{\rm e}$ \\

* \ion{Fe}{xviii} & 103.95  & 0.07 & 0.3 & 68 & 260  & 5.3 & 20 & 442 & 1.6$\times$10$^{3}$ &  \\

* \ion{Fe}{xxi} & 108.12   &  0.02 & 0.08 & 0.3 & 1.1 &  0.04 & 0.2 & 19 & 64  & hot weak $N_{\rm e}$ \\

* \ion{Fe}{xix} & 108.36  &  0.32 & 1.4 & 46 & 192 &  7.7 & 33 &  1030 & 4.4$\times$10$^{3}$ &  hot \\

* \ion{Fe}{xix} & 109.95  &  5.9$\times$10$^{-2}$ & & 7.3 & 1 & 1.4 & 0.2 & 194 & 26 & hot weak $N_{\rm e}$  \\


\ion{Fe}{xxii} & 117.15  &  0.21 & &  1.9  & &   0.1 & &  84 & &   hot \\


\ion{Fe}{xx} & 118.68  &  0.15 & &  5.1  & &   1.1 & &  254 & &   hot  \\

* \ion{Fe}{xix} & 119.98  & 0.09 & 0.07  & 12 & 10 &   2.1 & 1.7 & 279 & 230 & hot  \\

* \ion{Fe}{xxi} & 121.21 & 0.01 & 0.03 & 0.1 & 0.5 &  0.03 & 0.1 & 48 & 190 & hot weak $N_{\rm e}$ \\

* \ion{Fe}{xx} & 121.85  &  0.3 & 1.5 & 10 & 48 &  2.3 & 11 & 500 & 2.4$\times$10$^{3}$ &  hot \\

* \ion{Fe}{xxi} & 128.75  &  0.5 & 4.8 & 7 & 63 &  1.0 & 9.5 & 322 & 3.0$\times$10$^{3}$& hot  \\


* \ion{Fe}{xix} & 132.62  & 0.02 & 0.17 & 0.2 & 25 &  0.4 & 4 &  59  & 550 &  hot weak \\

* \ion{Fe}{xx} & 132.84   & 0.42 & 4.3 & 14 & 140 & 3.1 & 32 & 701 & 7.0$\times$10$^{3}$ &  hot (bl) \\

* \ion{Fe}{xxiii}& 132.91 &  0.22 & 2.2 & 2.4 & 24 & 0.03 & 0.3 & 40 & 400 & hot (bl)   \\

* \ion{Fe}{xxii} & 135.81 & 0.2 & 1.4 & 1.9 & 13 &  0.1 & 0.17 & 82 & 570 &  hot   \\



* \ion{Fe}{xxi} & 142.20  & 0.05 & 0.3 & 0.6 & 4 &  0.15 & 0.7 & 53  & 320 & hot weak $N_{\rm e}$ (sbl) \\

* \ion{Fe}{xxi} & 145.73 & 0.005 & 0.02 & 0.06 & 0.2 & 0.02  & 0.07& 37  & 120 & hot weak $N_{\rm e}$ \\

\noalign{\smallskip}\hline\noalign{\smallskip}

\ion{Ca}{xvii} & 192.85  &  0.04 & &  102 & &  6 & &  288 & &   (bl)  \\

\ion{Fe}{xxiii} & 263.76  &  0.02 & &  0.2 & &  0.002 & &  3.7  & &  hot weak   \\

\ion{Ca}{xviii} & 302.19  & 0.08 & &  32 & &  4.4 & &  337 & &   hot weak \\

\ion{Fe}{xx} & 384.21  &  0.03 & &  1.0 & &  0.2 & &  57 & &   hot weak  \\

\ion{Fe}{xix} & 424.27  &  0.02 & &  2.8 & &  0.5 & &  70 & &  hot weak  \\

\ion{Si}{xii} & 499.41  &  0.18 & &  5600 & &  16 & &   1020 & &   \\


\ion{Fe}{xx} & 567.87  &  0.11 & &  3.3 & &  0.7 & &   133 & &   hot   \\

\ion{Fe}{xxi} & 585.77  &  0.02 & &   0.3 & &  0.06 & &   31 & &  hot weak \\

\ion{Fe}{xix} & 592.23  &  0.17 & &  25 & &  4 & &  554 & &   hot \\

\ion{Fe}{xx} & 721.56  &  0.22 & &  7 & &  1.5 & &  347 & &   hot \\

\ion{Fe}{xxi} & 786.16  & 0.02 & &  0.3 & &  0.06  & &  30 & &  hot weak \\

\ion{Fe}{xxii} & 845.57  &  0.15 & &  1.4 & &  0.07 & &  55 & &   hot weak  \\

\ion{Fe}{xviii} & 974.86  &  0.24 & &  225 & &  17 & &  1370 & &   \\

\ion{Fe}{xix} & 1118.06  &  0.31 & &  43 & &  7.6 & &   1173 & &  hot \\


\ion{Fe}{xxi} & 1354.07  &  0.62 & &  8.1 & &  1.2 & &  369 & &  hot  \\

\end{tabular}
\end{center}
\end{table}

\end{document}


\onecolumn
\firstpage{1}

\title[Supplementary Material]{{\helveticaitalic{Supplementary Material}}}

\maketitle

\section{Observed and predicted SXR radiances in the quiet Sun}

\begin{figure*}[!htbp]
  \centerline{\includegraphics[width=16cm, angle=0]{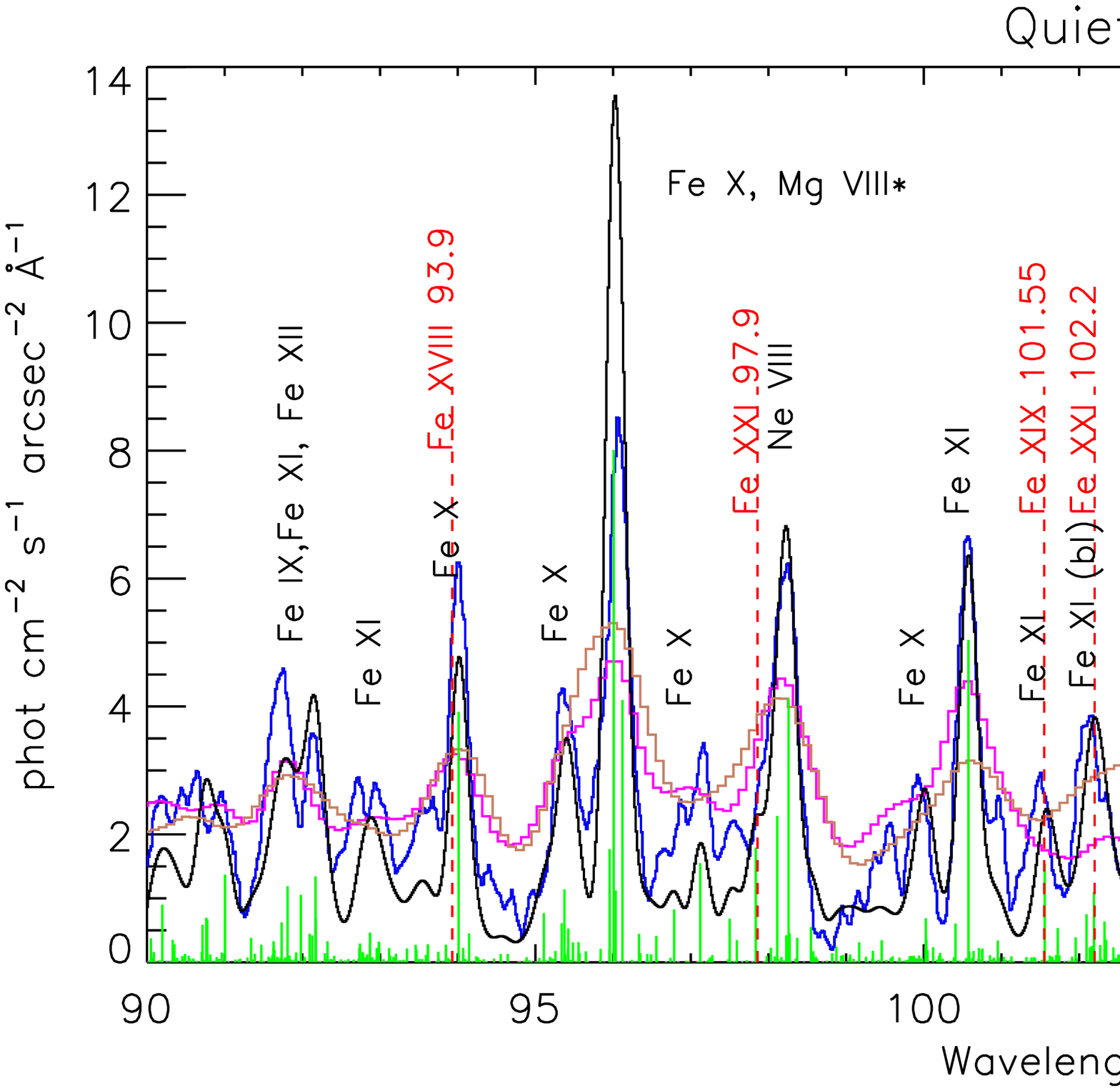}}
   \centerline{\includegraphics[width=16cm, angle=0]{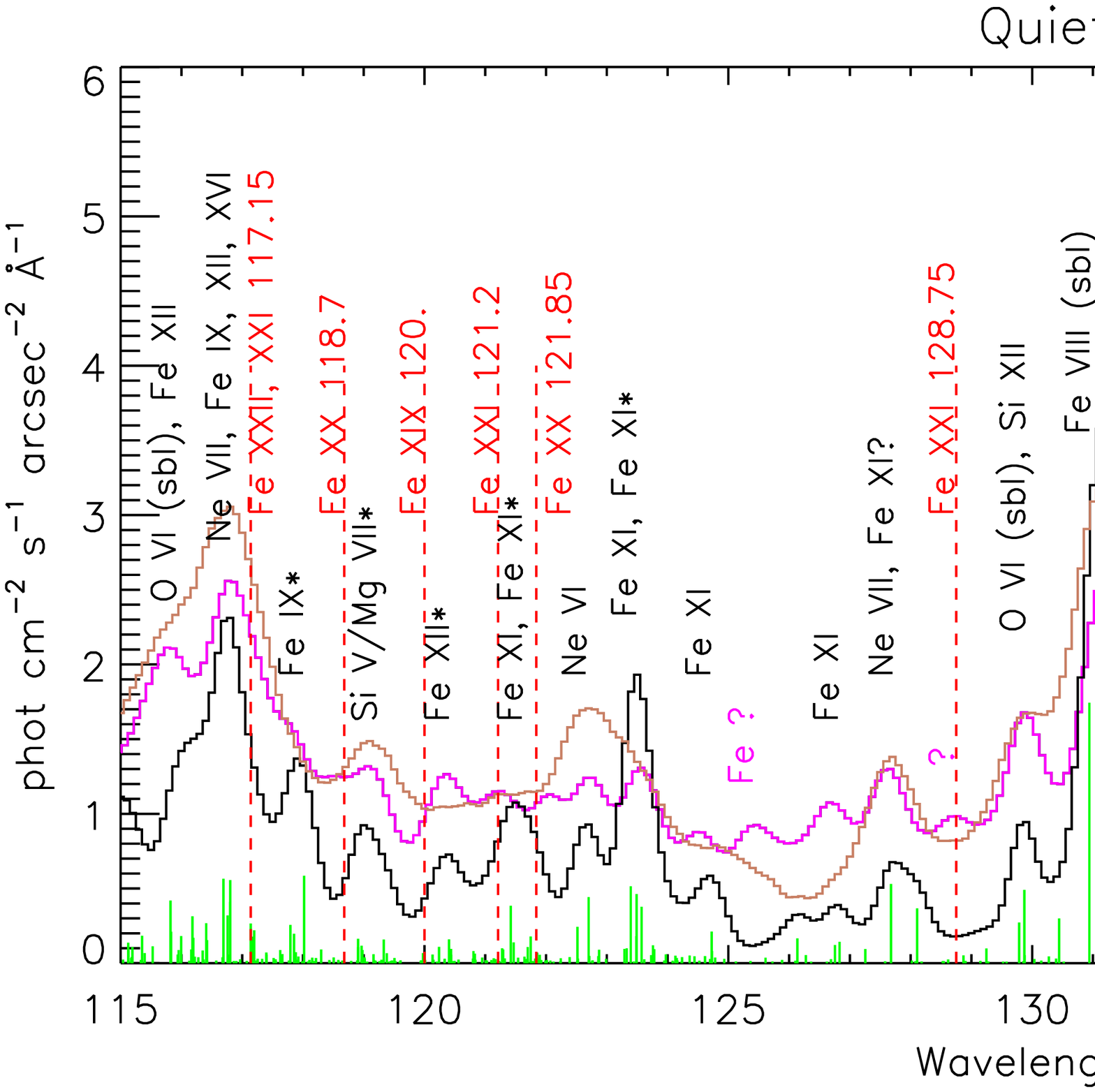}}
   \caption{Predicted SXR quiet Sun radiances, obtained from the 
irradiance spectra of 
\cite{manson:72}, \cite{malinovsky_heroux:73}, and prototype  EVE \citep{woods_etal:09},
with over-plotted a simulated quiet Sun
CHIANTI  spectrum (black).
The locations of the high-$T$ lines are shown in red.
The main ions contributing to the CHIANTI  lines are
labelled, while all those 
contributing to the CHIANTI spectrum are shown in green.
A few observed lines are missing in CHIANTI. Conversely,
a few CHIANTI lines are not at the correct wavelengths.
Those unidentified are labelled with an asterisk. 
}
\label{fig:qs} 
 \end{figure*}

The SXR lines are clearly visible
in irradiance spectra only for larger flares, due to the background radiation
of the solar disk. To assess the visibility of the hot lines
in active region cores, one would need SXR 
radiance measurements,  which are not available yet. 
We therefore need to rely on simulated spectra.
As we have mentioned,
the atomic data and line identifications in the whole SXR spectral region
are not complete, so we do have a problem.

We therefore show in this Section a benchmark of the latest CHIANTI v.10
atomic data against two of the best
 soft X-ray medium-resolution full-Sun spectra,  obtained
with rocket flights in the 1960s:
the quiet Sun spectrum from \cite{manson:72}, available up to 115~\AA, 
and the lower-resolution \cite{malinovsky_heroux:73}
irradiance spectrum. Both spectra were obtained from the
published plots, and 
radiometrically recalibrated by \cite{delzanna:12_sxr1} using as a reference the
2008 April 14 irradiances from the prototype  EVE instrument \citep{woods_etal:09},
when the Sun was at solar minimum.

We have converted the spectra to  radiances assuming a limb-brightening increase of
the average center-Sun radiances of a factor of 1.3 \citep{andretta_delzanna:2014}.
The spectral resolution of the  \cite{manson:72} spectra was about 0.2~\AA\ (FWHM).
The Sun probably had some flare
emission during the flight of the \cite{malinovsky_heroux:73} spectrum, as  significant emission in  
\ion{Fe}{xviii} (94, 104~\AA) was recorded. However, the bulk of the spectrum originated from
the quiet Sun, as in the \cite{manson:72} case. 

Fig.~\ref{fig:qs} shows a visual summary, with the observed spectra and
the main lines in the two key SXR regions discussed here.
There are some discrepancies between the \cite{manson:72}  and
\cite{malinovsky_heroux:73} spectra, with the latter generally agreeing
better with the lower-resolution PEVE.

Fig.~\ref{fig:qs}  also shows a completely independent quiet Sun (QS)
simulated spectrum, obtained from CHIANTI v.10, a $DEM$ obtained by 
\cite{andretta_etal:03} from SoHO CDS observations, and photospheric abundances.
An instrumental FWHM of 0.25~\AA\ was adopted for the 90--115~\AA\
spectral range, while a FWHM of 0.5~\AA\ was chosen for the
comparison at longer wavelengths.

In some spectral regions, surprisingly good agreement between
predicted and observed radiances is found. 
Further improvements will be achieved when the positions of the
many unidentified lines is adjusted.
 The unidentified lines are noted with an asterisk in the Figure.

 For the line identifications we have considered the \cite{behring_etal:72}
line list, the literature from EBIT plasma
\citep[see, e.g.][]{lepson_etal:2002,traebert_etal:2014_131},
the  \cite{delzanna:12_sxr1} identifications, B. Fawcett's plates for the iron ions,
and other sources such as various
compilations of laboratory measurements.
We have also considered stellar spectra such as those discussed in
\cite{beiersdorfer_etal:2014_procyon}, but note that
the  \cite{manson:72} solar spectra are far superior. 

Clearly, the atomic data are still not complete, as discussed in detail in
\cite{delzanna:12_sxr1}.   A few notable ions such as
\ion{Ni}{x}, still missing in CHIANTI, are noted.
The key ions present in  CHIANTI are marked in Fig.~\ref{fig:qs}, although we note that
many transitions (indicated with vertical green lines) generally contribute to
the spectrum, at such medium resolution.
Fig.~\ref{fig:qs} also shows in red the locations of the main hot lines,
which mostly fall  in regions relatively free of 'background' QS lines.

\section{Predicted SXR radiances in an active region}

\begin{figure*}[!htbp]
  \centerline{\includegraphics[width=16cm, angle=0]{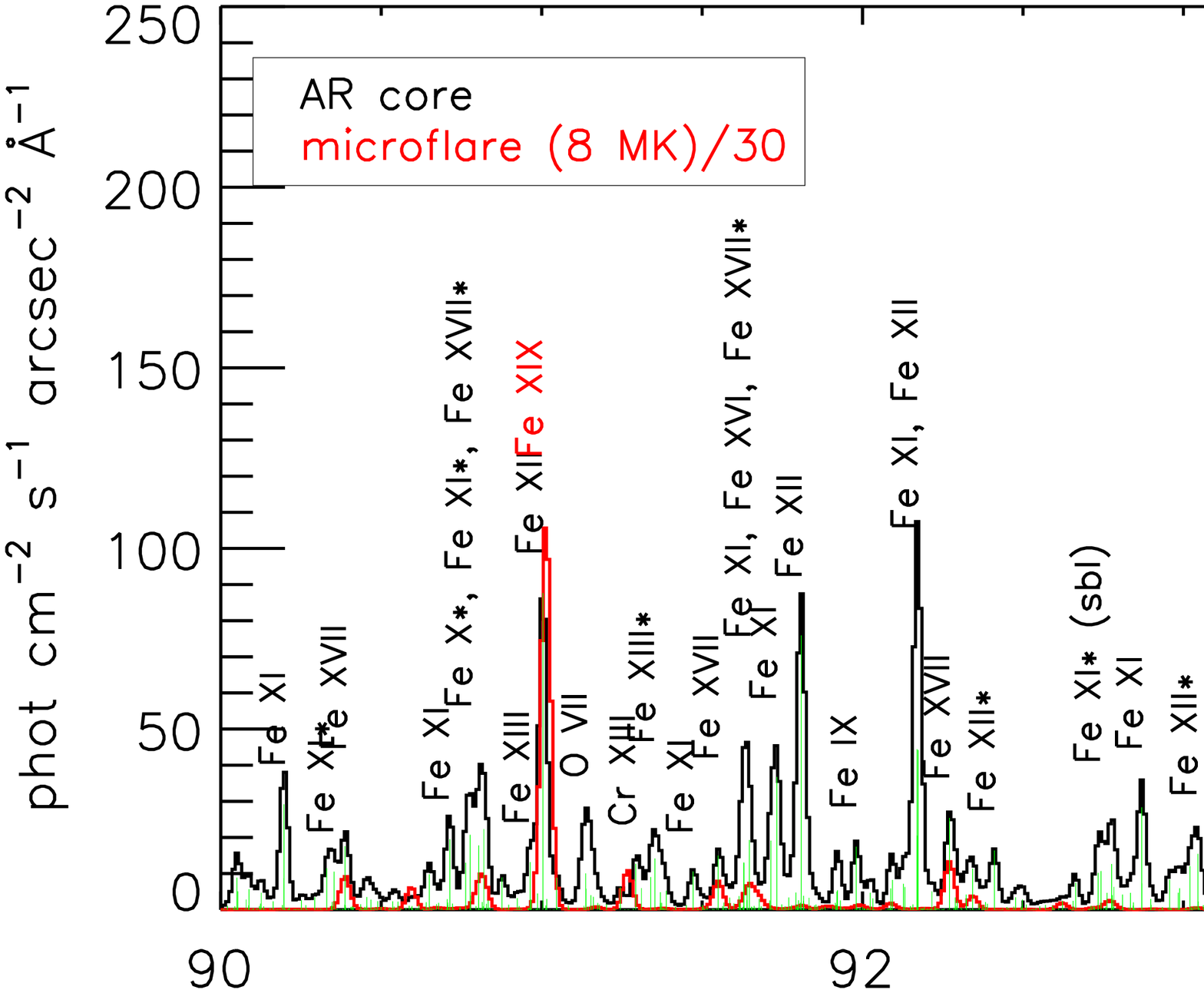}}
  \centerline{\includegraphics[width=16cm, angle=0]{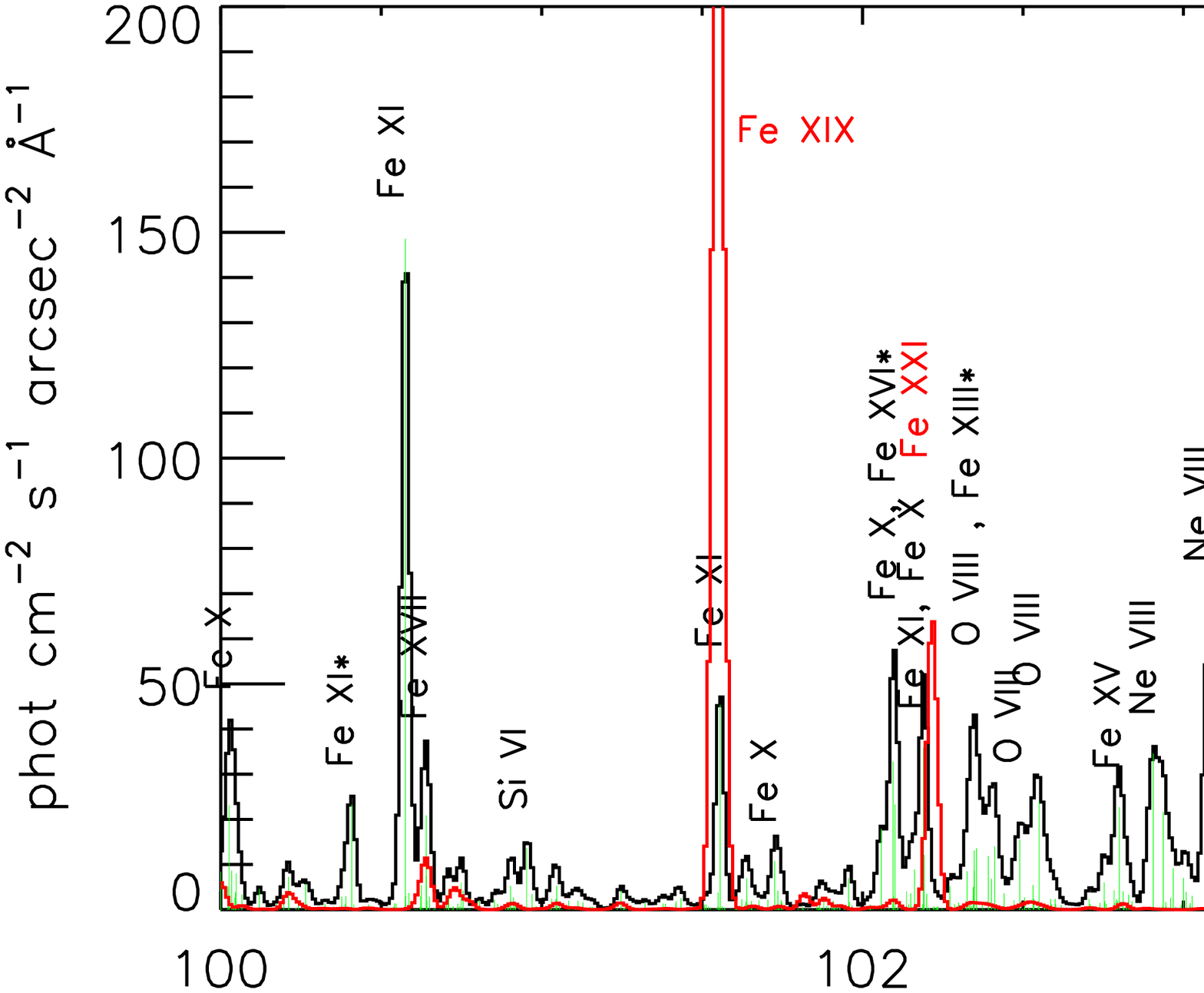}}
  \caption{Simulated  CHIANTI v.10 SXR spectra (90--110~\AA) of the core of an
    active region, with over-plotted  the spectrum of an A-class 8 MK  microflare,
    reduced by a factor of 30.   The  lines
    contributing to the spectrum are shown in green, while the main  ions
 are labelled. The unidentified lines are noted with an
asterisk. The pixel size is 0.01~\AA\ and the instrumental width is
0.025~\AA.
}
\label{fig:ar_core1} 
 \end{figure*}

\begin{figure*}[!htbp]
  \centerline{\includegraphics[width=16cm, angle=0]{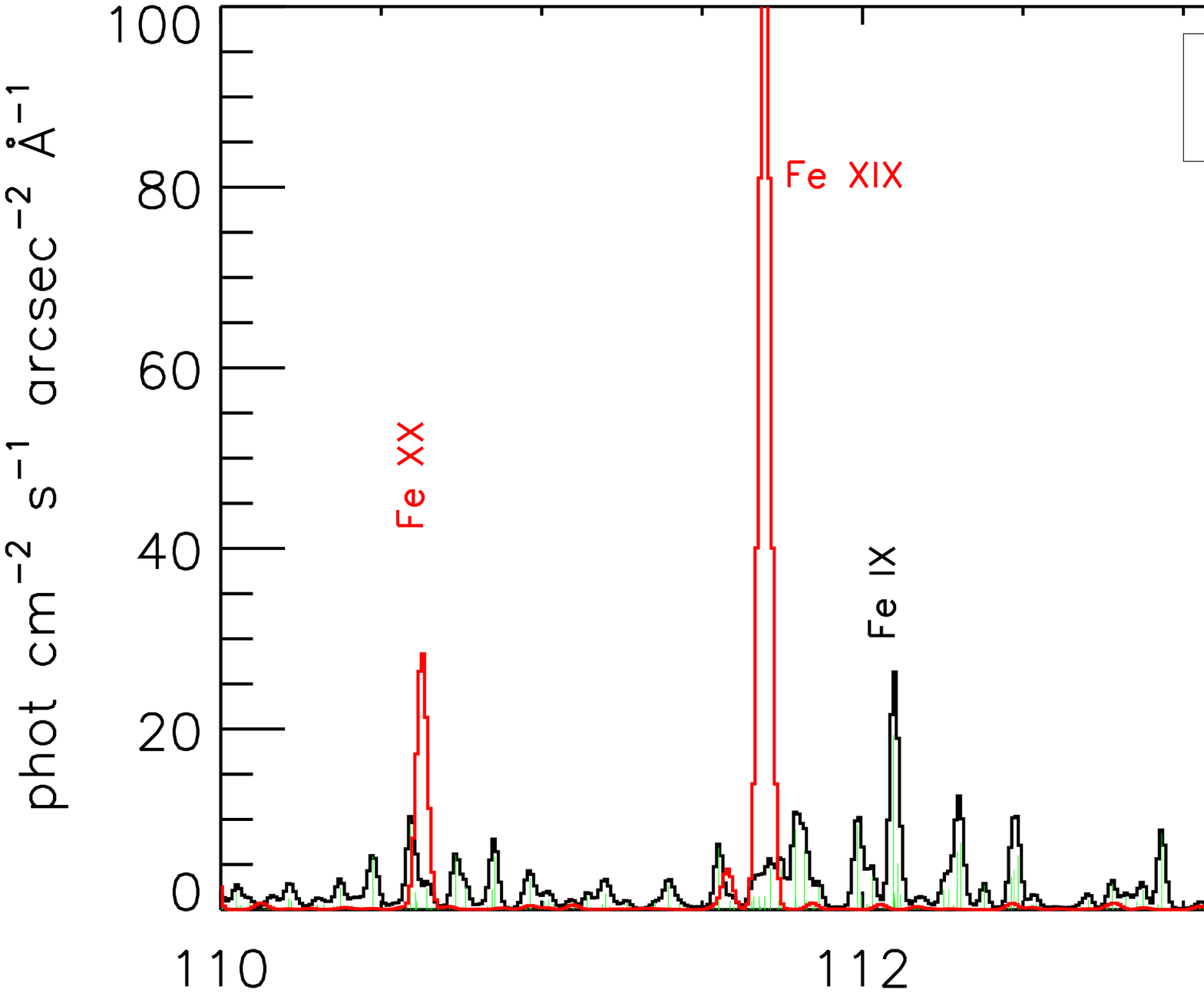}}
  \centerline{\includegraphics[width=16cm, angle=0]{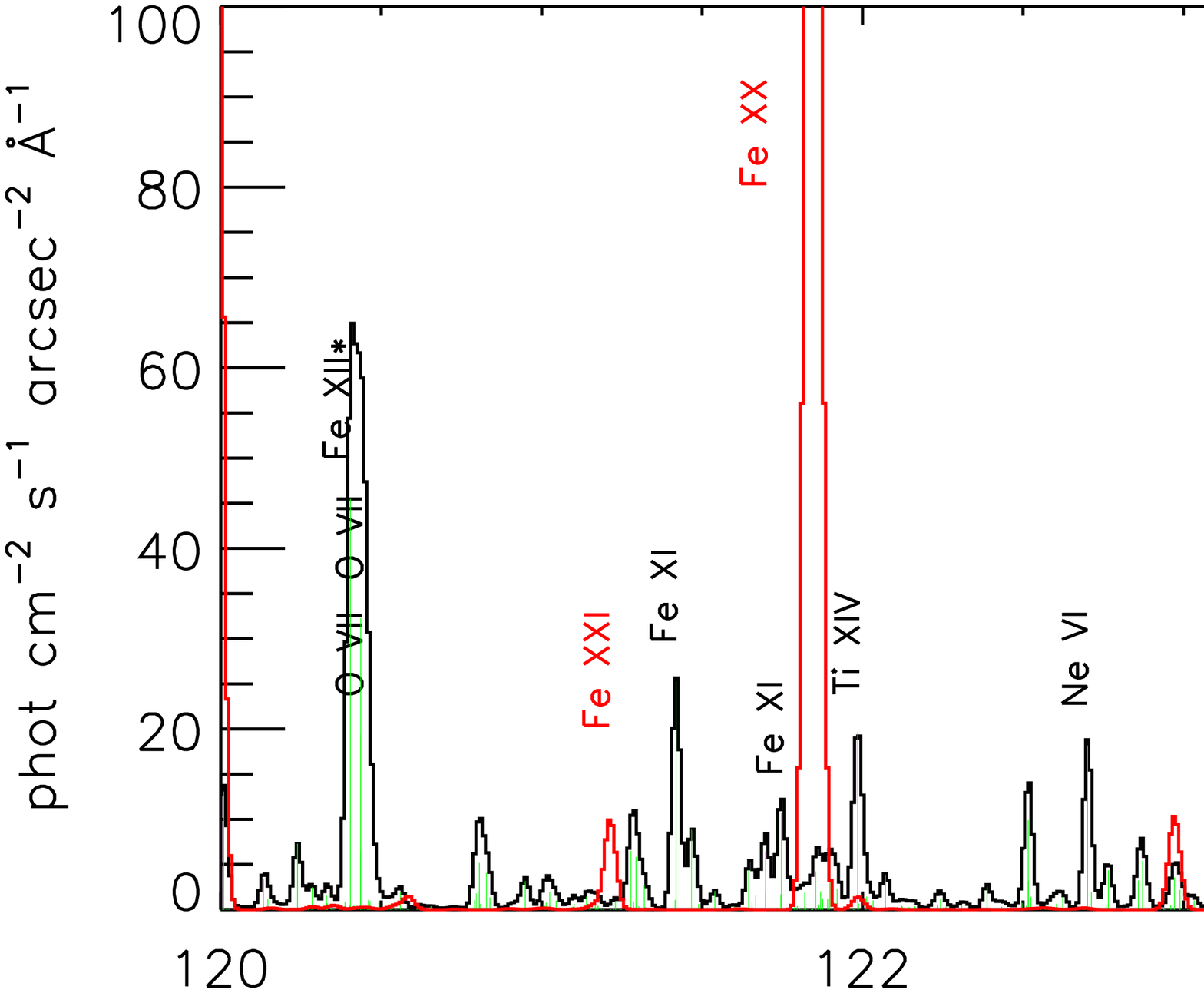}}
  \centerline{\includegraphics[width=16cm, angle=0]{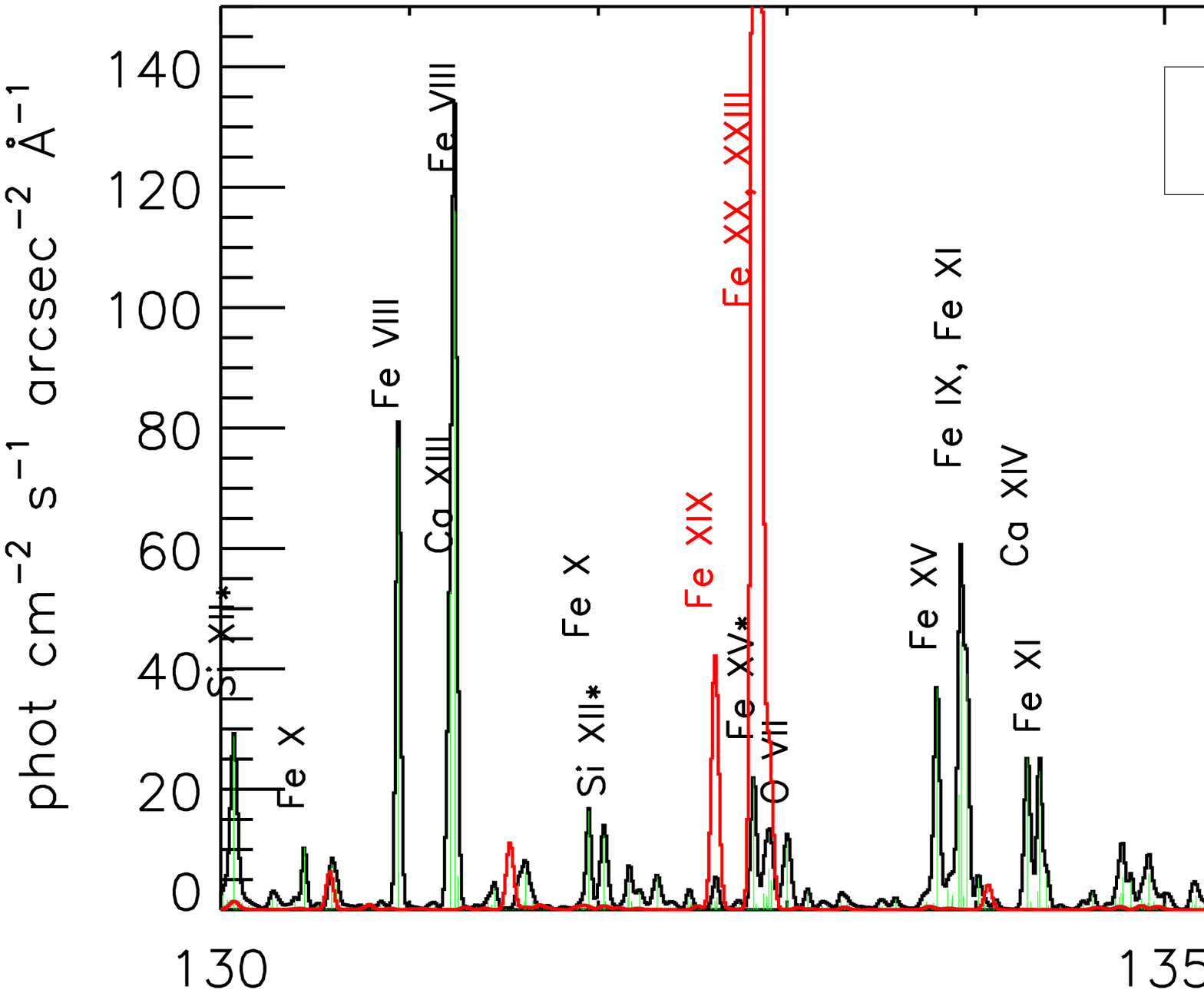}}
  \caption{Same as Fig.~\ref{fig:ar_core1} for the 110--147~\AA\ spectral region.
}
\label{fig:ar_core2} 
 \end{figure*}

The  comparison in Fig.~\ref{fig:qs} gives us some
confidence in presenting simulations of active region core spectra, in 
Figures~\ref{fig:ar_core1},\ref{fig:ar_core2}.  We have used the $DEM$ and coronal abundances
of an active region quiescent 3 MK loop, 
described in \cite{delzanna:2013_multithermal} and
shown in Figure~9 in that paper.
We adopted, following our straw-man design, a pixel size
of 0.01~\AA\ and an  instrumental width of 0.025~\AA.

With the increased iron  abundances by a factor of 3.2,
there is an increased signal in most  lines, compared to the quiet Sun case,
which had photospheric abundances.
Also, there is increased emission in 2--3 MK  spectral lines, especially in 
the  \ion{Fe}{xviii} 104~\AA\ line 
and several \ion{Fe}{xvi},  \ion{Cr}{xvi}, \ion{Si}{xii},
\ion{Ca}{xii}, \ion{Ca}{xiv}, and  \ion{Ca}{xv} lines.
The DEM was not constrained at 5-10 MK temperatures, and predicts a
weak emission  in the \ion{Fe}{xix} lines, which may or may not be present
in actual AR spectra. On the other hand, we expect some emission
from the \ion{Fe}{xviii} lines to always be present in AR observations.

A significant number of the weaker transitions are still not identified,
hence their wavelengths could be off by 1~\AA\ or more. They are noted with an
asterisk in Figures~\ref{fig:ar_core1},\ref{fig:ar_core2}.

There are dozens of cool lines in the SXR, from ions 
such as \ion{Mg}{v}, \ion{Mg}{vi}, \ion{Mg}{vii}, \ion{Mg}{viii},
\ion{O}{vi}, \ion{O}{vii},  \ion{O}{viii},
\ion{Si}{v}, \ion{Si}{vi}, \ion{Si}{vii}, \ion{Ne}{v}, \ion{Ne}{vi}, \ion{Ne}{vii}, \ion{Ne}{viii},
plus all ionisation stages of iron, from  \ion{Fe}{viii}.

The Figures also over-plot in red the simulated
A-class 8 MK  microflare spectrum, reduced by a factor of 30.
It is clear that most of the hot lines fall in region that are expected to be
relatively free of blends. The \ion{Fe}{xviii} and \ion{Fe}{xix}
microflare emission would be superimposed on the weak AR emission in the
lines from these ions.
A few hot lines are instead falling in regions where significant coronal
`background' emission is predicted to be present. 
However, we know from our previous Hinode EIS studies
\citep[see, e.g.][]{delzanna_etal:2011_flare,mitra-kraev_delzanna:2019}
that the lower-temperature coronal lines have little variations
during the peak of a flare, so we expect that even the weakest
hot lines will be easily detectable against the background.

\newcommand\aj{{AJ}}%
\newcommand\araa{{ARA\&A}}%
\newcommand\apj{{ApJ}}%
\newcommand\apjl{{ApJ}}%
\newcommand\apjs{{ApJS}}%
\newcommand\ao{{Appl.~Opt.}}%
\newcommand\apss{{Ap\&SS}}%
\newcommand\aap{{A\&A}}%
\newcommand\aapr{{A\&A~Rev.}}%
\newcommand\aaps{{A\&AS}}%
\newcommand\azh{{AZh}}%
\newcommand\baas{{BAAS}}%
\newcommand\jrasc{{JRASC}}%
\newcommand\memras{{MmRAS}}%
\newcommand\mnras{{MNRAS}}%
\newcommand\pra{{Phys.~Rev.~A}}%
\newcommand\prb{{Phys.~Rev.~B}}%
\newcommand\prc{{Phys.~Rev.~C}}%
\newcommand\prd{{Phys.~Rev.~D}}%
\newcommand\pre{{Phys.~Rev.~E}}%
\newcommand\prl{{Phys.~Rev.~Lett.}}%
\newcommand\pasp{{PASP}}%
\newcommand\pasj{{PASJ}}%
\newcommand\qjras{{QJRAS}}%
\newcommand\skytel{{S\&T}}%
\newcommand\solphys{{Sol.~Phys.}}%
\newcommand\sovast{{Soviet~Ast.}}%
\newcommand\ssr{{Space~Sci.~Rev.}}%
\newcommand\zap{{ZAp}}%
\newcommand\nat{{Nature}}%
\newcommand\iaucirc{{IAU~Circ.}}%
\newcommand\aplett{{Astrophys.~Lett.}}%
\newcommand\apspr{{Astrophys.~Space~Phys.~Res.}}%
\newcommand\bain{{Bull.~Astron.~Inst.~Netherlands}}%
\newcommand\fcp{{Fund.~Cosmic~Phys.}}%
\newcommand\gca{{Geochim.~Cosmochim.~Acta}}%
\newcommand\grl{{Geophys.~Res.~Lett.}}%
\newcommand\jcp{{J.~Chem.~Phys.}}%
\newcommand\jgr{{J.~Geophys.~Res.}}%
\newcommand\jqsrt{{J.~Quant.~Spec.~Radiat.~Transf.}}%
\newcommand\memsai{{Mem.~Soc.~Astron.~Italiana}}%
\newcommand\nphysa{{Nucl.~Phys.~A}}%
\newcommand\physrep{{Phys.~Rep.}}%
\newcommand\physscr{{Phys.~Scr}}%
\newcommand\planss{{Planet.~Space~Sci.}}%
\newcommand\procspie{{Proc.~SPIE}}%


\bibliographystyle{frontiersinSCNS_ENG_HUMS} 

\bibliography{SupplementaryMaterial}